\documentclass[11pt]{article}
\usepackage{graphicx}
\usepackage{amsmath}
\usepackage{amsfonts}
\usepackage{latexsym}
\usepackage{amsthm}
\usepackage{bm}
\usepackage[left]{lineno}
\usepackage{microtype}
\usepackage{parskip}
\usepackage{hyperref}
\usepackage{color}
\usepackage{mathrsfs}
\usepackage{cite}
\usepackage{pdfpages} 

\newcommand{\Pgg}{\ensuremath{\gamma}}

\newcommand {\ptg} {\ensuremath{p_{\mathrm{T}}^{\Pgg}}}

\providecommand{\ptg}{\ensuremath{p_{\mathrm{T},\Pgg}}}

\newcommand{\blue}[1]{{\color{blue}#1}}
\newcommand{\fp}[1]{{\textcolor{black}{#1}}} 

\newcommand{\qs}{\ensuremath{Q_{\mathrm{s}}}}
\newcommand{\qssqr}{\ensuremath{Q_{\mathrm{s}}^2}}
\newcommand{\as}{\ensuremath{\alpha_{\mathrm{s}}}}

\begin{document}
\pagenumbering{gobble}
\includepdf[offset=1.8cm -1.9cm,pages={1,{}},width=1.3\textwidth]{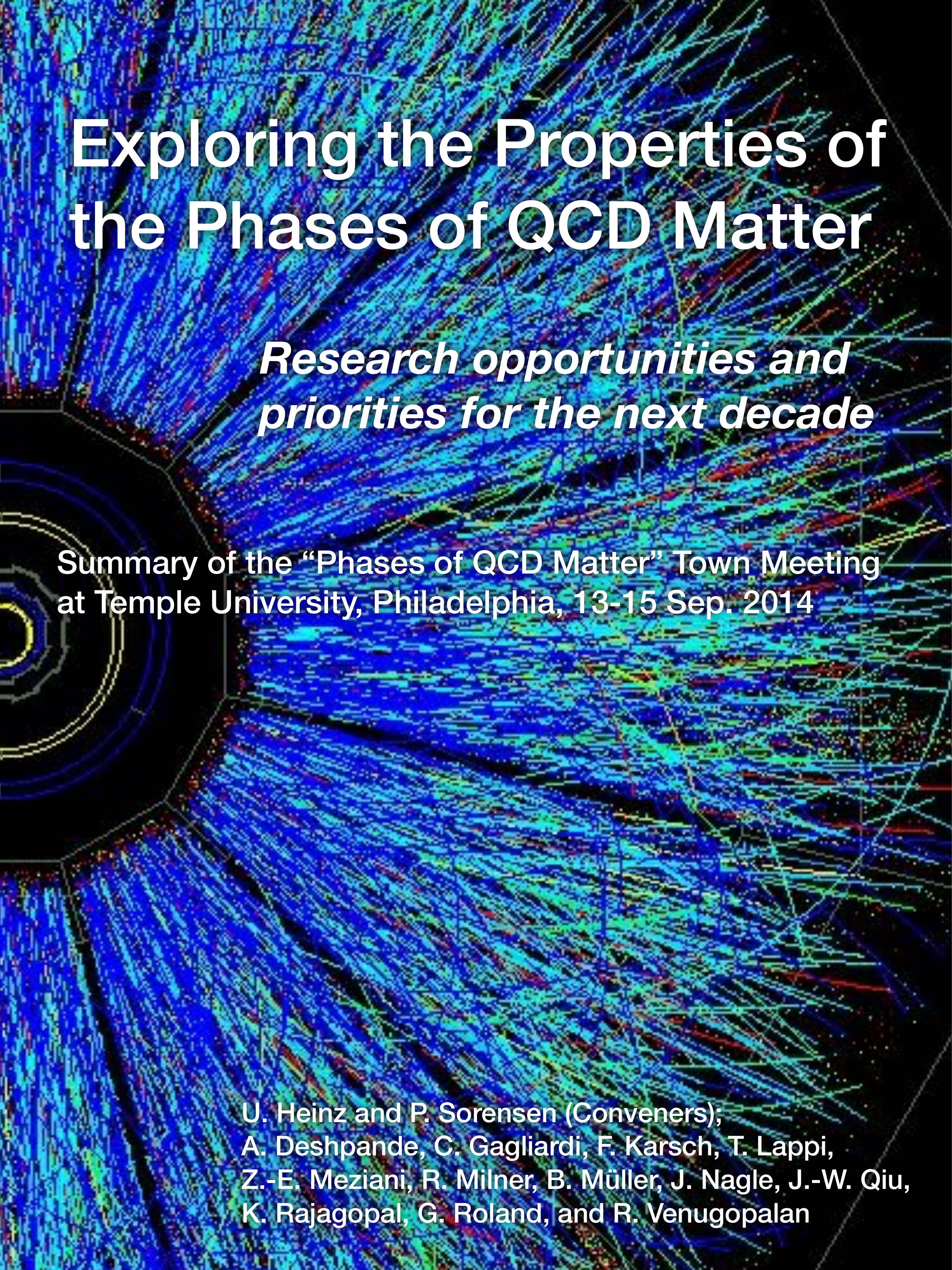}

\pagenumbering{arabic}
\setcounter{page}{1}
\centerline{\bf \Large Exploring the Properties of the Phases of QCD Matter}


\medskip

\centerline{\bf \large Research opportunities and priorities for the next decade}

\bigskip
\begin{center}
Summary of the ``Phases of QCD Matter'' Town Meeting \\
at Temple University, Philadelphia, 13-15 Sep. 2014\\

\bigskip

Prepared for consideration in the\\ 2015 NSAC Long Range Plan for Nuclear Physics\\

\bigskip
\bigskip

Ulrich Heinz (Ohio State University) and Paul Sorensen (BNL), co-conveners;\\
Abhay Deshpande (Stony Brook University),\\
Carl Gagliardi (Texas A\&M University),\\ 
Frithjof Karsch (BNL \& Universit\"at Bielefeld),\\ 
Tuomas Lappi (University of Jyv\"askyl\"a),\\
Zein-Eddine Meziani (Temple University),\\
Richard Milner (MIT),\\ 
Berndt M\"uller (BNL \& Duke University),\\ 
Jamie Nagle (University of Colorado),\\ 
Jianwei Qiu (BNL),\\
Krishna Rajagopal (MIT),\\ 
Gunther Roland (MIT),\\ 
Raju Venugopalan (BNL).

\vspace*{1cm}

\today

\end{center}

\vspace*{2cm}


\begin{abstract}
\noindent
This document provides a summary of the discussions during the recent joint QCD Town Meeting at Temple University of the status of and future plans for the research program of the relativistic heavy-ion community. A list of \fp{compelling questions is formulated, and a number of} recommendations outlining the greatest research opportunities and detailing the research priorities of the heavy-ion community, voted on and unanimously approved at the Town Meeting, \fp{are} presented. They are supported by a broad discussion of the underlying physics and its relation to other subfields. Areas of overlapping interests with the ``QCD and Hadron Structure'' (``cold QCD'') subcommunity, in particular the recommendation for the future construction of an Electron-Ion Collider, are emphasized. The agenda of activities of the ``hot QCD'' subcommunity at the Town Meeting is attached.
\end{abstract}

\eject


\pagebreak
\tableofcontents
\clearpage


\vspace*{-14mm}
\section{Executive Summary}
\label{sec1}
\vspace*{-3mm}

On September 13-15, 2014, the QCD Community within the APS Division of
Nuclear Physics met at Temple University, Philadelphia, for a
three-day Town Meeting to discuss the status and future priorities of
their research program, in preparation for a new Long Range Plan for
Nuclear Physics to be written and submitted to the NSF and DOE Nuclear
Science Advisory Committee in 2015. The U.S. Nuclear Physics QCD
Community consists of two, partially overlapping, subcommunities whose
activities focus on ``The Phases of QCD Matter'' (a.k.a. as ``hot QCD''
community) and ``QCD and Hadron Structure'' (a.k.a. as ``cold QCD''
community), respectively. Their joint Town Meeting featured a
daylong {\it joint session} (spread over two days, from 4pm on
Saturday to 4 pm on Sunday), with {\it separate sessions} of the
``hot'' and ``cold'' QCD subcommunities scheduled in parallel for the
rest of the time. The Town Meeting program schedule, including links
to the slides of all presentations, can be found at
\blue{\url{https://indico.bnl.gov/conferenceDisplay.py?confId=857}}. With
244 registrants (108 from ``hot QCD'', 136 from ``cold QCD''),
participation by the community in the 2014 joint QCD Town Meeting was
about 10\% stronger than for the corresponding meeting in early 2007
at Rutgers University, reflecting the health and strength of the
U.S. QCD Community in Nuclear Physics. Strong community interest in
the planning for the next decade was also reflected in a total of 49
short presentations submitted and delivered by participants, in
addition to the 47 invited talks (15 and 19 in the ``hot'' and
``cold'' QCD parallel sessions, respectively, and 13 in the joint
session) that were solicited, with detailed charges to the speakers to
ensure a well-rounded program, by the Town Meeting conveners (Haiyan
Gao, Ulrich Heinz, Craig Roberts, and Paul Sorensen). \fp{In this Executive 
Summary we list the four recommendations that were voted on and approved 
by the ``hot QCD'' community during the parallel and joint sessions.}

\fp{Members of the ``hot QCD'' community reported on recent progress in the field including the development of dynamical models accurate enough to provide quantitative access to emergent phenomena in Hot QCD. Heavy-ion data were shown to prefer an equation of state that agrees with that extracted from lattice QCD calculations and it was firmly established, through improvements in the determination of transport properties, that the hottest matter ever made by humans is also the most perfect liquid known in nature. Following on the significant progress made in heavy-ion studies probing zero baryon chemical potential, the Hot QCD community is seeking to 1) measure the temperature and chemical potential dependence of transport properties especially near the phase boundary, 2) explore the phase structure of the nuclear matter phase diagram, 3) probe the microscopic picture of the perfect liquid, and 4) image the high density gluon fields of the incoming nuclei and study their fluctuation spectrum.} 

\fp{These goals, which can be addressed with a targeted experimental program and continued strong support for theory, led to the unanimous approval of the ``hot QCD'' community's first and highest priority recommendation:}

\noindent \underline{\bf \large Recommendation \#1:}

{\bf The discoveries of the past decade have posed or sharpened
  questions that are central to understanding the nature, structure,
  and origin of the hottest liquid form of matter that the universe
  has ever seen. As our highest priority we recommend a program to
  complete the search for the critical point in the QCD phase diagram
  and to exploit the newly realized potential of exploring the QGP's
  structure at multiple length scales with jets at RHIC and LHC
  energies. This requires
\begin{itemize}
\item
implementation of new capabilities of the RHIC facility (a
state-of-the-art jet detector such as sPHENIX and luminosity upgrades
for running at low energies) needed to complete its scientific
mission,
\item
continued strong U.S. participation in the LHC heavy-ion program, and
\item
strong investment in a broad range of theoretical efforts employing
various analytical and computational methods.
\end{itemize}
}

\bigskip

\fp{A high-energy, high-luminosity, polarized Electron Ion Collider is considered a requirement to achieve a full understanding within QCD of the structure and properties of hadrons and nuclei. The construction of an EIC as a means to usher in a new era of precision measurements was unanimously approved as the {\it highest priority for future new construction} (after the completion of FRIB) by the entire U.S. QCD Community in Nuclear Physics:}

\noindent \underline{\bf \large Recommendation \#2:}

{\bf A high luminosity, high-energy polarized Electron Ion Collider
  (EIC) is the U.S. QCD community's highest priority for future
  construction \fp{after FRIB}. }

\bigskip

\fp{Recommendation \#3 is an amended version of a recommendation passed by
overwhelming majority in the joint session to ``endorse the new
initiatives and investments proposed in the Recommendation and Request
received from the Computational Nuclear Physics Town Meeting.'' The
amended version listed here was passed unanimously in a later vote
taken during the separate voting session of the ``hot QCD'' community:}

\noindent \underline{\bf \large Recommendation \#3:}

{\bf We endorse the new initiatives and investments proposed in the
  Recommendation and Request received from the Computational Nuclear
  Physics Town Meeting, at a level to be determined by the requested
  NSAC subcommittee. In addition, we recommend new funding to expand
  the successful ``Topical Collaborations in Nuclear Theory'' program
  initiated in the last Long Range Plan of 2007, to a level of at least one 
  new Topical Collaboration per year.}

\bigskip

\fp{The final recommendation regarding education and innovation was passed
unanimously in identical form by both subcommunities in their parallel
voting sessions:}

\noindent \underline{\bf \large Recommendation \#4:}

{\bf The QCD community endorses and supports the conclusions from the
  Education and Innovation Town Meeting. }

\bigskip

The rest of this document provides supporting arguments for these recommendations. It does not simply summarize the presentations and discussions at the Town Meeting, but tries to provide a broader context and to present, in the words of the authors listed on the cover, a coherent picture of the status and future of the field. In formulating this text, the authors made extensive use not only of the presentations at the Town Meeting, but also of additional material collected for the community white papers by the "Hot QCD" and "Electron Ion Collider" communities \cite{Akiba:2015jwa,Accardi:2012qut} whose scope is significantly more comprehensive and to which the interested reader is directed for further details and additional projects. Section~\ref{sec2} offers an overview of the status, recent achievements, major open questions and future goals of research on the ``Phases of QCD Matter'', and describes how the above recommendations arise from such an overarching view. Section~\ref{sec3} discusses in greater depth the science behind these future goals and what is needed to successfully address the open questions listed in Section~\ref{sec2}. Section~\ref{sec4} provides brief descriptions of the science at and the design options for an Electron Ion Collider. Like our Recommendation \#2 above, the material provided in that section, although slightly rearranged, is identical in physical content with a corresponding section in the ``cold QCD'' summary of the joint QCD Town Meeting, to reflect the unanimous and strong support for the EIC project by the entire U.S. QCD Community. References and an Appendix with the Town Meeting schedule are found at the end of this document.

\clearpage
\section{Overview and Recommendations}
\label{sec2}

\subsection{Where we stand: recent insights and open questions}

The bulk of the mass of the visible matter in the universe comes from
energy stored in the strong interaction between its fundamental
constituents. At a basic level, the strong interaction is well
understood and described by Quantum Chromodynamics (QCD). Strongly
interacting multi-particle systems feature numerous emergent phenomena
that are difficult to predict from the underlying QCD theory, just
like in condensed matter and atomic systems where the interactions are
controlled by QED theory.

We now know that, as the universe evolved to its present state,
strongly interacting matter existed in at least two distinct forms, a
quark-gluon plasma (QGP) phase containing deconfined color charges
that filled the universe homogeneously during its first few
microseconds, and a clustered phase in which colored degrees of
freedom are permanently confined into color-neutral objects called
hadrons which make up the nuclei of today's atoms. Additional forms of
strongly interacting matter that include a color superconducting phase
may still exist in the cores of some compact stars or may have briefly
existed in others before they collapsed into black holes. QCD matter
is expected to have a complex phase diagram, possibly including one or
more critical points.

The Relativistic Heavy Ion Collider (RHIC) at Brookhaven National
Laboratory and the Large Hadron Collider (LHC) at CERN enable us to
study strongly interacting matter at extreme temperatures in the
laboratory. Conjectured over 35 years ago, the existence of the QGP
phase was established unambiguously by experiments at RHIC which also
made the surprising discovery that the QGP is strongly coupled and
behaves like an almost perfect liquid. The successful implementation
of luminosity and detector upgrades at the RHIC facility, recommended
in the last Long Range Plan, and the beginning of the higher-energy
heavy-ion program at the LHC in 2010, made additional studies possible
that have begun to yield precise estimates for some of its key
properties. Decisive theoretical progress at several fronts has been
instrumental for these achievements. In response to a recommendation
in the last Nuclear Physics Long Range Plan, the DOE made a dedicated
investment into three Topical Collaborations in Nuclear Theory, among
them the highly successful JET Collaboration. Both DOE and NSF have
provided strong support for high performance computing. Facilitated by
these initiatives, a standard framework has been developed and
implemented that describes, with quantitative predictive power, the
complete dynamical evolution of and the interaction of penetrating
diagnostic probes with the expanding QCD matter created in heavy-ion
collisions.

The unparalleled flexibility of the RHIC facility to collide atomic
nuclei of different sizes over a wide range of energies, complemented
by p+p, p+Pb and Pb+Pb collisions at the LHC with about 15 times the
top RHIC energy, provides the experimental leverage necessary to
clarify the nature of QCD matter. New discoveries made over the
past decade have sharpened some questions and posed several new ones
that address the core of our understanding of the nature, structure
and origin of the QGP liquid. These questions frame our research
program for the coming decade. To address them requires, in the short
term, a suite of facility and detector upgrades at RHIC and the LHC
and a series of new experiments that exploit these upgrades. In the
long term they necessitate the construction of an Electron Ion
Collider (EIC). The questions, in part motivated by recent discoveries
summarized in the following subsection, are listed here and expanded
upon in the science sections of this document:

\begin{itemize}

{\bf

\item What are the transport properties of the QGP? How do they change
  when the plasma is heated or doped with excess quarks?

\item How do the collective properties of the QGP liquid, one of the
  most strongly coupled forms of matter now known, emerge from the
  interactions among the individual quarks and gluons that we know
  must be visible if the liquid is probed with sufficiently high
  resolution?

\item What is the precise nature of the initial state from which this
  liquid forms, and how does it reach approximate local thermal
  equilibrium in the short time and rapidly expanding environment
  provided by heavy-ion collisions?

\item Can dense systems of quarks and gluons act like strongly coupled
  liquids without thermalizing? Does the Color Glass Condensate state
  that manifests itself when fast-moving atomic nuclei are probed at
  very small longitudinal momentum fraction exhibit collective
  behavior?

\item What is the structure of the QCD phase diagram? Does it, like
  that of water, feature a critical end point separating a line of
  first-order phase transitions at large baryon density from the rapid
  but continuous crossover found by lattice QCD at low baryon
  density?

\item How does the observed structure of the QGP change when it is
  probed at different length scales, with photons, jets, and heavy
  quark flavors? What is the shortest length scale on which the plasma
  liquid looks liquid-like?

\item What is the smallest size and density of a droplet of QCD matter
  that behaves like a liquid?  }
\end{itemize}

\subsection{Accomplishments and future goals}

We briefly summarize some recent achievements that play decisive roles
in defining the research priorities for the coming decade. Additional
accomplishments and related future developments are described in the
science sections of this document.
\medskip

{\bf Flow fluctuations and correlations}

One of the most important recent breakthroughs has been the
realization that each collision event (``Little Bang'') exhibits its
own unique flow pattern, with measurable strength up to the 7th 
or 8th multipole. It was recently discovered that these ripples in 
the near-perfect QGP liquid bring information about initial-state 
nucleonic and, likely, sub-nucleonic gluon fluctuations into the final 
state. This has opened new possibilities to study the dense gluon fields 
and their quantum fluctuations in the colliding nuclei via correlations 
between final state particles. Precise measurements of the complete 
set of measurable flow coefficients, and their event-by-event fluctuations
and correlations in both magnitude and flow angle, for a number of
identified hadron species will make it possible (in conjunction with the 
following item) to map the transverse and longitudinal spatial dependence 
of the initial gluon fluctuation spectrum. This will provide a test for QCD
calculations in a high gluon density regime.
\medskip

{\bf Quantifying the fluidity of quark-gluon plasma}    

Precise measurements of the flow coefficients up to the 5th and 6th
multipole for charged hadrons, and of elliptic and triangular flow for
several identified hadron species, have made it possible to determine
the shear viscosity to entropy density ratio, $\eta/s$, of the QGP to
within less than a factor of 2. This was only possible with recently
converged results from lattice QCD for the QCD equation of state and
due to the development of a comprehensive and sophisticated
theoretical model for the dynamical evolution of the collision, with
relativistic viscous fluid dynamics to describe the QGP liquid at its
core. A simultaneous analysis of RHIC and LHC data has provided first
evidence for an increase of $(\eta/s)_\mathrm{QGP}$ with rising
temperature. The present uncertainty on $(\eta/s)_\mathrm{QGP}$
(conservatively estimated as $\pm50\%$) is not so much limited by the
quality of the available experimental data (which would be sufficient
to determine the viscosity with relative precision of 5-10\%) as by
the incompleteness of the presently available set of measurements. 
It has recently been understood (see Sec. \ref{sec:SM}) that 
measurement of a complete set of correlation functions between
the momentum-dependent anisotropic flow coefficients and their 
associated flow angles, for several particle species covering a wide 
range of masses, can not only tightly constrain the spatial dependence 
of the initial gluon fluctuation spectrum but will also result in a large 
improvement in our knowledge of the QGP transport coefficients.

The recent unexpected discovery of collective, anisotropic-flow signatures in p+Pb collisions at the LHC and in \fp{d+Au and $^3$He+Au} collisions at RHIC suggests that similar signatures seen in very-high-multiplicity p+p collisions might also be of collective origin. How collectivity develops in such small systems cries out for explanation. \fp{The additional running this year of p+Au and p+Al at RHIC augmented the d+Au and $^3$He+Au data and added a crucial comparison at the same collision energy.} The inescapable last question on the list above can only be answered systematically by exploiting RHIC's flexibility to collide atomic nuclei of any size over a wide range of energies.
\medskip
 
{\bf Color opacity and temperature evolution of QCD matter from hard
  and penetrating probes}

{\bf Parton energy loss and jet quenching:} Hard probes yield
information about how energetic partons diffuse in transverse momentum
space and lose energy as they slice through strongly coupled
QGP. State-of-the-art analysis of the energy loss of leading hadrons
in jets, together with significant recent advances in theoretical
modeling, both within perturbative QCD and by introducing insights
from strong coupling calculations, have increased the precision of our
knowledge of the transverse momentum diffusion parameter, $\hat
q/T^3$, by about an order of magnitude, to within a factor 2-3. The
parton mass dependence of jet modification and energy loss in heavy
flavor jets will make it possible to separately quantify the
contributions from different energy-loss mechanisms. More generally,
detailed studies of jet modification by strongly coupled plasma over a
wide range of angular and energy scales will connect its macroscopic
hydrodynamic description to a microscopic description in terms of
quarks and gluons. As such, these jet measurements will provide unique
microscopic tools to move closer to a fundamental understanding of how
a strongly coupled liquid can arise in an asymptotically free gauge
theory. Such measurements require high luminosity operation and new
instrumentation at RHIC and the LHC, and their quantitative
interpretation rests on further development of theoretical tools for a
direct comparison of calculations to the data.

{\bf Quarkonium thermometry:} Precise and systematic measurements of
quarkonium production can determine the screening length of the static
QCD force in a QGP. Screening effects are expected to be easier to
discern in bottomonium production, due to the absence of diluting
effects from bottom quark recombination. An initial observation of a
significant suppression of the three $\Upsilon$ states in Pb+Pb
collisions at the LHC, indicating a sequential suppression pattern,
was recently made by CMS; a precise measurement will be available from
the LHC by 2023. Low resolution and low statistics measurements have
been made by PHENIX and STAR and are consistent with the suppression
of higher $\Upsilon$ states but the different states could not be
individually resolved. \fp{With increased beam luminosity and the newly installed 
Muon Telescope Detector the statistical significance of STAR's
measurements and its ability to separate the different $\Upsilon$ states 
will be improved over the next several runs,} and by 2021
the sPHENIX experiment will be making $\Upsilon$ measurements at RHIC
with both excellent mass resolution and much better statistical
precision. The combination of these data sets at quite different
initial temperatures, together with the different sizes of the three
$\Upsilon$ states which provide measurements at three different length
scales, will provide strong constraints on the screening length in hot
QCD matter.

{\bf Electromagnetic probes:} Electromagnetic radiation from the
Little Bangs integrates over the electromagnetic spectral function of
hot QCD matter as it changes with position and time. This provides
information on the temperature evolution of the expanding fireball and
opens a direct window on how the degrees of freedom in the vector
channel change with temperature. Pioneering measurements by PHENIX
have recently been augmented by first results from Pb+Pb collisions at
the LHC and dilepton measurements at several collision energies by the
STAR collaboration. These measurements were central in proving that
the temperature achieved in heavy-ion collisions is the hottest ever
man-made temperature. More precise future determinations of the
low-mass dilepton spectrum are expected to lead to an improved
understanding of chiral symmetry restoration at high temperature. 
Total yield and spectral slope measurements in the RHIC
Beam Energy Scan (BES) program will help to quantitatively determine
the changing fireball lifetime and temperature history at decreasing
collision energy. Recently, unexpectedly large elliptic flows of
direct photons measured by PHENIX and ALICE have presented the theory
community with a puzzle. Its resolution requires more precise future
measurements of the yields, slopes and anisotropic flow coefficients
of direct photons at RHIC and LHC collisions.
\medskip
 
{\bf Mapping the QCD phase diagram}

Theoretical models suggest a phase diagram for QCD matter that 
rivals that of water in complexity. It is the only experimentally accessible 
phase diagram of matter that is controlled directly by the non-Abelian 
gauge field interactions in the fundamental forces of nature. Heavy-ion 
collisions at top RHIC and LHC energies produce strongly coupled 
plasma with a low value of $\mu_B$ where lattice QCD predicts a 
smooth crossover between the QGP liquid and a hadron resonance 
gas. Experimentally mapping the QCD phase diagram is
one of the big unsolved challenges in the field. A first Beam Energy
Scan (BES-I), with Au+Au collisions at center-of-mass energies between 
39 and 7.7 GeV to explore QCD matter at baryon chemical potentials 
110\,MeV${\,\leq\,}\mu_B{\,\leq\,}420$\,MeV, was completed in 2014. 
BES-I led to a number of intriguing observations of non-monotonic beam 
energy dependences of several flow and fluctuation observables which 
might be connected with the appearance of a first-order phase transition 
at large $\mu_B$. State-of-the-art lattice QCD calculations combined with
dynamical modeling, using a hybrid approach that couples viscous fluid 
dynamics for the QGP liquid with a microscopic approach to the critical 
phase transition dynamics and the subsequent evolution of the hadronic 
phase, will be required to test these interpretations. A second Beam 
Energy Scan (BES-II) planned for 2018-2019, with significantly improved 
beam luminosity and upgraded detector capabilities, and concurrent 
improvements of the theoretical modeling at lower beam energies are 
needed to solidify the suggestive results from BES-I with precision 
measurements in the targeted energy region identified in BES-I. 
Unambiguous discovery of a critical point in BES-II would warrant 
additional measurements at a later time to further quantify its properties.
\medskip

{\bf Forward rapidity studies at high energy and the Color Glass
  Condensate}

Both RHIC and the LHC are capable of probing new, unmeasured physics
phenomena at low longitudinal momentum fraction $x$. Data from the
2013 p+Pb run at the LHC will make it possible to study previously
unreachable phase-space in the search for parton saturation effects
(the Color Glass Condensate). Forward-rapidity detector upgrades in
STAR and PHENIX will open the door to studies of saturation physics at
RHIC. However, a complete exploration of parton dynamics at low $x$
will require an Electron Ion Collider (EIC). While a future EIC will deliver 
crucially missing precise information on the nuclear parton distribution 
functions in a kinematic regime where in heavy-ion collisions saturation 
effects are difficult to separate from QGP physics, forward rapidity studies 
in p+A and A+A collisions at RHIC and LHC provide access to low-$x$ 
physics in a complementary kinematic range.

\subsection{Recommendations}

\subsubsection{New questions and opportunities in relativistic heavy-ion collisions}

Over the past decade, through a panoply of measurements made in
heavy-ion collisions at the Relativistic Heavy Ion Collider (RHIC) and
the Large Hadron Collider (LHC), in concert with theoretical advances
coming from calculations done using many different frameworks, we have
obtained a broad and deep knowledge of what hot QCD matter does, but
we still know little about how it works. These collisions create
exploding little droplets of the hottest matter seen anywhere in the
universe since it was a few microseconds old. We have increasingly
quantitative empirical descriptions of the phenomena manifest in these
explosions, and of some key material properties of the matter created
in these ``Little Bangs'' which turns out to be a strongly coupled
liquid. However, we still do not know the precise nature of the
initial state from which this liquid forms, and know very little about
how the properties of this liquid vary across its phase diagram or
how, at a microscopic level, the collective properties of this liquid
emerge from the interactions among the individual quarks and gluons
that we know must be visible if the liquid is probed with sufficiently
high resolution.

Answering these and other questions requires an intensive modeling and
computational effort to simultaneously determine the set of key
parameters needed for a multi-scale characterization of the QGP medium
and the initial state from which it emerges. This phenomenological
effort requires broad experimental input from a diverse set of
measurements, including 1) the completion of the heavy quark program
to measure the diffusion coefficient of heavy quarks, 2) energy scans
to map the phase diagram of QCD and the dependence of transport
coefficients on the temperature and baryon number chemical potential, 
3) collisions of nuclei with varied sizes, including p+A and very high 
multiplicity p+p collisions, to study the emergence of collective phenomena, 
4) the quantitative characterization of the electromagnetic radiation emitted
by the Little Bangs and its spectral anisotropies, and 5) a detailed
investigation at RHIC and LHC of medium effects on the production
rates and internal structure of jets of hadrons, for multi-scale
tomographic studies of the medium. These considerations lead us to our
\bigskip

\begin{center}

\framebox[0.98\linewidth]{\parbox{0.93\linewidth}{

\smallskip

\noindent \underline{\bf \large Recommendation \#1:}\\[-0.8ex]

{\bf The discoveries of the past decade have posed or sharpened
  questions that are central to understanding the nature, structure,
  and origin of the hottest liquid form of matter that the universe
  has ever seen. As our highest priority we recommend a program to
  complete the search for the critical point in the QCD phase diagram
  and to exploit the newly realized potential of exploring the QGP's
  structure at multiple length scales with jets at RHIC and LHC
  energies. This requires\\[-1.5ex]
\begin{itemize}
\item
implementation of new capabilities of the RHIC facility (a
state-of-the-art jet detector such as sPHENIX and luminosity upgrades
for running at low energies) needed to complete its scientific
mission, \\[-1.5ex]
\item
continued strong U.S. participation in the LHC heavy-ion program, and
\\[-1.5ex]
\item
strong investment in a broad range of theoretical efforts employing
various analytical and computational methods.
\end{itemize}
\smallskip
}
}}
\end{center}
\bigskip

RHIC and the LHC, together, provide an unprecedented opportunity to
study the properties of QCD matter.  While collisions at the LHC
create temperatures well above those needed for the creation of QGP
and may thus be able to explore the expected transition from a
strongly coupled liquid to a weakly coupled gaseous phase at higher
temperatures, the RHIC program enables unique research at temperatures
close to the phase transition. Moreover, the unparalleled flexibility
of RHIC makes possible collisions between a variety of different ion
species over a broad range in energy. The combined programs permit a
comprehensive exploration of the QCD phase diagram, together with
precise studies of how initial conditions affect the creation and
dynamical expansion of hot QCD matter and of the microscopic structure
of the strongly coupled QGP liquid. \fp{There is no single facility in the short- or long-term future that could come close to duplicating what RHIC and the LHC, operating in concert, will teach us about Nature.}

\subsubsection{A precision femtoscope to study the glue that binds us all}

It is generally believed that, at high energies, interacting gluon fields from the colliding hadrons dominate the energy deposition in the collision zone and the subsequent thermalization processes that eventually lead to the creation of a quark-gluon plasma. The possibility to precisely measure the gluon wave functions of the incoming nuclei, in order to complement the empirical determination of the gluonic initial state from flow fluctuations and correlations in heavy-ion collisions, is expected to further solidify the determination of the key parameters characterizing the thermodynamic and transport properties of the QGP. The construction of an Electron Ion Collider (EIC) will address this need.

The EIC will image the gluons and sea quarks in the proton and nuclei
with unprecedented precision and probe their many-body correlations in
detail, providing access to novel emergent phenomena in QCD. It will
definitively resolve the proton's internal structure, including its
spin, and explore a new QCD frontier of ultra-dense gluon fields in
nuclei at high energy. These advances are made possible by the EIC's
unique capability to collide polarized electrons with polarized
protons and light ions at unprecedented luminosity and electrons with
heavy nuclei at high energy. An EIC will be absolutely essential to
maintain U.S. leadership in fundamental nuclear physics research in
the coming decades:
\bigskip

\begin{center}

\framebox[0.98\linewidth]{\parbox{0.93\linewidth}{
\smallskip

\noindent \underline{\bf \large Recommendation \#2:}\\[-0.8ex]

{\bf A high luminosity, high-energy polarized Electron Ion Collider
  (EIC) is the U.S. QCD community's highest priority for future
  construction after FRIB.
\smallskip}
}}
\smallskip
\end{center}

\subsubsection{New initiatives to further strengthen nuclear theory}

Due to the complexity of heavy-ion collision dynamics, the success of
the research program outlined in this document hinges on continued
strong support of a broad range of theoretical activities. In addition
to a healthy base program, the field requires specific support in
high-performance computing and of collaborative efforts focused on
addressing complex issues that need contributions from several
scientists or groups of scientists with complementary
expertise. Examples for the first need are the determination of
thermal equilibrium properties and response functions of the QGP from
lattice QCD, the dynamical simulation of the thermalization processes
in the initial stage of heavy-ion collisions, and comprehensive
model/data comparisons involving large data sets and complex dynamical
models that aim at extracting key model parameters with quantified
uncertainties. An example for the latter is the DOE-funded JET Topical
Collaboration that addresses the problem of turning measurements of
the medium modification of jets and electromagnetic probes into
precise tomographic tools that yield quantitative information about
the properties and dynamical evolution of the dense QCD matter created
in heavy-ion collisions.

The high-performance computing needs of the nuclear theory community
were addressed at a Town Meeting in Washington, D.C., on July 14-15,
2014, which passed the following recommendation and request:
   
{\sl Recommendation:} Realizing the scientific potential of current
and future experiments demands large-scale computations in nuclear
theory that exploit the US leadership in high-performance
computing. Capitalizing on the pre-exascale systems of 2017 and beyond
requires significant new investments in people, advanced software, and
complementary capacity computing directed toward nuclear theory.

{\sl Request:} To this end, we ask the Long-Range Plan to endorse the
creation of an NSAC subcommittee to develop a strategic plan for a
diverse program of new investments in computational nuclear theory. We
expect this program to include:

\begin{itemize}
\item
new investments in SciDAC and complementary efforts needed to maximize
the impact of the experimental program;
\item
development of a multi-disciplinary workforce in computational nuclear
theory;
\item
deployment of the necessary capacity computing to fully exploit the
nations leadership-class computers.
\end{itemize}

At the QCD Town Meeting at Temple University, the ``Phases of QCD
Matter'' subcommunity endorsed this resolution and amended it as
follows:\\

\begin{center}

\framebox[0.98\linewidth]{\parbox{0.93\linewidth}{

\smallskip

\noindent \underline{\bf \large Recommendation \#3:}\\[-0.8ex]

{\bf We endorse the new initiatives and investments proposed in the
  Recommendation and Request received from the Computational Nuclear
  Physics Town Meeting, at a level to be determined by the requested
  NSAC subcommittee. In addition, we recommend new funding to expand
  the successful ``Topical Collaborations in Nuclear Theory'' program
  initiated in the last Long Range Plan of 2007, to a level of at
  least one new Topical Collaboration per year.}
\smallskip
}}

\end{center}

\subsubsection{Educating and mentoring the next generation of scientists}

A continuous pipeline of highly and broadly educated young scientists
is not only the lifeblood of the U.S. Nuclear Science program, but
also a guarantor of our nation's continued technological and economic
strength and its ability to innovate. At a separate Town Meeting
dedicated to Education and Innovation (Michigan State University,
August 6-8, 2014) the following resolutions were passed:

\begin{enumerate}
\item
Education and mentoring of the next generation nuclear scientists as
well as dissemination of research results to a broad audience should
be recognized by all researchers as an integral part of the scientific
enterprise.
\item
Nuclear science is an active and vibrant field with wide applicability
to many societal issues. It is critical for the future of the field
that the whole community embraces and increases its promotion of
nuclear science to students at all stages in their career as well as
to the general public.
\item 
Researchers in nuclear physics and nuclear chemistry have been
innovative leaders in the full spectrum of activities that serve to
educate nuclear scientists as well as other scientists and the general
public in becoming informed of the importance of nuclear science. The
researchers are encouraged to build on these strengths to address some
of the challenges in educating an inclusive community of scientists as
well as those on the path to future leadership in nuclear science.
\item
The interface between basic research in nuclear physics and exciting
innovations in applied nuclear science is a particularly vital
component that has driven economic development, increased national
competitiveness, and attracts students into the field. It is critical
that federal funding agencies provide and coordinate funding
opportunities for innovative ideas for potential future applications.
\end{enumerate}

In separate sessions at the QCD Town Meeting at Temple University, the
``Phases of QCD Matter'' and ``QCD and Hadron Structure"
subcommunities voted unanimously for the following

\bigskip

\begin{center}

\framebox[0.98\linewidth]{\parbox{0.93\linewidth}{
\smallskip

\noindent \underline{\bf \large Recommendation \#4:}\\[-0.8ex]

{\bf The QCD community endorses and supports the conclusions from the
  Education and Innovation Town Meeting.
\smallskip}
}}
\end{center}

\clearpage
\section[Quantifying the Properties of QCD Matter - Now and Tomorrow]{Quantifying the Properties of QCD Matter --\\Present Status and Future Opportunities}
\label{sec3}
%
\subsection{Completing the Little Bang Standard Model}
\label{sec:SM}

One of the most important recent discoveries in heavy-ion collisions is that density fluctuations from the initial state of heavy-ion collisions survive through the expansion of the fireball showing up as correlations between produced particles \cite{Mishra:2007tw,Voloshin:2003ud,Takahashi:2009na,Sorensen:2010zq,Alver:2010gr,ATLAS:2012at,ALICE:2011ab,Adare:2011tg,Adamczyk:2013waa}. Prior work had mostly approximated the incoming nuclei as smooth spheres and the initial overlap region as an ellipse. The survival of density and geometry fluctuations was first hinted at in measurements of cumulants related to shape of the $v_2$ distribution \cite{Adler:2002pu,Miller:2003kd}. The picture started to become more clear after measurements were made in Cu+Cu collisions where the fluctuations were more prominent in the smaller system \cite{Alver:2008zza}. Ultimately, a new paradigm emerged as the structure of the initial state was found to play a central role in two-particle correlation functions and the previous measurements of $v_2$ were generalized to $v_n$, a spectrum carrying information about both the initial densities in the collision as well as the dissipative properties of the subsequent plasma phase \cite{Heinz:2013th}. The survival of the initial state fluctuations is intimately related to the earlier finding that the QGP discovered at RHIC is the most perfect fluid known \cite{Teaney:2003kp,Romatschke:2007mq,Song:2010mg}, with a viscosity to entropy density ratio $\eta/s$ near the string theory bound \cite{Kovtun:2004de}. The low viscosity plasma phase acts as a lens (albeit of strongly non-linear character), faithfully transferring the geometric structure of the initial density distributions, with its associated distribution of pressure gradients which act as a hydrodynamic force, into the final state. There it shows up most prominently as correlations between produced particles. Quantum fluctuations in the initial state cause these correlations to fluctuate from event to event.

\fp{\bf A standard dynamical framework for heavy-ion collisions}

\begin{figure}[t!]
\centering
\includegraphics[width=1.0\textwidth]{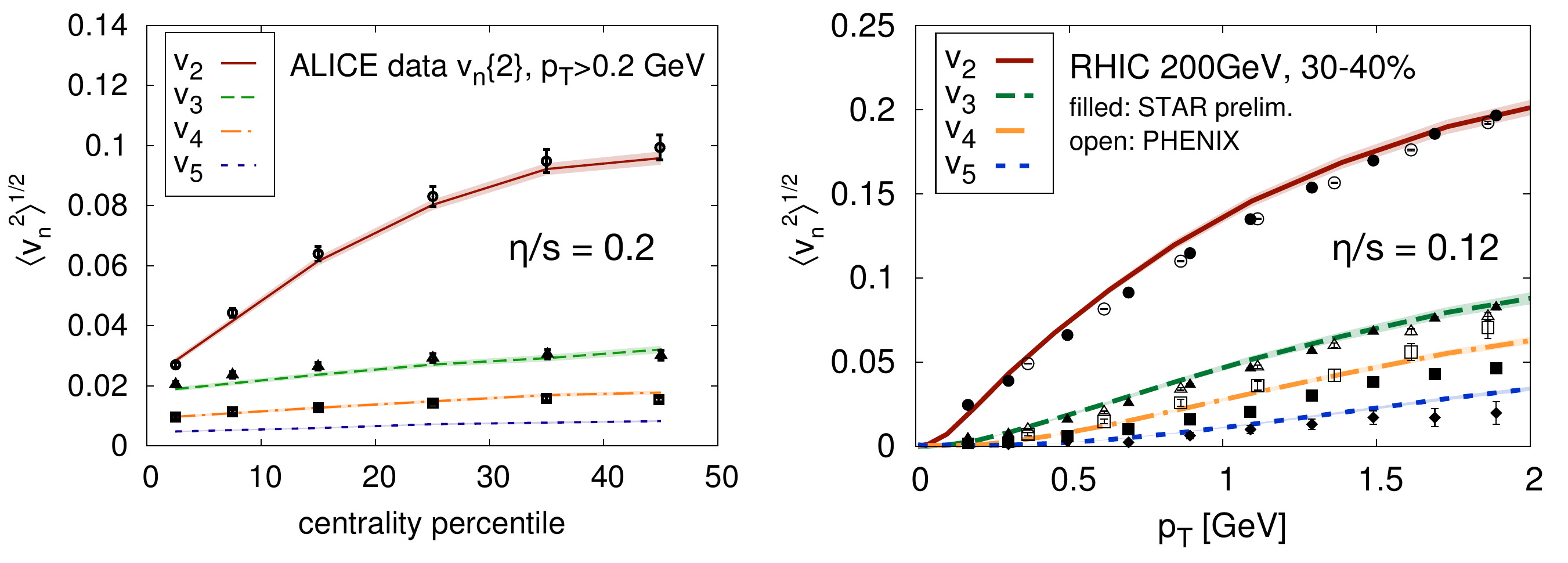}
\caption{Model calculations compared to measurements of the harmonic decomposition of azimuthal correlations produced in heavy-ion collisions. The left panel shows model calculations and data for $v_n$ vs. collision centrality in Pb+Pb collisions at $\sqrt{s_\mathrm{NN}}=2.76$ TeV. The right panel shows similar studies for the $p_T$ dependence of $v_n$ in 200 GeV Au+Au collisions. The comparison of the two energies provides insight on the temperature dependence of $\eta/s$. }
\label{fig:vn}
\end{figure}

Descriptions of these new phenomena have required the development of a new dynamical framework for heavy-ion collisions. It includes i) modeling of initial-state quantum fluctuations of nucleon positions and sub-nucleonic color charges and the color fields generated by them, ii) a description of the pre-equilibrium dynamics that evolves the initial energy-momentum tensor by solving either the (2+1)-dimensional Yang-Mills equations for the gluon fields (weakly-coupled approach) or Einstein's equations of motion in five-dimensional anti-deSitter space (strongly-coupled approach), followed by iii) the rapid transition, event-by-event, to second-order viscous relativistic fluid dynamics, and iv) a late-stage hadron phase described by microscopic transport calculations. While there is widespread agreement on the general structure of such a standardized dynamical approach, it has not yet reached the level of uniqueness that would justify calling it the ``Little Bang Standard Model'' \cite{Heinz:2013wva}. Model comparisons with experimental data that illustrate the state of the art in dynamical modeling can be found in \cite{Schenke:2010rr,Song:2010mg,Song:2010aq,Schenke:2011bn,Schenke:2012wb,Gale:2012rq,Song:2013tpa,Song:2013qma,vanderSchee:2013pia,Habich:2014jna}. With the existence of a reliable equation of state from lattice QCD calculations \cite{Bazavov:2009zn,Borsanyi:2010cj,Borsanyi:2013bia,Bazavov:2014pvz} a crucial degree of uncertainty in hydrodynamic modeling could be eliminated, enabling the development of a complete hydrodynamic space-time model. With this full space-time picture in hand, the comparisons of model calculations to harmonic decompositions of correlation functions ($\sqrt{v_{n}^{2}}$) at RHIC and the LHC (shown in Fig.~\ref{fig:vn}) have reduced the uncertainty on $\eta/s$ by a factor of 10 \cite{Gale:2013da}. With this newfound precision, studies suggest that $\eta/s$ is smaller for RHIC collisions (right panel of Fig.~\ref{fig:vn}) than it is at the LHC (left panel), consistent with a temperature dependent $\eta/s$ with a minimum near the critical temperature. In the next phase of study we seek to 1) accurately determine the temperature dependence of $\eta/s$ (aided by the Beam Energy Scan Program at RHIC) and 2) develop a clearer picture of the high density gluon fields that form the precursor of the plasma phase (aided by the p+A program and ultimately by an Electron Ion Collider).

\fp{\bf The path towards a Standard Model for the Little Bang}

So what is needed to turn this standard dynamical framework into the ``Little Bang Standard Model''? One fundamental challenge along the way is the need to determine {\em simultaneously} the space-time picture of the collective expansion and the medium properties that drive this expansion \cite{Heinz:2013wva}. A unique and reliable determination of these two unknowns is aided by measurements of multiple flow observables sensitive to medium properties in different stages of the evolution~\cite{Bhalerao:2011yg,Heinz:2013th,Jia:2014jca}. Due to the large event-by-event fluctuations in the initial state collision geometry, in each collision the created matter follows a different collective expansion with its own set of flow harmonics (magnitude $v_n$ and phases $\Phi_n$). Experimental observables describing harmonic flow can be generally given by the joint probability distribution of the magnitudes $v_n$ and phases $\Phi_n$ of flow harmonics:
\begin{equation}
\label{eq:flow1}
p(v_n,v_m,..., \Phi_n, \Phi_m, ...)=\frac{1}{N_{\mathrm{evts}}}\frac{dN_{\mathrm{evts}}}{dv_n dv_m \dots d\Phi_{n} d\Phi_{m} }.
\end{equation}
Specific examples include the probability distribution of individual harmonics $p(v_n)$, flow de-correlation in transverse and longitudinal directions, and correlations of amplitudes or phases between different harmonics ($p(v_n,v_m)$ or $p(\Phi_n,\Phi_m)$). The latter are best accessed through measurements of correlations with three or more particles. The joint probability distribution (\ref{eq:flow1}) can be fully characterized experimentally by measuring the complete set of moments recently identified in Ref.~\cite{Bhalerao:2014xra}. With the added detail provided by these measurements, hydrodynamic models can be fine-tuned and over-constrained, thus refining our understanding of the space-time picture and medium properties of the heavy-ion collisions. Initial measurements of some of these observables~\cite{Aad:2013xma,Aad:2014fla,CMS:2013bza} and comparison to hydrodynamic models~\cite{Gale:2012rq,Heinz:2013bua,Qiu:2012uy,Bhalerao:2013ina} already provided unprecedented insights on the nature of the initial density fluctuations and dynamics of the collective evolution. \fp{However, at this point none of the state-of-the-art hydrodynamic models properly accounts for the dynamical fluctuations generated {\em during} the evolution by thermal noise \cite{Kovtun:2011np,Kapusta:2011gt,Murase:2013tma,Young:2013fka} -- future quantitative work will need to address these, too.}

\begin{figure}[hbt]
\centerline{\includegraphics[width=0.85\textwidth]{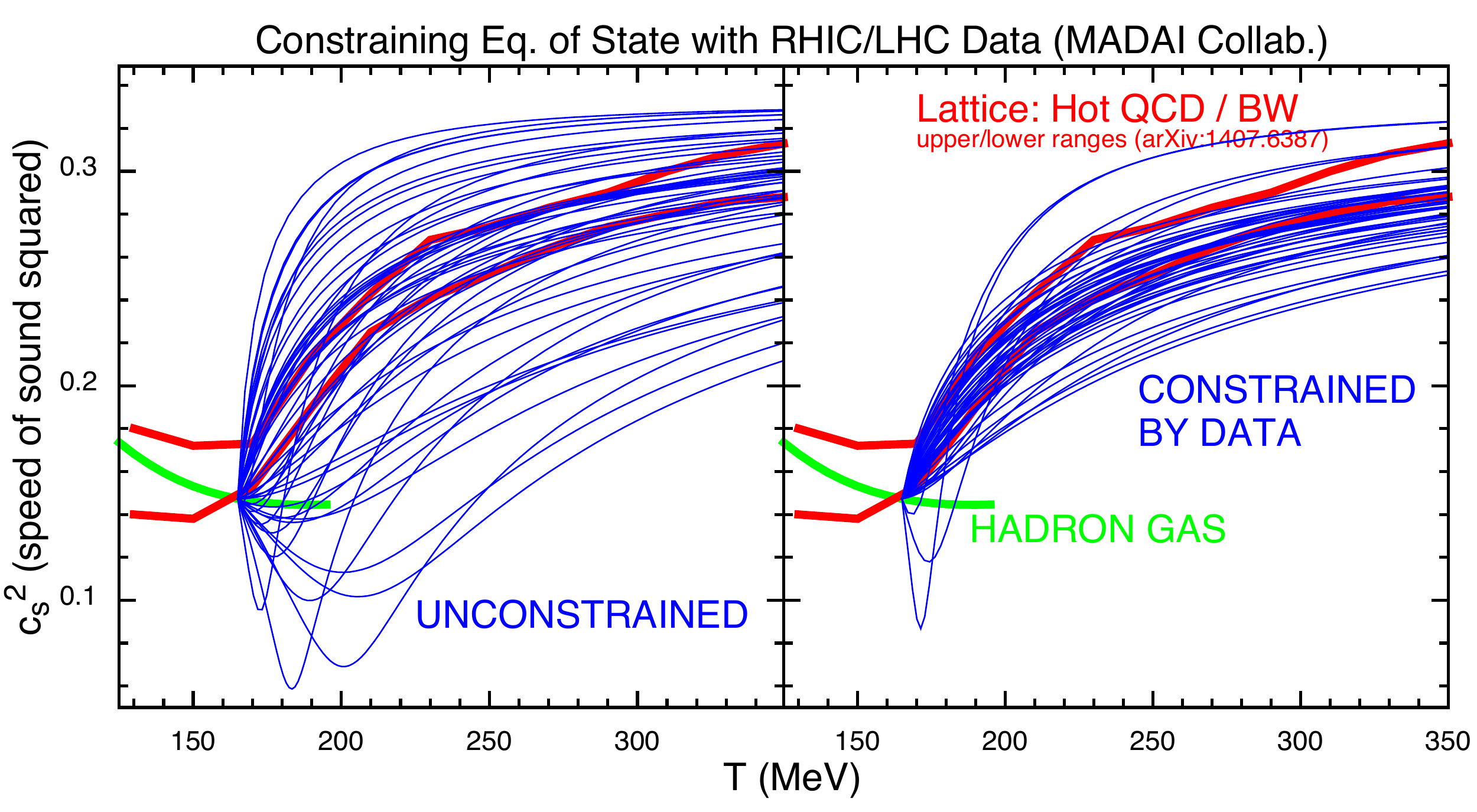}}
\caption{The QCD Equation of State (more precisely: the squared speed of sound, $c_s^2$, as a function of temperature) from lattice QCD calculations and from models constrained by data from RHIC and the LHC. The red lines delimit the present uncertainty range of $c_s^2$ from lattice QCD. The blue lines show a number of parametrizations of $c_s^2(T)$; in the left panel, they are constrained only by their asymptotic values at $T{\,=\,}165\,\mathrm{MeV}\simeq T_c$ and $T=\infty$, in the right panel they are additionally required to provide an acceptable fit to experimental data as described in \cite{Novak:2013bqa,Pratt:2015zsa}. The right panel shows that experimental data give preference to an Equation of State consistent with lattice QCD. This demonstrates that our model of the collision dynamics is good enough to allow us to study emergent properties in QCD matter.
\label{fig:EOS}
}
\end{figure}

\fp{\bf Precision determination of key QGP parameters}

\begin{figure}[!htb]
\begin{center}
\begin{minipage}{0.95\textwidth}
\begin{minipage}{0.6\textwidth}
\centerline{\includegraphics[width=\textwidth]{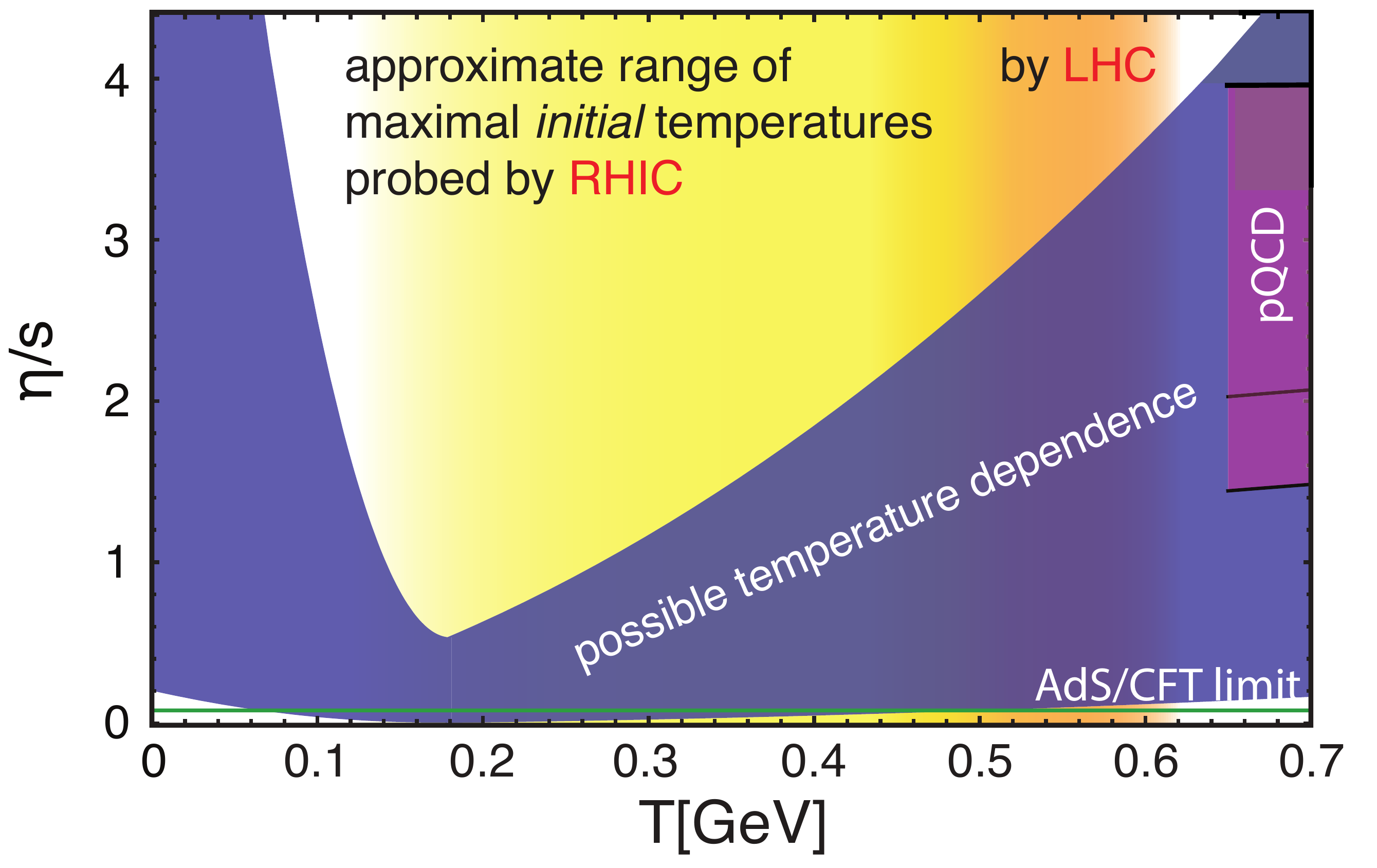}}
\end{minipage}
\hspace*{3mm}
\begin{minipage}{0.35\textwidth}
\caption{The temperature dependence of the specific viscosity $\eta/s$. The blue band represents the range allowed by our current understanding based on comparing experimental data to models with a minimum at the transition temperature. pQCD calculations and the string theory lower bound are also shown. The shaded vertical regions represent the ranges of initial temperatures probed by RHIC and the LHC.
\label{fig:visc}}
\end{minipage}
\end{minipage}
\end{center} 
\vspace*{-3mm} 
\end{figure}

The agreement between the model and the data shown in Figure \ref{fig:vn} suggests that the essential features of the dynamic evolution of heavy-ion collisions are well described by current models. These model calculations depend on a significant number of parameters that are presently poorly constrained by fundamental theory, and a reliable determination of the QGP properties requires a systematic exploration of the full parameter space. An example of such an exploration \cite{Novak:2013bqa,Pratt:2015zsa} is shown in Figure~\ref{fig:EOS} where the shape of the QCD EOS is treated as a free parameter. The left panel shows a random sample of the thousands of possible Equations of State, constrained only by results on the velocity of sound obtained by perturbative QCD at asymptotically high temperature and by lattice QCD at the crossover transition temperature. They are compared to the EOS determined from lattice QCD \cite{Bazavov:2014pvz}.The right panel shows a sample of the Equations of State allowed by experimental data. The results of this study suggest that data at RHIC and the LHC require an EOS consistent with that expected from QCD. This demonstrates that our model of heavy-ion collisions describes the dynamics of the collisions well enough that we can extract information on the emergent properties of finite temperature QCD from the experimental traces left by the tiny droplet of QGP created in the collisions. These state-of-the-art models can therefore be used to both determine properties of finite temperature QCD currently inaccessible to lattice calculations and to provide an accurate space-time profile needed for modeling other processes like jet quenching. Figure~\ref{fig:visc} shows a schematic representation of our current uncertainty on the temperature dependence of $\eta/s$ in QCD matter. While many of the existing measurements are accurate enough, as seen in Fig.~\ref{fig:vn}, to determine $\eta/s$ with much greater precision {\it if all other model parameters were already known}, the non-linear simultaneous dependence of the observables on multiple parameters does not yet allow one to translate the high quality of these experimental data into a more precise estimate of $\eta/s$. The studies shown in Figures~\ref{fig:EOS} and \ref{fig:vn} suggest, however, that a more complete set of measurements of the moments of the joint probability distribution (\ref{eq:flow1}) at the LHC and RHIC (particularly in the Beam Energy Scan), coupled with extensive quantitative modeling, will provide the desired access to $(\eta/s)(T)$ in and around the transition temperature where hadrons melt into a Quark Gluon Plasma, and strongly reduce the width of the blue uncertainty band in Fig.~\ref{fig:visc}.

The temperature dependence of $\eta/s$ can be further constrained by measuring directly emitted photons and dileptons. Their lack of strong interactions makes them penetrating probes that reflect the medium properties at the time of their emission which, on average, precedes that of the much more abundant strongly interacting hadrons. The slopes of their transverse momentum spectra and their anisotropic flow coefficients therefore show sensitivity to the hydrodynamic flow and dissipative medium properties at higher average temperature than the corresponding hadronic observables \cite{Shen:2013vja,Shen:2014cga}. Recent elliptic and triangular flow measurements of direct photons in Au+Au collisions at RHIC \cite{Adare:2011zr} and Pb+Pb collisions at the LHC \cite{Lohner:2012ct} have presented a puzzle to theorists because the measured flow anisotropies are much larger than theoretically predicted \cite{Shen:2013vja,Shen:2014cga}. This may point to stronger electromagnetic radiation from the critical crossover region between the QGP and hadronic phases \cite{vanHees:2014ida} than presently thought. Future higher-luminosity runs and improved detector capabilities will provide much more precise data for these very difficult measurements. In concert with state-of-the-art dynamical modeling, this is expected to not only yield a resolution of this ``direct photon flow puzzle'' but also lead to tighter constraints on the transport properties of QCD matter at higher temperatures. 

\fp{\bf The origin of collective behavior: Understanding the initial state and its thermalization}

Completion of the Little Bang Standard Model will also require a better 
understanding of the gluon densities in the initial overlap
regions. As discussed in sections~\ref{Sec:Saturation} and~\ref{sec4},
this topic can be studied in p+A collisions and ultimately at an
electron ion collider (EIC). The physics of p+A collisions has
proven to be richer than originally anticipated. While collective flow
phenomena are firmly established in sufficiently central collisions of
heavy nuclei, these effects were not expected in p+p and p+A
collisions. It was widely assumed that, as the mean-free-path of the
matter approaches the characteristic size of the system, the effects
of viscous damping would become too strong and invalidate a
hydrodynamic description. Surprisingly, correlations that are
long-range in rapidity and similar to those measured in A+A collisions
have now also been observed at the LHC in rare high-multiplicity
p+p collisions~\cite{Khachatryan:2010gv} (corresponding to high
gluon-density initial states). Subsequent measurements revealed
similar phenomena at both the LHC and RHIC in high multiplicity
p+Pb~\cite{CMS:2012qk,Abelev:2012ola,Aad:2012gla} and
d+Au~\cite{Adare:2013piz} collisions. In particular, the observed
mass ordering of the spectral slopes \cite{Chatrchyan:2013eya,Abelev:2013haa}
and of $v_n(p_T)$ \cite{Aad:2013fja,Chatrchyan:2013nka,ABELEV:2013wsa,Abelev:2014mda,Aad:2014lta,Khachatryan:2014jra} is
reminiscent of the effects arising from a common radial flow boost in
A+A collisions. Multi-particle measurements~\cite{CMS-PAS-HIN-14-006,CMS_cumulants_pPb}
show unambiguously that the correlations in high-multiplicity p+Pb
collisions are collective.

Understanding the relationship between initial and final-state effects
in these systems appears to be non-trivial and has now developed into
a very active research area. While hydrodynamic models with strong
final-state interactions may provide a natural interpretation for many
of the observed features, their applicability in such small systems,
including the process of rapid thermalization, requires additional
scrutiny \cite{Niemi:2014wta}. Meanwhile, other novel mechanisms,
exploiting quark and gluon momentum correlations in the initial state,
have been proposed as alternative interpretations of the observed
long-range correlations and even predicted the correlation results in
p+p collisions. Recently collected data from $^3$He+Au collisions 
will shed additional light on this question, as will improved correlation
measurements at forward rapidity, enabled by current and future
detector upgrades. These programs play an important role in 1)
understanding the initial-state component of our Little Bang Standard 
Model and its subsequent modification by final-state interactions, 2) 
testing how small a system can be and still exhibit fluid-like phenomena,
3) providing new opportunities to probe the spectrum of fluctuations
in high gluon density matter, and 4) mapping the transition to a
classical description of gluonic matter at high density. 

It is suspected that the rapid formation of almost perfectly liquid hot QCD matter in heavy-ion collisions may be related to the emergence of universal characteristics in high-density gluon matter at zero temperature that is predicted to dominate the low-$x$ component of the nuclear wave function when probed at high energy. To explore this connection, precision measurements of the nuclear wave function at an EIC will be required to complement nuclear collision experiments with small and large nuclei.  
\subsection{Mapping the QCD phase diagram}
\label{sec:PD}

\begin{figure}[htb]
\begin{center}
\begin{minipage}{\textwidth}
\begin{minipage}{0.55\textwidth}
\includegraphics[width=\textwidth]{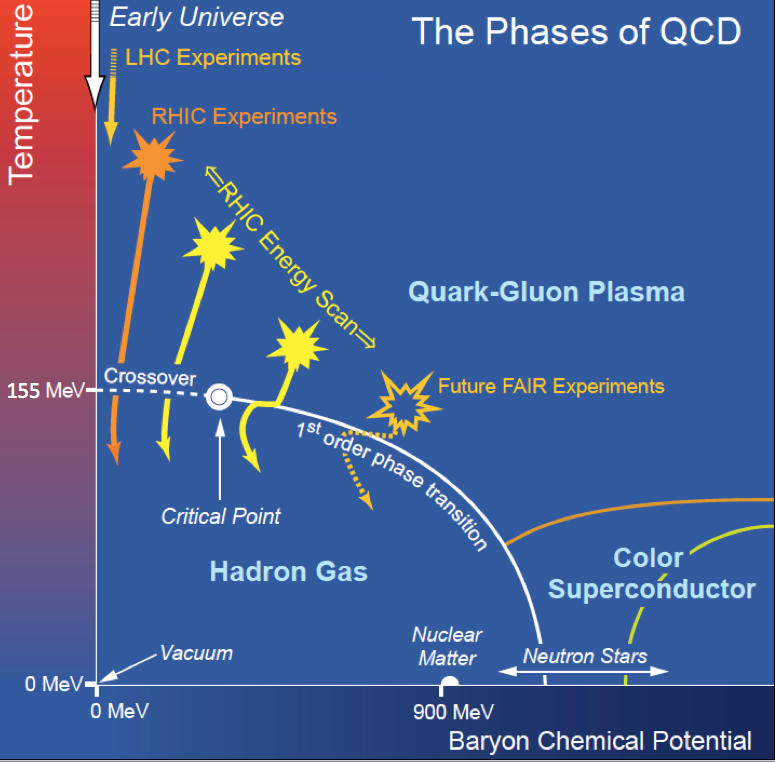}
\end{minipage}
\hspace*{4mm}
\begin{minipage}{0.4\textwidth}
\caption{A sketch illustrating the experimental and theoretical
  exploration of the QCD phase diagram. Although experiments at
  highest energies and smallest baryon chemical potential are known to
  cross from a QGP phase to a hadron gas phase through a smooth
  crossover, lower energy collisions can access higher baryon chemical
  potentials where a first order phase transition line is thought to
  exist.
\label{F-PD1}
}
\end{minipage}
\end{minipage}
\end{center}
\end{figure}

When the first protons and neutrons and pions formed in the microseconds-old universe, and when they form in heavy-ion collisions at the highest RHIC energies and at the LHC, they condense out of liquid quark-gluon plasma consisting of almost as much antimatter as
matter. Lattice calculations \cite{Aoki:2006we,Aoki:2009sc,Bazavov:2011nk} 
show that QCD predicts that, in such an environment, this condensation 
occurs smoothly as a function of decreasing temperature, with many 
thermodynamic properties changing dramatically but continuously within 
a narrow temperature range around the transition temperature 
$T_c\in [145\,\mathrm{MeV},163\,\mathrm{MeV}]$
\cite{Bazavov:2011nk,Bazavov:2014pvz}, referred to as the crossover
region of the phase diagram of QCD, see Fig.~\ref{F-PD1}.
%
%
In contrast, quark-gluon plasma doped with a sufficient excess of
quarks over anti-quarks may instead experience a sharp first order
phase transition as it cools, with bubbles of quark-gluon plasma and
bubbles of hadrons coexisting at a well-defined critical temperature,
much as bubbles of steam and liquid water coexist in a boiling
pot. The point where the doping of matter over antimatter
(parametrized by the net baryon number chemical potential $\mu_B$)
becomes large enough to instigate a first order phase transition is
referred to as the QCD critical point. It is not yet known whether QCD
has a critical point
\cite{Stephanov:1998dy,Fodor:2004nz,Allton:2005gk,Gavai:2008zr,deForcrand:2008zi},
nor where in its phase diagram it might lie. Lattice calculations
become more difficult or more indirect or both with increasing $\mu_B$
and, although new methods introduced within the past decade have
provided some hints \cite{Fodor:2004nz,Gavai:2008zr,Datta:2012pj}, at
present only experimental measurements can answer these questions
definitively. The theoretical calculations are advancing, however,
with new methods and advances in computational power both anticipated.

The phase diagram of QCD, with our current knowledge schematically
shown in Fig.~\ref{F-PD1}, is the only phase diagram of any form of
matter in Nature that we have a chance of both mapping experimentally
and relating directly and quantitatively to our fundamental
description of Nature, the Standard Model. With QCD the only strongly
interacting theory in the Standard Model, mapping the transition
region of its phase diagram is a scientific goal of the highest
order. In the long term, successfully connecting a quantitative,
empirical understanding of its phases and the transitions between
phases to theoretical predictions obtained from the QCD Lagrangian
could have ramifications in how we understand phases of strongly
coupled matter in many other contexts.

{\bf RHIC's unique capability to probe the QCD phase diagram}

\begin{figure}[h!]
\vspace*{-3mm}
\begin{minipage}{\textwidth}
\begin{center}
\begin{minipage}{0.5\textwidth}
\hspace*{-12mm}
\includegraphics[width=1.22\textwidth]{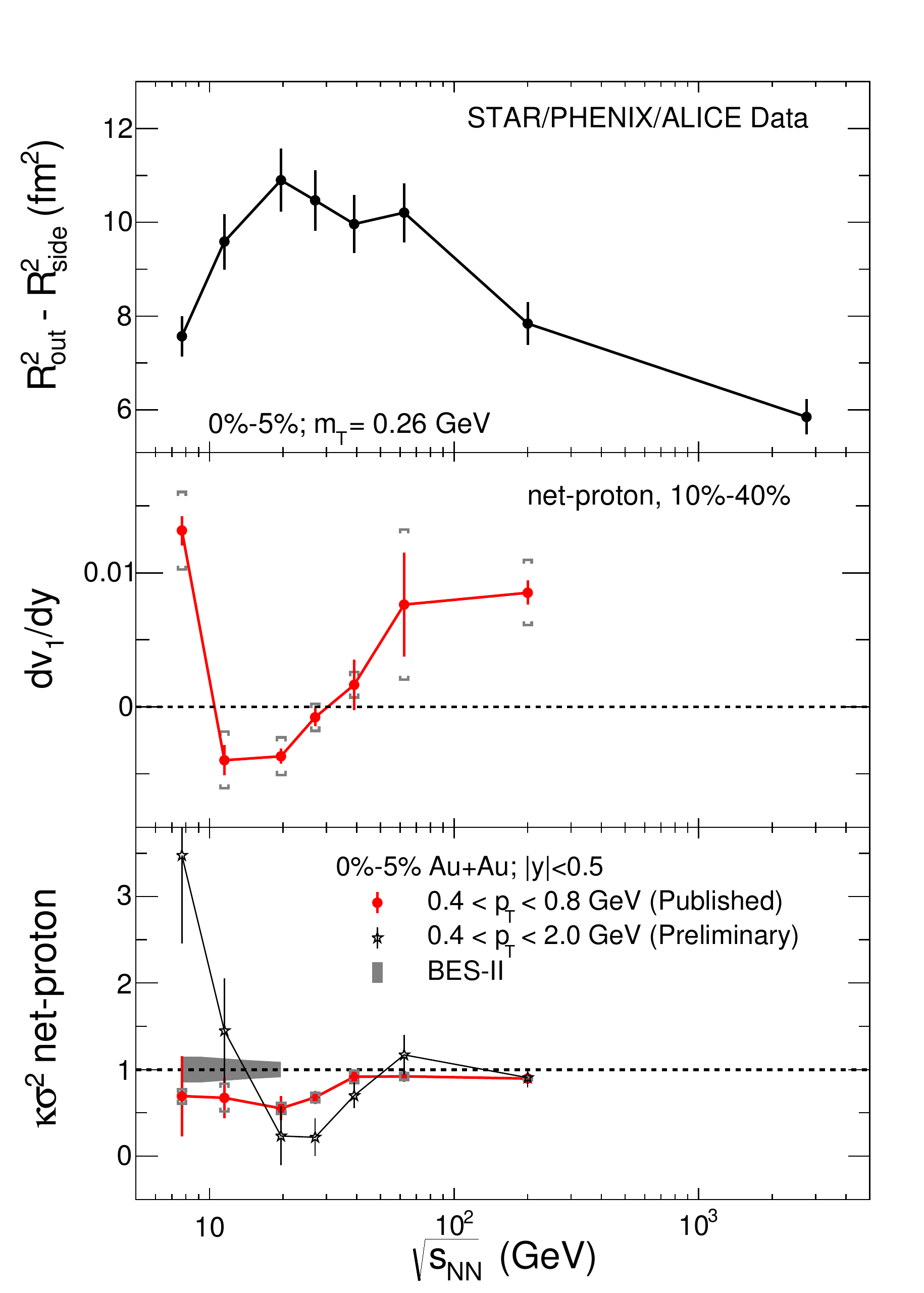}
\end{minipage}
\hspace*{2mm}
\begin{minipage}{0.47\textwidth}
\caption{Three selected observables that all show interesting non-monotonic behavior
     as functions of collision energy around $\sqrt{s_\mathrm{NN}}{\,\sim\,}15{-}20$\,GeV.
     \newline 
{\bf Top panel:} $R_{out}^{2}{-}R_{side}^{2}$, measured via 
     two-pion interferometry by STAR \cite{Adamczyk:2014mxp}, PHENIX
     \cite{Adare:2014qvs}, and ALICE \cite{Aamodt:2011mr}, reflects the lifetime of 
     the collision fireball. $R_{out}^{2}{-}R_{side}^{2}$ was predicted 
     \cite{Rischke:1996em} to reach a maximum for collisions in which the 
     hydrodynamic fluid forms at temperatures where the equation of state is softest.
     \newline 
{\bf Middle panel:} The rapidity-slope of the net proton directed flow $v_1$, 
     $dv_1/dy$. This quantity is sensitive to early pressure fields in the medium. 
{\bf Bottom panel:} The kurtosis of the event-by-event distribution of the net proton
     (i.e. proton minus antiproton) number per unit of rapidity, normalized such that 
     Poisson fluctuations give a value of $1$. In central collisions, published results 
     in a limited kinematic range \cite{Adamczyk:2013dal} show a drop below the 
     Poisson baseline around $\sqrt{s_\mathrm{NN}}=$27 and 19.6 GeV. New preliminary 
     data over a larger $p_T$ range \cite{CPODKurtosis}, although at present with 
     substantial uncertainties, hint that the normalized kurtosis may, in fact, rise 
     above 1 at lower $\sqrt{s_\mathrm{NN}}$, as expected from critical fluctuations 
     \cite{Stephanov:2011pb}. The grey band shows the much reduced uncertainties 
     anticipated from BES-II in 2018-2019, for the 0-5\% most central collisions.
\label{F-PD2}
\vspace*{-5mm}
}
\end{minipage}
\end{center}
\end{minipage}
\end{figure}

A major effort to use heavy-ion collisions at RHIC to survey the phase
diagram of QCD is now underway. The excess of matter over antimatter
in the exploding droplet produced in a heavy-ion collision can be
increased by decreasing the collision energy, which reduces the
production of matter-antimatter symmetric quark-antiquark pairs and
gluons relative to the quarks brought in by the colliding nuclei, thus
increasing $\mu_B$. Decreasing the collision energy also decreases the
maximum, i.e. initial, temperature reached by the matter produced in
the collision. A series of heavy-ion collision measurements scanning
the collision energy \cite{BESII} can therefore explore the properties
of matter in the crossover region of the phase diagram, matter that is
neither quark-gluon plasma nor hadronic or both at the same time, as a
function of the doping $\mu_B$. Such a program can scan the transition
region of the QCD phase diagram out to $\mu_B$ values that correspond
to collision energies below which the initial temperature no longer
reaches the transition. If the crossover region narrows to a critical
point within this experimentally accessible domain, an energy scan can
find it. RHIC completed the first phase of such an energy scan in
2014, taking data at a series of energies ($\sqrt{s_\mathrm{NN}}=$ 200, 62.4,
39, 27, 19.6, 14.5, 11.5 and 7.7 GeV) corresponding to values of
$\mu_B$ that range from 20 to 400 MeV. Data from these experiments at
RHIC \cite{Kumar:2012fb,BESII} and from previous experiments confirm 
that lower-energy collisions produce QGP with higher $\mu_B$, as
anticipated. A selection of data from BES-I that exhibit interesting 
non-monotonic behavior as a function of collision energy is shown in 
Fig.~\ref{F-PD2}.

%
%

RHIC is, and will remain, the optimal facility in the world for
studying matter in the crossover region and searching for a possible
critical point in the so far less well understood regions of the phase
diagram with larger $\mu_B$. What makes RHIC unique is both its wide
reach in $\mu_B$ and that it is a collider, meaning that the
acceptance of detectors, and hence the systematics of making
measurements, change little as a function of collision
energy. Accelerator and detector performance has been outstanding
during the first phase of this program, referred to as Beam Energy
Scan I or BES-I. Measurements of all the important observables
targeted in the planning of this campaign have now been made in
collisions with energies varying by a factor of 25, allowing for a
first look at a large region of the phase diagram of QCD.

{\bf The need for BES-II and accompanying advances in theory}

We still await with interest the results from the most recent run in
this program at $\sqrt{s_\mathrm{NN}}=14.5$\,GeV, where data were taken only
a few months ago; and for a number of important observables the
measurements made at and below $\sqrt{s_\mathrm{NN}}=19.6$\,GeV have quite
limited statistics. Nevertheless, it is already possible to see trends
and features in the data that provide compelling motivation not only
for a strong and concerted theoretical response aiming at quantitative
precision, but also for much higher statistics data at the lower end
of the energy range (i.e. at the largest achievable values of $\mu_B$)
that will be provided by the second phase of the Beam Energy Scan
program (BES-II) in 2018 and 2019. The goals of BES-II, as described
in more detail below, are to follow through and turn trends and
features into definitive scientific conclusions. To this end, the
accelerator physicists at RHIC are planning a machine upgrade to
provide electron cooling to increase the beam luminosity at these
energies by about a factor 10 \cite{BESII}. \fp{Targeted new detector
capabilities (an Inner Sector Upgrade for the STAR TPC and a new event plane
detector) will also increase the sensitivity to signals already observed in
BES-I and open the door to new measurements described below in the 
BES-II campaign \cite{BESII}.}

Experimental discovery of a first order phase transition or a critical
point on the QCD phase diagram would be a landmark achievement. The
first goals of the BES program, however, relate to obtaining a quantitative 
understanding of the properties of matter in the crossover region of the 
phase diagram as it changes with increasing $\mu_B$. Available tools
developed over the last few years now make a quantitative comparison
between theory and experiment tractable in the $\mu_B$-range
below any QCD critical point. Success in this, in and of itself, would
constitute a major and lasting impact of the RHIC program. Questions
that can be addressed in this regime include quantitative study of the
onset of various signatures associated with the presence of
quark-gluon plasma and of the onset of chiral symmetry restoration as
one traverses the crossover region. Data now in hand from BES-I
provide key inputs and impetus toward this goal. Here we give {\bf four
examples}, intended to be illustrative, of areas where a coherent
experimental and theoretical effort is expected to have substantial
impact on our understanding of QCD. In each case we also note the
substantial impact expected from the additional measurements
anticipated during the BES-II:

{\bf 1.} The directed flow observable $dv_1/dy$ for net protons has been 
found to feature a dip as a function of collision energy (see middle panel 
in Fig.~\ref{F-PD2}), with a minimum at energies somewhere between 
$\sqrt{s_\mathrm{NN}}=11.5$ and 19.6\,GeV \cite{Adamczyk:2014ipa}. This has 
long been predicted in qualitative terms as a consequence of the 
softening of the equation of state in the transition region of the phase 
diagram \cite{Brachmann:1999mp,Stoecker:2004qu}. Several theoretical 
groups around the world have now begun hydrodynamic calculations with 
nonzero baryon density, deploying all the sophistication that has been
developed very recently in the analysis of higher energy collisions,
including initial fluctuations and a hadronic afterburner, in
applications to these lower energy collisions. These
hydrodynamic+hadronic cascade calculations will be used to compare the
$dv_1/dy$ data with equations of state in the crossover region of the
phase diagram obtained from lattice calculations via Taylor expansion
in $\mu_B/T$ \cite{Huovinen:2014woa}. This is a program where a
quantitative comparison, successful or not, will be of great interest,
since failure to describe the data could signal the presence of a
first order phase transition. The precision of a comparison like this
will be substantially improved in 2018-19 when BES-II data will allow
$dv_1/dy$ to be measured for the first time with tightly specified
centrality; the statistics available in the BES-I data sets limit
present measurements to averages over collisions with widely varying
impact parameters \cite{Adamczyk:2014ipa}.

{\bf 2.} A second goal of the hydrodynamic calculations referred to
above will be to use identified particle BES-I $v_2$ data to map, in
quantitative terms, where and how hydrodynamics starts to break down
at lower collision energies, and where, to an increasing extent, $v_2$
develops during the hadron gas phase when viscosities are not small,
i.e. where the contribution of the partonic phase to observed measures
of collectivity decreases in importance. A key future experimental
input to this program is the measurement of the elliptic flow $v_2$ of
the $\phi$-meson, which will be obtained with substantially greater
precision in the BES-II program. The first measurements of $v_2$ of
$\Omega$ baryons at these collision energies, also anticipated in
BES-II, will represent a further, substantial advance. Seeing $\phi$
mesons flowing like lighter mesons and $\Omega$ baryons flowing like
lighter baryons in collisions at a given energy would indicate that
the dominant contribution to the collective flow in those collisions
was generated during the partonic phase \cite{Abelev:2007rw}.

This component of the BES program, together with the following one,
will yield guidance as to what the lowest collision energies are at
which temperatures in the transition region of the phase diagram can
be explored. That is, they will tell us out to what $\mu_B$ it will be
possible for heavy-ion collisions, anywhere, to study matter in the
crossover region and search for a possible critical point.

{\bf 3.} Heavy-ion collisions at top RHIC energies and at the LHC have
now seen several experimental phenomena
\cite{Abelev:2009ac,Abelev:2012pa,Adamczyk:2013kcb} that may be
related to the chiral magnetic effect (CME
\cite{Fukushima:2008xe,Kharzeev:2010gr}, see Sec.\ref{sec:open}). In
each case, alternative explanations are also being considered
\cite{Bzdak:2009fc,Pratt:2010zn}. One of the intriguing BES-I results
is that the three-particle correlations that are related to charge
separation across the reaction plane, possibly induced by the CME, are
robustly observable over most of the BES range but then seem to turn
off at $\sqrt{s_\mathrm{NN}}=$ 7.7 GeV \cite{Adamczyk:2014mzf}, where the
elliptic flow $v_2$ is still robust. This is an indication that
$v_2$-induced backgrounds alone do not explain the observed
correlations. The observation that these three-particle correlations
disappear at the lowest energy could prove crucial to understanding
their origin and how they are related to the formation of QGP. On the
theoretical side, lattice QCD calculations probing the response of the
equation of state and transition temperature to the presence of
external magnetic fields \cite{DElia:2010nq,Bali:2011qj,Bali:2014kia},
as well as hydrodynamic calculations incorporating magnetic fields and
chiral effects, are needed and are being pursued by several groups. On
the experimental side, higher statistics BES-II data will make it
possible to determine with much greater precision the $\sqrt{s_\mathrm{NN}}$
at which this effect turns off and will also make it possible to
measure the (related but theoretically more robust) chiral magnetic
wave phenomenon \cite{Kharzeev:2010gd,Burnier:2011bf}, which has also
been seen at top RHIC energy and at the LHC
\cite{Wang:2012qs,Belmont:2014lta}, and which should turn off at the
same $\sqrt{s_\mathrm{NN}}$ if these interpretations are correct.

{\bf 4.} Theoretical developments over the past decade have identified
specific event-by-event fluctuation observables most likely to be
enhanced in collisions that cool in the vicinity of the critical point
\cite{Stephanov:2008qz,Athanasiou:2010kw}. Higher moments of the
event-by-event distribution of the number of protons, or the net
proton number, are particularly sensitive 
\cite{Ejiri:2005wq,Athanasiou:2010kw,Karsch:2010ck}. STAR has now 
measured the first four moments (mean, variance, skewness and kurtosis) 
of the event-by-event distribution of net proton number and net charge at 
the BES-I energies \cite{Adamczyk:2013dal,Adamczyk:2014fia}. At the 
lowest collision energies, although the statistics are at present rather
limiting, there are interesting trends, including e.g. the drop in the
kurtosis of the net-proton distribution at $\sqrt{s_\mathrm{NN}}{\,=\,}27$
and 19.6 GeV (see bottom panel in Fig.~\ref{F-PD2}). This drop in and 
of itself can be at least partially reproduced via prosaic effects captured 
in model calculations that do not include any critical point. Theoretical 
calculations of the contributions from critical fluctuations predict  
\cite{Stephanov:2011pb} that if the freezeout $\mu_B$ scans past a 
critical point as the beam energy is lowered, this kurtosis should first 
drop below its Poisson baseline and then rise above it. Both the drop 
and the rise should be largest in central collisions in which the 
quark-gluon plasma droplet is largest and therefore cools most slowly, 
allowing more time for critical fluctuations to develop
\cite{Berdnikov:1999ph}. A recent and still preliminary analysis
\cite{CPODKurtosis} of the data at $\sqrt{s_\mathrm{NN}}{\,=\,}11.5$ and 
7.7\,GeV over a larger range in $p_T$ than measured before 
\cite{Adamczyk:2013dal}, also shown in the bottom panel of 
Fig.~\ref{F-PD2}, shows intriguing hints of a rise in the net proton 
kurtosis in central collisions, but the uncertainties are at present too 
large to draw conclusions. If this kurtosis does rise at $\sqrt{s_\mathrm{NN}}$ 
values below 19.6 GeV, this would be difficult to understand in 
conventional terms and thus would be suggestive of a contribution 
from the fluctuations near a critical point. Determining whether this 
is so requires the higher statistics that BES-II will provide, as illustrated
by the grey band in the bottom panel of Fig.~\ref{F-PD2}. 

The present data on moments of both the net proton number and the 
net charge at the higher BES-I energies are already very useful, as
they can be compared to lattice calculations of the Taylor expansions 
(in $\mu_B/T$) of the baryon number and charge susceptibilities
\cite{Karsch:2012wm}. First versions of this comparison have been
reported recently and are being used to provide an independent
determination of how the freeze-out values of $\mu_B$ and $T$ change
with collision energy \cite{Bazavov:2012vg,Mukherjee:2013lsa,Borsanyi:2013hza,Borsanyi:2014ewa}. However, looking ahead, theoretical 
calculations will need to faithfully
account for the dynamical evolution of the medium formed in the
collision for a full quantitative exploitation of the experimental
data. For the higher statistics BES-II data on the net proton
kurtosis, skewness, and other fluctuation observables at low collision
energies to determine the location of the critical point on the phase
diagram of QCD, if one is discovered, or to reliably exclude its
existence within the experimentally accessible region of the phase
diagram, a substantial theoretical effort will be needed that couples
the sophisticated hydrodynamic calculations referred to above with a
fluctuating chiral and dynamically evolving order parameter.

As the following {\bf fifth example} illustrates, BES-II will also open the
door to measurements that were not yet accessible in the first phase
of the BES program:

{\bf 5.} \fp{Dileptons are unique penetrating probes to study the chiral properties of hot and dense matter \cite{Rapp:2009yu,Hohler:2013eba}. The dielectron invariant mass distributions measured in the BES-I (in data taken at $\sqrt{s_\mathrm{NN}}=$ 200, 62.4, 39, 27 and 19.6 GeV) have shown that there is a significant enhancement of low mass dileptons below 1 GeV relative to a hadronic cocktail~\cite{Huck:2014mfa,Adamczyk:2013caa,Adamczyk:2015bha,Rapp:2013nxa}. The data to date are consistent with model predictions of a $\rho$ resonance that broadens through interactions with the hadronic medium \cite{Rapp:2009yu} and melts in a partonic medium. This picture captures the essential features of the low-mass excess at small chemical potentials \cite{Rapp:2013nxa}. Data at lower energies with higher statistics are crucial in order to test the predicted dependence of dilepton yields on baryon density and draw firm conclusions. The dilepton measurements at and below $\sqrt{s_\mathrm{NN}}{\,=\,}19.6$\,GeV that BES-II will provide will advance our understanding of the chiral properties of QCD matter with significant net baryon density. In addition, the low-mass excess can serve as a chronometer for the fireball lifetimes as the QCD phase diagram is scanned \cite{Rapp:2014hha}. At higher masses (above 1\,GeV) chiral restoration can be probed via the mixing between vector and axial-vector currents in the hot and dense environment, and there the slope of the dilepton mass spectrum provides access to the early temperature in the system, unaffected by collective flow \cite{Rapp:2014hha}.}

Each of these five examples makes it clear that in order to maximize
the physics outcome from BES-I and BES-II, a coherent effort between
experimentalists and theorists working on QCD at nonzero $T$ and
$\mu_B$ is essential and must be organized and supported. Indeed,
there has been considerable progress in lattice QCD recently on the
calculation of various QCD susceptibilities
\cite{Borsanyi:2011sw,Bazavov:2012jq} and the QCD equation of state in
the regime where $\mu_B$ is nonzero but sufficiently small compared to
$3\,T_c$ \cite{Borsanyi:2012cr,Hegde:2014wga}. These lattice
calculations provide the necessary inputs for extending to nonzero
$\mu_B$ the kind of sophisticated hydrodynamic calculations (including
initial fluctuations and a late stage hadron cascade) that have been
developed over the past few years. For some purposes, these
calculations additionally require dynamical coupling to a fluctuating
chiral order parameter.

In concert, such developments will provide the critical tools for
obtaining from BES-I and BES-II data answers to fundamental questions
about the phases, the crossover, and perhaps the critical point and
first order transition, in the QCD phase diagram. Quantitative
understanding of the properties of matter in the crossover region
where QGP turns into hadrons will come first. If there is a critical
point with $\mu_B<400$~MeV, BES-II data together with the theoretical
tools now being developed should yield quantitative evidence for its
presence. The span in $T$ and $\mu_B$ that the flexibility of RHIC
makes accessible, along with the mentioned technical advantages of
measuring fluctuation observables at a collider and recent and planned
detector and facility upgrades (e.g. low energy electron cooling), put
RHIC in a globally unique position to discover a critical point in the
QCD phase diagram if Nature has put this landmark in the
experimentally accessible region. Late in the decade, the FAIR
facility at GSI will extend this search to even higher $\mu_B$ if its
collision energies continue to produce matter at the requisite
temperatures.

\subsection{Probing hot QCD matter at multiple length scales}
\label{sec:probes}

Hard probes, such as jets and bottomonia, provide the unique opportunity to test the microscopic structure of hot QCD matter at characteristic momentum and length scales. Below we describe the qualitative and quantitative insights into QGP properties these probes have already provided and outline a future program that aims to understand how the QGP properties arise from its microscopic nature.

\subsubsection{Jets as microscopic probes of QGP}
\label{Sec:Jets}

The properties of jets~\cite{Sterman:1977wj} and their emergence from pQCD have been extensively studied in high-energy physics~\cite{Feynman:1978dt,Field:1977fa}. One of the earliest discoveries at RHIC was the phenomenon of \em jet quenching \em, observed as a suppression of high-$p_T$ hadrons and di-hadron correlations in Au+Au collisions~\cite{Adcox:2001jp,Adler:2002tq}. This can be understood as a medium-modification of jet showers \cite{Bjorken:1982tu} through an enhanced rate of gluon bremsstrahlung, resulting in a depletion of high momentum parton fragmentation products.

The virtuality of a hard parton within a jet establishes the intrinsic scale at which it resolves fluctuations in the medium.  As partons cascade down to lower virtualities, they probe the medium over a multitude of length scales. As long as this resolution scale is much larger than $\Lambda_{QCD}$, the parton will be weakly coupled with the medium and pQCD can be applied to describe its propagation~\cite{Gyulassy:1993hr,Wang:1991xy,Baier:1996kr,Baier:1996sk,Zakharov:1996fv,Zakharov:1997uu,Gyulassy:2000er,Wang:2001ifa, Arnold:2002ja,Majumder:2009ge}. Some partons within a jet may lose sufficient amounts of energy to encounter non-perturbatively strong coupling~\cite{Chesler:2008uy,Chesler:2008wd,Friess:2006aw,CasalderreySolana:2006rq}. 

\noindent {\bf Jet transport parameters from single-hadron suppression}

The medium-induced changes to the shower radiation pattern can be described as longitudinal drag/diffusion, transverse diffusion, and enhanced splitting of the propagating partons. The transport coefficients $\hat{q}$~\cite{Baier:2002tc} and $\hat{e}$~\cite{Majumder:2008zg} quantify the transverse diffusion and longitudinal drag, respectively.

\begin{figure}[!h]
\vspace*{-2mm}
\centerline{
\hspace*{-4mm}\includegraphics[width=1.08\textwidth]{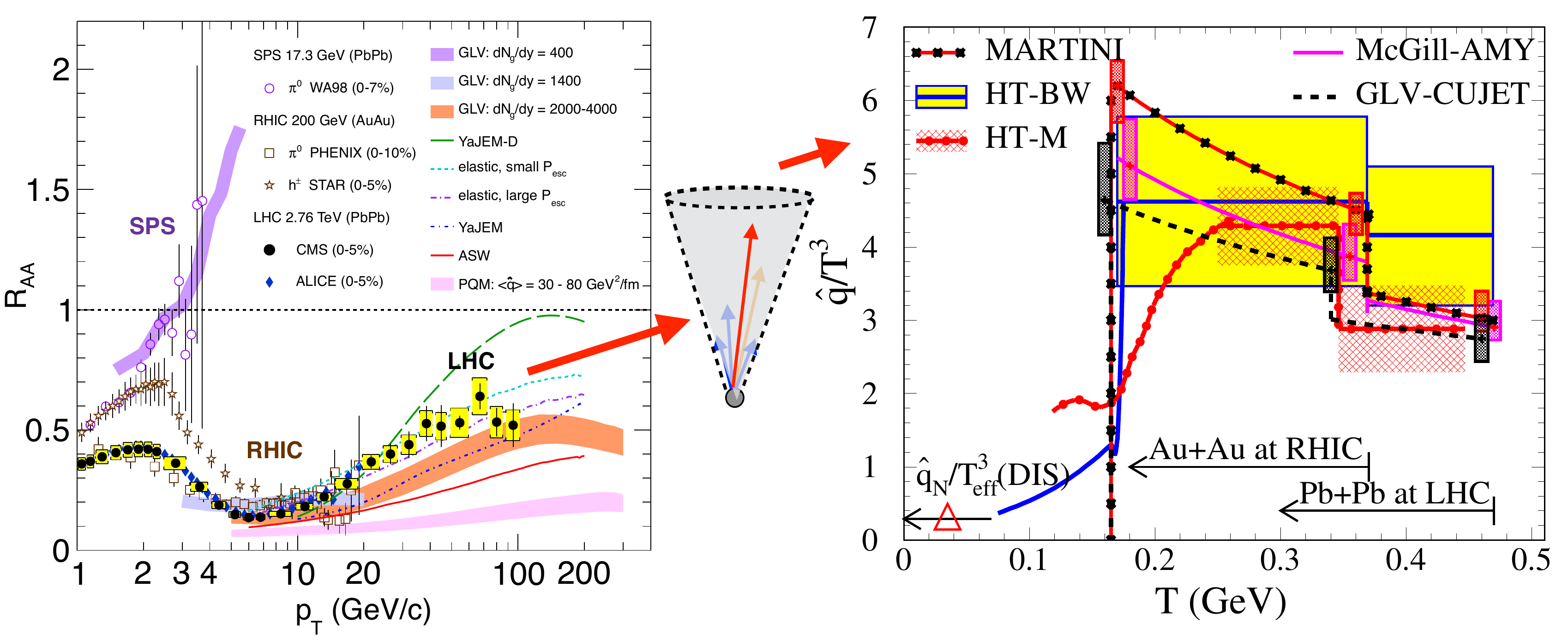}}
\vspace*{-4mm}
\caption{A comparison of several different pQCD based energy loss energy loss schemes to the measured leading hadron suppression in central events at RHIC and LHC \cite{CMS:2012aa}, and the extracted scaled (i.e. dimensionless) transport coefficient $\hat{q}/T^3$ along with its dependence on temperature \cite{Burke:2013yra}.
\vspace*{-3mm}}
\label{fig:JetProgressFig1}
\end{figure}

Since the last NP long-range plan, enormous progress has been achieved in the quantitative determination of $\hat{q}$ and $\hat{e}$. This has been made possible by new high precision data from RHIC and LHC, as well as coordinated theory efforts, notably by the JET collaboration. One now obtains a very good description of the combined RHIC and LHC data on single hadron suppression, as shown in the left panel of Fig.~\ref{fig:JetProgressFig1}. The extracted values and temperature dependence of the dimensionless ratio $\hat{q}/T^3$ are plotted in the right panel of Fig.~\ref{fig:JetProgressFig1}. Using identical initial states and hydro simulations, the JET collaboration has carried out a systematic analysis of a wide range of pQCD based energy loss schemes~\cite{Burke:2013yra}. Compared to earlier studies~\cite{Bass:2008rv}, where the extracted $\hat{q}$ varied by an order of magnitude, the values in the new analysis \cite{Burke:2013yra} differ at most by a factor of 2. This makes it possible for the first time to discern the medium temperature dependence of $\hat{q}/T^3$ and demonstrates that quantitative properties of the QGP can be extracted from jet modification data. 

\begin{figure}[t]
\vspace*{-10mm}
\begin{center}
\hspace*{-7mm}\includegraphics[width=1.05\textwidth]{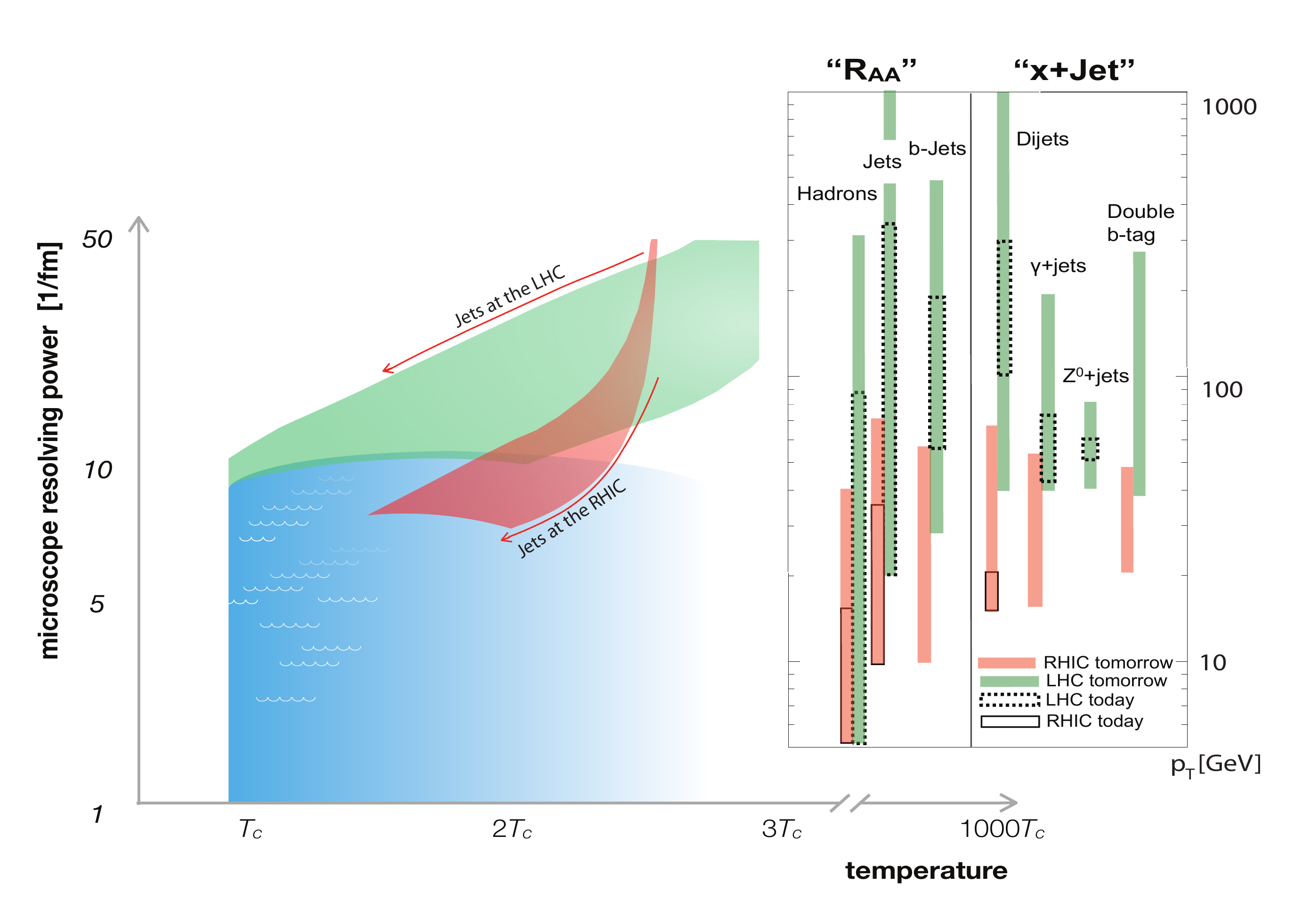}
\end{center}
\vspace*{-8mm}
\caption{Shown on the left are jet virtuality evolution paths simultaneous with the QGP temperature evolution for central Au+Au and Pb+Pb collisions at RHIC and the LHC. The vertical axis shows the inverse length scale probes in the medium and the blue region where the medium is substantially altering the vacuum parton splitting~\cite{sPHENIX}. This displays the complementarity of RHIC and LHC jet probes in surveying distinct regions of possible QGP excitations and temperature. The right panel shows the momentum regions covered by penetrating probe measurements at RHIC, enabled by high luminosity and the sPHENIX upgrade, and at the LHC with full Run~II and Run~III statistics.}
\label{Fig:HardProbesFuture}
\end{figure}

\noindent {\bf Fully reconstructed jets}

Fully reconstructed jets have been a crucial tool in high energy physics for precision tests of pQCD~\cite{Sterman:1977wj}. Experimental progress now allows one to isolate and reconstruct the entire jet shower from the high-multiplicity underlying event in heavy-ion collisions. Some of the first LHC heavy-ion results included the observation \cite{Aad:2010bu,Chatrchyan:2011sx} of highly asymmetric dijet events \cite{Qin:2010mn,Lokhtin:2011qq,Young:2011qx,Renk:2012cb}, which provided a striking visual demonstration of the energy loss. Since these initial measurements, experimental control over the measured jet energies and the understanding of the role of underlying event fluctuations have improved substantially, resulting in precise measurements of jet suppression and the properties of quenched jets. Photon-jet correlations \cite{Dai:2012am,Qin:2012gp,Wang:2013cia} have made it possible to select the parton kinematics and flavor before energy loss, providing direct evidence for the degradation of the parton energy as they traverse the medium~\cite{Chatrchyan:2012gt,Steinberg:2014xha,Adare:2012qi}.

These measurements at the LHC, and pioneering studies at RHIC \cite{Adamczyk:2013jei}, have contributed qualitatively new information \cite{Vitev:2008rz,Vitev:2009rd,Renk:2014lza} on jet modification by the medium. Studies of the cone-size dependence of jet spectra \cite{He:2011pd}, jet-hadron correlations \cite{Renk:2012hz} and the overall energy flow in dijet events demonstrate that the energy transported out of the hard core of the jet, as seen in hadron suppression, is not contained within typical jet cone radii~\cite{Aad:2012vca,Chatrchyan:2012gw}. Rather, the energy is recovered in low momentum modes (few GeV or less) at large angles to the jet direction. The jet structure within the cone reveals only a moderate modification, undergoing a softening and broadening of the fragmentation pattern.
  
\noindent {\bf Heavy-quark jets and the mass-dependence of parton energy loss}

The LHC also made new tests of the mass dependence of energy-loss possible using heavy-quark measurements. Data on D-meson production and non-prompt $J/\psi$ from B-meson decays (for $p_T$ up to 20 GeV/c) exhibit the predicted mass hierarchy \cite{ALICE:2012ab,Chatrchyan:2012np}, with the heaviest quarks losing significantly less energy than the lighter flavors.  At jet energies of 80 GeV and above, i.e. much higher than the quark mass, tagged $b$-quark jets show a similar suppression pattern as inclusive jets, \fp{as theoretically expected. At RHIC, pioneering studies of electrons from heavy flavor decays \cite{Adare:2006nq}, of directly reconstructed $D^0$ mesons \cite{Adamczyk:2014uip}, and of bottom/charm-separated non-photopic electrons \cite{Adare:2009ic,Aggarwal:2010xp} have shown considerable heavy-flavor suppression, with nuclear modification factors approaching those of light-flavor hadrons in the range $5\,\mathrm{GeV}/c{\,<\,}p_T{\,<\,}10$\,GeV/$c$.} \\[-0.5ex]

\noindent {\bf Future plans -- facilities, detectors and measurements}

Future jet-based studies built on the achievements at RHIC and LHC will address fundamental questions about the nature of QGP. These include precise measurements of QGP transport coefficients as a function of temperature, a detailed characterization of the QGP response to the parton energy loss and studies of the modification the jets'\ angular and momentum structure as a function of angular and momentum scale. In combination, the goal of these studies is to understand the microscopic (or quasi-particle) nature of QGP at varying scales, and to understand how the macroscopic QGP liquid emerges from the underlying QCD degrees of freedom. A schematic sketch of the present and expected future resolving power for the structure of QCD matter at different temperatures in RHIC and LHC experiments is shown in Fig.~\ref{Fig:HardProbesFuture}~\cite{sPHENIX}.

This program will be enabled by the evolution of the RHIC and LHC accelerator facilities, upgrades to the existing experiments and the construction of a state-of-the-art jet detector at RHIC, sPHENIX \cite{sPHENIX}. In parallel, the recently emerged experiment/theory collaborations will be strengthened and expanded, to fully utilize the increased precision and range of experimental observables. 

\begin{figure}[!hbt]
\vspace*{-5mm}
   \centering
   \begin{minipage}[b]{0.59\textwidth}
   \hspace*{-1mm}\includegraphics[width=1.0\textwidth]{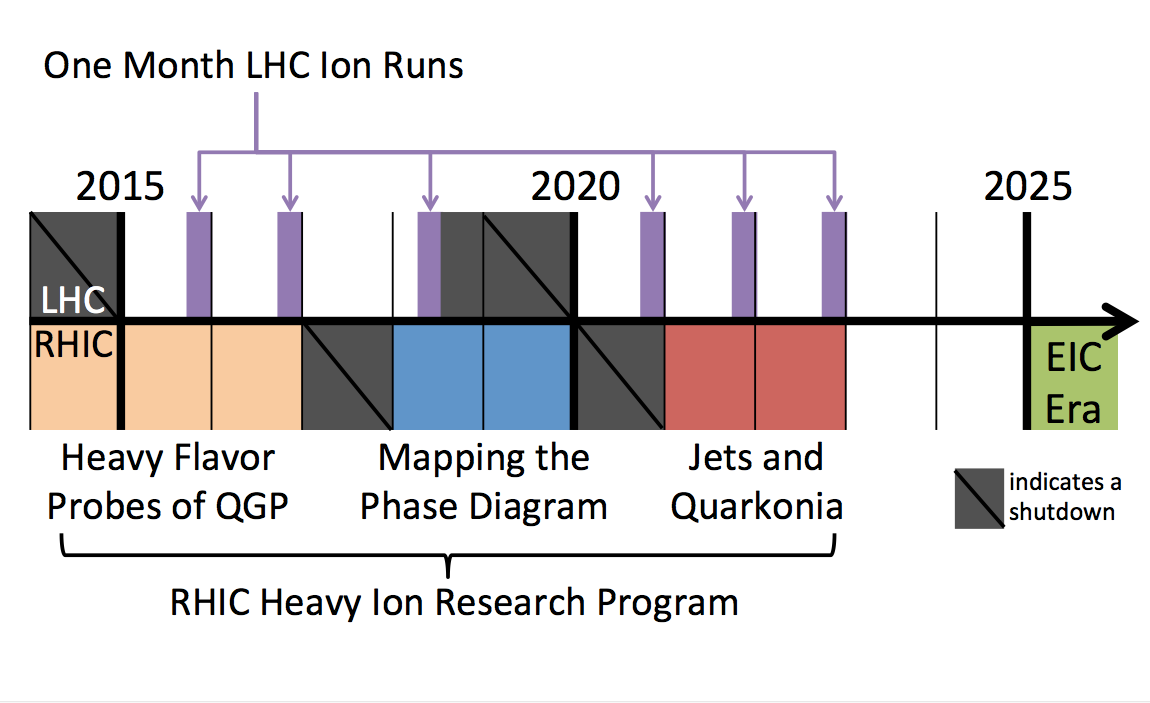}
   \end{minipage}
   \begin{minipage}[b]{0.4\textwidth}
   \hspace*{3mm}\includegraphics[width=\textwidth]{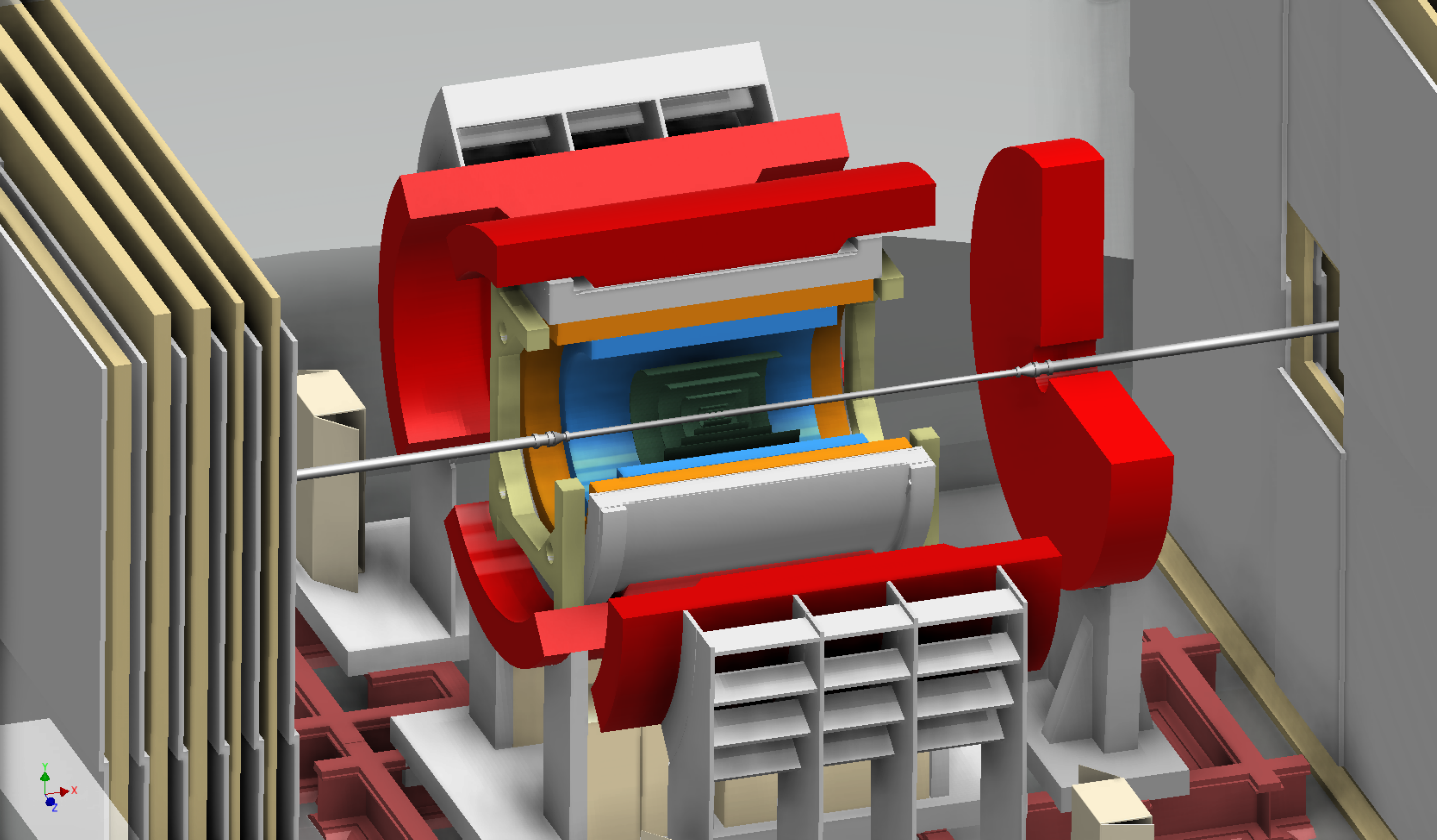}
   \vspace*{5mm}
   \end{minipage}
   \vspace*{-12mm}
     \caption{(Left) The timeline for future RHIC and LHC heavy-ion running.
        (Right) A cut-away view of the proposed sPHENIX upgrade to the PHENIX 
        experiment, showing the inner tracking system, the electromagnetic calorimeter, 
        the BaBar solenoid and the hadronic calorimeter.}
   \label{Fig:Timeline}
\end{figure}

{\bf At BNL,} \fp{three heavy-ion running campaigns are currently envisioned to complete the RHIC mission,} as shown in Fig.~\ref{Fig:Timeline} (left). A key goal of the 2014-16 period is to measure heavy flavor probes of the QGP using the newly installed silicon vertexing detectors in PHENIX and STAR. For 2018-19, Phase II of the RHIC Beam Energy Scan program (BES-II) is foreseen. In 2021-22, precision jet quenching and quarkonia measurements will be made possible by the installation of sPHENIX.

The proposed sPHENIX detector, shown in the right panel of Fig.~\ref{Fig:Timeline}, would dramatically extend the range of jets measurable at RHIC, as well as provide precision spectroscopy of quarkonia, in particular the upsilon states. The program relies on a very high data acquisition bandwidth, which combined with RHIC II luminosities enables the measurements of jet energies up to 70~GeV. The sPHENIX design takes advantage of various technological advances to minimize costs. sPHENIX also provides one route for a for a smooth evolution to a full-capability EIC detector.

{\bf The LHC} is now preparing for Run II, foreseen to include p+p, p+Pb and heavy-ion data taking from 2015 to 2018. Run II will be followed by a shutdown from 2018 to 2020 (LS2) and Run III from 2020 to 2023. For both the p+p and Pb+Pb data taking, the LHC upgrades during the current shutdown should make collisions possible at close to the design energy, i.e., about 5~TeV per nucleon pair for Pb+Pb. Collision rates are expected to exceed the design values by up to one order of magnitude, with a combined Run II and III goal to deliver about 10~nb$^{-1}$ per experiment. In combination, the increased collision energy and luminosity will increase statistics for high $p_T$ probes by about a factor of 200.

To exploit this improved accelerator performance, ALICE, ATLAS and CMS are undergoing significant upgrades during the current shutdown and LS2. For ATLAS and CMS, these upgrades are mostly driven by the needs of the p+p program. Of particular importance for future heavy-ion running are upgrades to the inner tracker detectors, extending the reach of charged particle measurements, and upgrades to the trigger systems, allowing the experiments to fully sample the expected collision data for hard probes.

ALICE is preparing for Run II with an expansion of the calorimetric coverage (EMCAL) which will make possible dijet studies and improved jet triggering. During LS2 the experiment's data taking capabilties will be significantly enhanced with major upgrades to detector readout and data aquisition systems. In addition a new silicon inner tracker will be installed. While precision measurements of low $p_T$ open heavy flavor are the main physics driver for these upgrades, they also benefit the intermediate and high $p_T$ jet quenching program.

These RHIC and LHC facility and experiment upgrades benefit jet physics studies in three major ways. The statistical precision and kinematic reach for commonly used jet physics observables is vastly increased. For single charged hadrons and reconstructed jets, the $p_T$ reach will be extended by a factor of 2--3, up to 40~GeV for hadrons and 70~GeV for jets at RHIC and 300~GeV and 1~TeV, respectively, at the LHC. This will further improve the precision of the extraction of e.g. $\hat{q}$ and its temperature dependence from model comparisons.

A second benefit is improved access to rare, highly specific observables, such as isolated photon + jet correlations at RHIC and LHC, as well as $Z^0$+jet correlations at the LHC. For example, LHC experiments will record more than $5 \times 10^5$ photons with $\ptg > 60$~GeV, compared to about 3000 in the Run I data sets. 

Finally, the very high statistics jet samples enable the measurement of a new generation of jet shape observables. There is intense activity in developing generalized jet structure variables for the $pp$ program at the LHC to maximize the efficiency of discovery measurements by improving quark/gluon discrimination and the tagging of boosted objects. Heavy-ion studies of the modification of the jet momentum and angular structure through medium interactions will benefit greatly from these developments.
\smallskip

\noindent {\bf Jet physics outlook}

In combination, these developments will enable a coherent RHIC and LHC physics program employing well-calibrated common observables. As shown in Fig.~\ref{Fig:HardProbesFuture}, future RHIC and LHC measurements will both overlap with and complement each other. This program will have three main components: 

First, a combined global analysis of RHIC and LHC data will lead to a precise determination of the values and temperature dependence of QGP transport coefficients. Both the RHIC and LHC final states represent an integral of jet-medium interactions over the evolution of both jet and medium from initial to final state. To disentangle the temperature dependence from this evolution it will be essential to deploy directly comparable observables (theoretically and experimentally) in different QGP temperature regimes, with particular focus on  the fraction of their evolution spent in vicinity to the phase transition region that will be passed by all collisions.

The second component relies on the increased systematic and statistical precision afforded by new probes (e.g. photon-jet) to identify the medium response to the modified jet radiation and further elucidate the liquid nature of the medium in its response to local perturbations.

Third, precision measurements of the angular and momentum structure of jets can be used to characterize the microscopic structure of the QGP, using hard partons as probes in ``Rutherford scattering'' off effective QGP constituents or quasi-particles. Studies include potential modifications to the back-to-back jet scattering distributions, as well as modifications of the intra-jet angular structure. For the latter, the correlated angular and momentum evolution of the jet from the initial scattering to the final hadronic structure probes a  wide range of scales, opening a window to interactions of jet and QGP constituents between vacuum-like and in-medium cascade regimes. To elucidate the physics of this intermediate window, a systematic variation of both the jet conditions and medium conditions and dynamics is necessary. Combining RHIC and LHC measurements will make it possible to control the initial density and temperature (in particular in respect to their vicinity to the critical temperature) and expansion dynamics of the system. The different energy regimes and tagging of particular initial states (photon+jet, $b$-tagged jets, multi-jet events) will enable the selection of different or common jet populations in relation to different medium conditions. Success in this long-term endeavor will require a global analysis of a diverse set of RHIC and LHC data in an improved, well controlled theoretical framework matched to the experimental observables.

\subsubsection{Probing the QGP with bottomonia}

A key probe of the strength and nature of the color interactions in the QGP are precision measurements of bottomonia states. The three lowest bound states of $b$-$\bar{b}$ ($\Upsilon$\,1s, 2s, 3s) have a range of binding energies and length scales that allow one to understand the temperature dependence of plasma screening~\cite{Brambilla:2010cs,Brambilla:2010vq} that leads to a characteristic sequential suppression pattern \cite{Karsch:1987pv} for in-medium quarkonium bound states. Compared to charmonia, bottomonia have the advantage that \fp{recombination in the QGP and at the phase boundary is expected to be smaller because of the smaller numbers of $b$-quarks in the system, even at LHC energies, thus providing a more direct measure of color screening through the suppression mechanism \cite{Strickland:2011mw,Strickland:2011aa}. On the other hand, in p+p collisions only 0.1\% of $b$-$\bar{b}$ pairs form bottomonia (compared to ${\sim\,}1\%$ for charm) \cite{Grandchamp:2005yw}. Transport calculations suggest that bottomonium regeneration at both RHIC and LHC is relatively small \cite{Emerick:2011xu}, which facilitates the interpretation of observed medium modifications of bottonium production yields in terms of color screening and dissociation effects \cite{Strickland:2011mw,Emerick:2011xu,Strickland:2011aa}.}

First detailed bottomonium measurements at the LHC from CMS show the expected ordering by their size and binding energies in the suppression of these states as a function of collision centrality \cite{Chatrchyan:2011pe}. \fp{At LHC energies regeneration may affect bottomonium production, making bottomonium measurements at RHIC an important benchmark measurement.}
First measurements at RHIC without resolving the three states have also been completed \cite{Adare:2014hje,Adamczyk:2013poh}. Critical to the effort to understand the screening interactions as a function of temperature are simultaneous precision measurements at RHIC and the LHC (probing cooler and hotter plasmas) in p+p, p+A, and A+A, with sufficient statistics to map out the centrality and $p_T$ dependencies. Increased luminosities at the LHC in Run II and Run III will provide these data. At RHIC, additional statistics will be provided by the STAR MTD upgrade; ultimately, for matched precision at RHIC and the LHC, the sPHENIX upgrade is required.

\subsection{A polarized p+A program for initial-state and low-x phenomena}
\label{Sec:Saturation}

Four dominant questions have been identified that can be explored in a
polarized p+p and p+A program: 1) What is the nature of the spin of
the proton? 2) How do quarks and gluons hadronize into final-state
particles? 3) How can we describe the multidimensional landscape of
nucleons and nuclei? 4) What is the nature of the initial state in
nuclear collisions? The first question is discussed in the ``cold
QCD'' Town Meeting summary and \fp{the RHIC-Spin White Paper
  \cite{Aschenauer:2015eha}.}  The remaining questions share
particularly strong synergies with the goals of the hot QCD community
and are discussed below.

At the high collision energies of RHIC and the LHC, the available phase
space for radiating small-$x$ gluons and quark-antiquark pairs is very
large. Since each emitted parton is itself a source of additional
gluons, an exponentially growing cascade of radiation is created which
would eventually violate unitarity. However, when the density of
partons in the transverse plane becomes large enough, softer partons
begin to recombine into harder ones and the gluon density
saturates. This limits the growth of the cascade and preserves the
unitarity of the $S$-matrix.  The transverse momentum scale below
which these nonlinear interactions dominate is known as the
\emph{saturation scale} \qs. The saturation scale grows with energy,
but also with the size of the nucleus as $\qssqr \sim A^{1/3}$. For high 
enough energies $\qs$ is large and the corresponding QCD coupling is
weak: $\as(\qs) \ll 1$. This makes it possible to calculate even bulk
particle production using weak coupling methods, although the
calculation is still nonperturbative due to the large field strengths. 
Because the gluonic states have large occupation numbers, the fields 
behave classically. The classical field theory describing the saturation 
domain is known as the ``Color Glass Condensate'' (CGC) \cite{Gelis:2010nm}.

The ideal probe of the CGC description are dilute-dense collisions,
where a simple small projectile collides with a large nucleus. At RHIC
and the LHC this makes p+A collisions important for understanding
saturation.  Significant progress has been made in describing the
systematics of particle production as a function of transverse
momentum and rapidity in p+p and p+A collisions with CGC calculations,
which match the collinearly factorized perturbative QCD description at
high transverse momenta. The case of saturation effects in
multiparticle correlations as a function of azimuthal angle and
rapidity remains more open \fp{(for more discussion of the ridge in
  p+p and p+A collisions see Sec.~\ref{sec:SM}).} While there are
contributions to these correlations that originate already in the
nuclear wave functions~\cite{Dusling:2013oia}, experimental evidence
points to strong collective behavior also in the final states of p+A
and even p+p collisions. The versatility of RHIC to systematically
change the size of the projectile nucleus and complement p+A with d+A,
$^3$He+A etc. collisions over a wide range of collision energies is
unparalleled and a key to exploring where these collective effects
turn on.  Precise and controlled access to the high energy nucleus can
be used to disentangle the effects of strong collectivity in the
initial wave functions and the final state.

As noted elsewhere, an EIC~\cite{Accardi:2012qut} which can measure
the transverse and longitudinal structure of the small-$x$ gluons in
nuclei is crucial for a deep understanding of hadron wave
functions. However, measurements of forward photon, J/$\psi$,
Drell-Yan, inclusive and di-jet, and hadron/jet correlation probes in
p+A collisions \cite{starloi,PHENIXpppA} provide a unique opportunity
to make timely progress in our understanding and to complement the
eventual EIC measurements. \fp{These measurements take advantage of
  the unique capabilities afforded by the RHIC accelerator and
  detector complex, complemented with detector upgrades including the
  forward calorimetric system and forward tracking system required to
  carry out the proposed physics program at forward rapidities.} In
particular, the observables mentioned are related to different
transverse momentum dependent gluon distributions (TMDs) that have
previously been studied at higher $x$. Recent theoretical advances
\cite{Dominguez:2011wm} have clarified the relation of these
distributions to each other via the CGC picture, in both p+A and e+A
collisions, and they are opening up new exciting connections to
polarized observables. One particular example of these connections is
the possibility of extracting the saturation scale $\qs$ in nuclei by
comparing transverse single spin asymmetries measured in polarized p+p
and polarized p+A collisions, for different nuclei and at different
beam energies \cite{Kang:2011ni,Kovchegov:2013cva}. Another exciting
opportunity in polarized p+A collisions is the possibility of
extracting \fp{first information about} generalized parton
distributions (GPDs) for gluons. This can be achieved for instance in
exclusive J/$\Psi$ production via photon-gluon fusion of quasi-real
ultraperipheral photons from the nucleus with gluons from the
polarized proton \cite{Aschenauer:2015eha}.

The study of these novel TMDs and GPDs will deepen our understanding
of the momentum and spatial structure of polarized and unpolarized
quarks and gluons in hadrons. These studies require as key ingredients
establishing factorization and universality that help separate the
structure of the hadron wave function from the dynamics of the
probe. From this perspective, even if measurements at a future EIC
with an electron probe provides unmatched precision, a polarized
proton-nucleus program provides a complementarity that may prove
essential.

\begin{figure}[!ht]
\begin{center}
\begin{minipage}{0.98\textwidth}
\begin{minipage}{0.6\textwidth}
\vspace*{-3mm}
\centerline{\includegraphics[width=\textwidth]{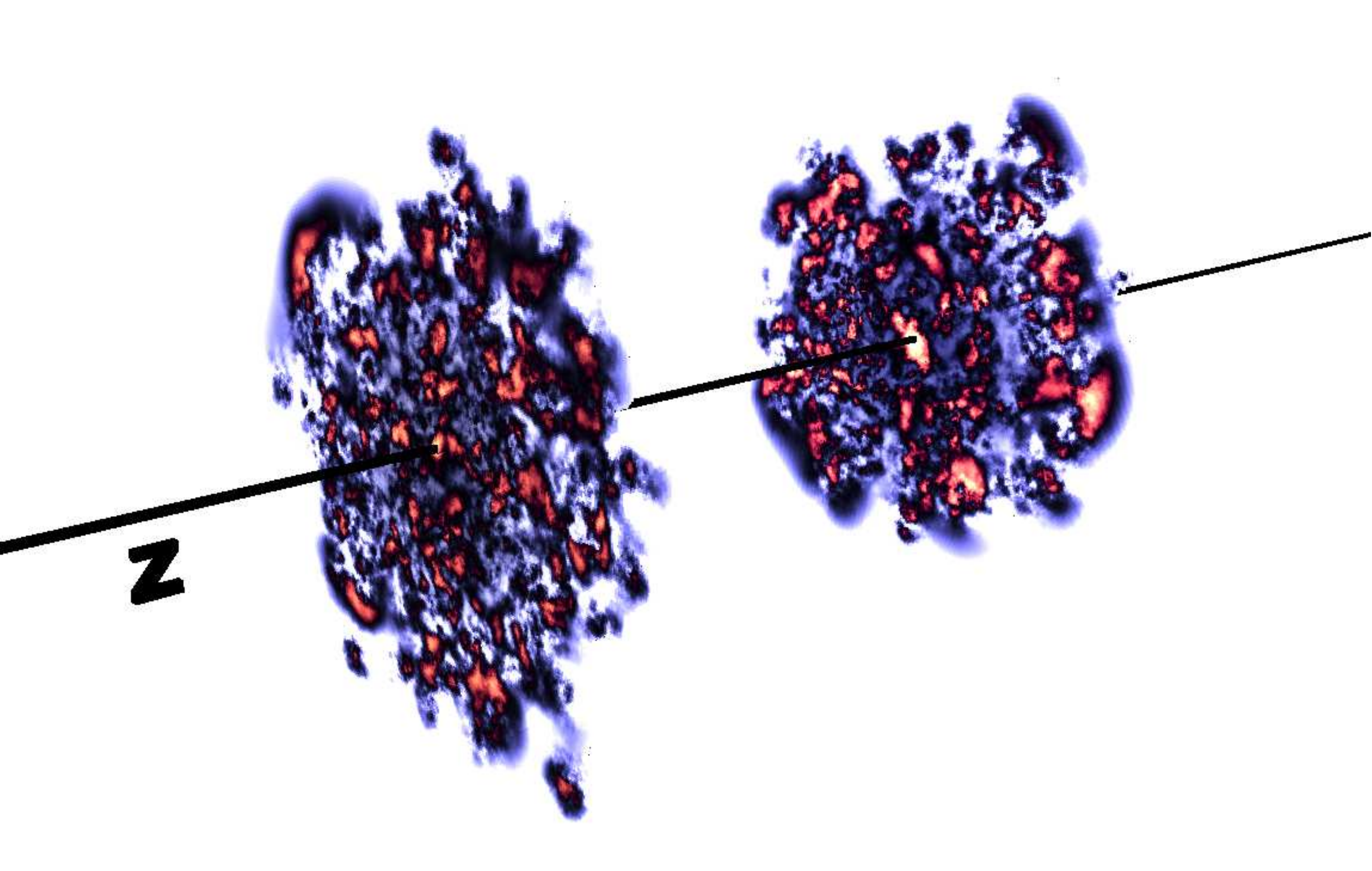}}
\end{minipage}
\hspace*{2mm}
\begin{minipage}{0.35\textwidth}
\caption{Color fields of the two nuclei before the collision,
from~\cite{Schenke:2012fw}.}
\label{fig:colorfield}
\end{minipage}
\end{minipage}
\end{center}
\end{figure}
\vspace*{-0.5cm}

The theoretical description of the initial stage of quark gluon plasma
formation has become increasingly detailed. State of the art
calculations \cite{Gale:2012rq,Paatelainen:2013eea} now combine a
fluctuating nuclear geometry with a microscopic QCD description of the
dynamics of matter formation (see Fig.~\ref{fig:colorfield}), going
beyond the Glauber model descriptions of the geometry. As discussed in
Sec.~\ref{sec:SM}, these initial-state calculations, combined with
detailed measurements of correlations and fluctuations in the observed
flow patterns have helped to significantly improve the precision of
the first quantitative experimental determinations of e.g. the
viscosity/entropy ratio $\eta/s$. \fp{In this context, p+A collisions provide
crucial consistency checks for our understanding of heavy-ion
collisions: Clearly, theoretical descriptions developed for partonic
interactions in a hot and dense QCD medium must also be consistent
with the effects of the cold QCD medium on hadronization of a colored
probe that can be studied in p+A collisions. Such collisions, across a
range of beam energies and target species, will therefore provide
important control experiments for our theoretical understanding of jet
quenching in heavy-ion collisions, on a time scale that precedes the
construction of an EIC.} As noted previously, p+A collisions have
provided surprises in their own right--we now understand the
resolution of these to be sensitive to the detailed spatial structure
of partons in both protons and heavier nuclei. The importance of a
polarized p+A program is therefore two-fold: (i) It will provide
unique and essential information on the parton structure of proton and
nuclear wave functions. (ii) The implementation of this information in
models of heavy-ion collision will provide more sensitive tests of and
precise extraction of the parameters of the Little Bang Standard
Model.


\subsection{Opportunities and challenges in theory}
\label{ssec:theory}

\subsubsection{Progress since the 2007 Nuclear Physics Long Range Plan}

Part of Recommendation IV of the 2007 Long Range Plan was the appreciation that {\it ``achieving a quantitative understanding of the properties of the quark-gluon plasma also requires new investments in modeling of heavy-ion collisions, in analytical approaches, and in large-scale computing.''} Since then there has been tremendous progress along these lines. Convergence has been reached in lattice QCD calculations of the temperature for the crossover transition in strongly interacting matter which has now been established at $145\,\mathrm{MeV}{\,<\,}T_c{\,<\,}163\,\mathrm{MeV}]$ \cite{Aoki:2006br,Aoki:2009sc,Bazavov:2011nk,Bazavov:2014pvz}. Continuum extrapolated results for the equation of state, the speed of sound and many other properties of strong interaction matter have also been provided \cite{Borsanyi:2013bia,Bazavov:2014pvz}. The modeling of the space-time evolution of heavy-ion collisions has become increasingly reliable. (2+1)-dimensional, and subsequently, (3+1)-dimensional relativistic viscous fluid dynamics computations have been performed. All such computations use an equation of state extracted from lattice QCD. A paradigm shift occurred with the broad appreciation of the importance of fluctuations.  For heavy-ion collisions at the highest RHIC and at LHC energies, the Color Glass Condensate (CGC) effective theory of QCD provides a framework to compute the properties of the fluctuating initial state. Under the aegis of the JET topical collaboration, a successful effort was undertaken to consolidate the results of different approaches to computing the transport properties of jets as they traverse the strongly correlated quark-gluon plasma. Much progress has been made towards a systematic understanding from first principles of the properties of strongly interacting matter at non-zero baryon number density. Such studies rely heavily on the development of theoretical concepts on critical behavior signaled by conserved charge fluctuation \cite{Stephanov:1998dy,Ejiri:2005wq,Stephanov:2011pb}. They are accessible to lattice QCD calculations which opens up the possibility, via dynamical modeling, for a systematic comparison of experimental fluctuation observables with calculations performed in QCD \cite{Karsch:2012wm,Bazavov:2012vg,Mukherjee:2013lsa,Borsanyi:2013hza,Borsanyi:2014ewa}. This will greatly profit from the steady development of computational facilities which are soon expected to deliver sustained petaflop/s performance for lattice QCD calculations.

An outcome of these efforts has been the development of a standard dynamical framework for heavy-ion collisions, as described Sec.~\ref{sec:SM}. Relativistic viscous fluid dynamics forms the core of this dynamical framework and, at high RHIC and LHC energies, it describes the largest part of the evolution history of the Little Bang -- the explosive expansion of the hot and dense QCD medium formed in the collisions. Much recent theoretical work has led to increasingly complete formulations of this theory and improved our understanding of its limits of applicability to heavy-ion collisions, and further improvements to optimize the framework for such applications are ongoing.

As described earlier in Sec.~\ref{sec:SM}, this standard framework does a good job of describing the wealth of data obtained on bulk spectra and event-by-event distributions of anisotropy coefficients, both at RHIC and the LHC. To turn this framework into a Little Bang Standard Model requires fixing all the model parameters that cannot (yet) be directly computed theoretically, and further refining our descriptions of the initial energy deposition and thermalization stages and the interfaces between the different collision stages (see following subsection). This process has started but is by no means complete. Finalizing the Little Bang Standard Model also requires new sophisticated model/data comparison tools and technologies that are just now being developed. Still, with the limited data/theory comparison tools that have so far been brought to bear on the large sets of experimental data collected at RHIC and LHC, the specific shear viscosity $\eta/s$ of QCD matter created at RHIC could be constrained to be approximately 50\% larger than the conjectured lower bound of $1/4\pi=0.08$, and to be about 2.5 times larger than this bound at the LHC. 

Concurrent efforts on extracting the jet quenching parameter $\hat{q}/T^3$ from similar theory-data comparisons have narrowed the range of values for this parameter to
$2{\,<\,}\hat{q}/T^3{\,<\,}6$ within the temperature range probed by RHIC and the LHC, nearly an order of magnitude lower than some previous estimates for this quantity. 

\subsubsection{Open questions and future goals}
\label{sec:open}

The development of a broad consensus on the elements of a Standard Model paradigm provides the basis for a deeper exploration of each of these elements, as well as the opportunity to further solidify the overall picture. Some of the challenging issues over which we need to get better control include i) the pre-equilibrium ``Glasma" dynamics of coherent gluon fields, and the approach to thermalization; ii) the extraction of the values and temperature dependences of transport parameters that reflect the many-body QCD dynamics in deconfined matter; iii) the initial conditions at lower collision energies where the Glasma framework breaks down; iv) the proper inclusion of the physics of hydrodynamic fluctuations; v) an improved treatment of hadron freeze-out and the transition from hydrodynamics  to transport theory, in particular the treatment of viscous corrections that can influence the extraction from data of the physics during the earlier collision stages. 

Quantitative improvements in these aspects of the dynamical modeling of a heavy-ion collision will lead to increased precision in the extraction of the underlying many-body QCD physics that governs the various collision stages. Additional conceptual advances in our understanding of QCD in matter at extreme temperatures and densities are required to answer a number of further outstanding questions. We here list a few of them, in chronological order as seen by an observer inside a heavy-ion collision: 

{\bf 1.}  A complete quantitative understanding of the properties of the nuclear wave functions that are resolved in nucleus-nucleus and proton-nucleus collisions so far remains elusive. Progress requires the extension of computations of the energy evolution of these wave functions in the Color Glass Condensate (CGC) framework to next-to-leading logarithmic accuracy, and matching these to next-to-leading order perturbative QCD computations at large momenta. Simultaneously, conceptual questions regarding the factorization and universality of distributions need to be addressed for quantitative progress. These ideas will be tested in upcoming proton-nucleus collisions at RHIC and the LHC, and with high precision at a future EIC. 

{\bf 2.} How the Glasma thermalizes to the quark-gluon plasma is not well understood. There has been significant progress in employing classical statistical methods and kinetic theory to the early stage dynamics -- however, these rely on extrapolations of weak coupling dynamics to realistically strong couplings. Significant insight is also provided from extrapolations in the other direction -- to large couplings -- using the holographic AdS/CFT correspondence between strongly coupled ${\cal N}{\,=\,}4$ supersymmetric Yang-Mills theory in four dimensions and weakly coupled gravity in an AdS$_5{\times}$S$_5$ space. Significant numerical and analytical progress can be anticipated in this fast evolving field of non-equilibrium non-Abelian plasmas. 

{\bf 3.} A novel development in recent years has been the theoretical study of the possible role of quantum anomalies in heavy-ion collisions. A particular example is the Chiral Magnetic Effect (CME), which explores the phenomenological consequences of topological transitions in the large magnetic fields created at early times in heavy-ion collisions. How the sphaleron transitions that generate topological charge occur out of equilibrium is an outstanding question that can be addressed by both weak coupling and holographic methods. Further, the effects of these charges can be propagated to late times via anomalous hydrodynamics. While there have been hints of the CME in experiments, conventional explanations of these exist as well. For a future beam energy scan, quantifying the predictions regarding signatures of quantum anomalies is crucial. This requires inclusion of the aromalies into the standard dynamical framework. We note that the study of the CME has strong cross-disciplinary appeal, with applications in a number of strongly correlated condensed matter systems. 

{\bf 4.} Noteworthy progress has been made in thermal field theory computations of photon and di-lepton production in heavy-ion collisions, where NLO computations are now available. \fp{The melting of the $\rho$ meson observed in dilepton spectra has been shown to be compatible with chiral symmetry restoration.  A challenging problem remains the determination of the axial-vector spectral function in the medium, to identify the mechanism underlying chiral restoration in the hadron spectrum. The discrepancy between theory and experiment in the ``photon $v_2$ puzzle'' mentioned in Sec.~\ref{sec:SM} may require additional sources of electromagnetic emissivity, likely in the near-$T_c$ region, or arise from more unconventional mechanisms in the early collision stages, e.g. transient Bose-Einstein condensates or} effects arising from the coupling of the conformal anomaly to external magnetic fields. 

{\bf 5.} As observed in Sec.~\ref{Sec:Jets}, progress has been made in quantifying the jet quenching parameter $\hat{q}$, which characterizes an important feature of the transverse response of the quark-gluon medium. \fp{Significant advances were made in calculating next-to-leading order corrections to $\hat{q}$, its QCD evolution and proving its factorization from the hard jet production processes \cite{Kang:2013raa,Liou:2013qya,Blaizot:2014bha}. We are beginning to understand the flow of the energy radiated by hard partons into a hot medium and its effect on jet shapes \cite{Vitev:2008rz,Blaizot:2012fh,Blaizot:2013hx,Blaizot:2014rla,Fister:2014zxa}. Soft Collinear Effective Theory (SCET), developed a decade ago in high energy physics, has led to improved accuracy of the theoretical predictions of jet cross sections and jet substructure observables \cite{Almeida:2014uva}. It has been generalized to describe jet propagation in nuclear matter \cite{Idilbi:2008vm,Ovanesyan:2011xy} and applied to parton transverse momentum broadening \cite{DEramo:2010ak}.} However, significant challenges persist. \fp{A general proof of factorization of the transport coefficients from the hard partons in an arbitrary medium is still outstanding. A second important transport parameter $\hat{e}$, characterizing the longitudinal drag of the medium on the hard or heavy probe, also needs to be quantified. For confrontation with the available jet fragmentation data, the mentioned recent insights into how parton showers develop in the quark-gluon plasma must be implemented into Monte-Carlo shower evolution codes that are coupled to a realistic dynamically evolving medium. Finally, novel directions for extracting information on the non-perturbative dynamics of the strongly correlated quark-gluon plasma from first principles have been pointed out with recent attempts to compute the jet quenching parameter using lattice techniques \cite{Majumder:2012sh,Panero:2013pla}.}

{\bf 6.} Quarkonia and heavy flavor, like jets, are hard probes that provide essential information on the quark-gluon plasma on varied length scales. Further, the two probes find common ground in studies of $b$-tagged and $c$-tagged jets \cite{Huang:2013vaa}. \fp{Open heavy flavor has attracted considerable theoretical attention~\cite{Moore:2004tg,vanHees:2004gq,Adil:2006ra,Sharma:2009hn,Kang:2011rt,Djordjevic:2013pba}, but the mechanisms of $D$- and $B$-meson production need to be further elucidated using input from the new experimental capabilities at RHIC and LHC. Non-relativistic QCD (NRQCD) has been the standard framework for describing
heavy quarkonium production in p+p collisions, and much progress has recently been made to extend it to p+A \cite{Kang:2013hta,Ma:2014mri} and A+A collisions \cite{Liu:2009nb,Sharma:2012dy}. However, this approach has difficulties in describing the polarization of quarkonium produced at collider energies. Recently, a new approach based on perturbative QCD factorization \cite{Kang:2011zza,Kang:2011mg,Kang:2014tta,Kang:2014pya} and soft-collinear effective theory \cite{Fleming:2012wy,Fleming:2013qu}, called ``fragmentation-function approach'' \cite{Brambilla:2010cs}, was developed to address these and other issues.} 

\fp{Combinations of lattice QCD and effective field theory approaches (NRQCD, pNRQCD and HQET) have furthered our understanding of in-medium quarkonium properties \cite{Laine:2006ns,Brambilla:2008cx,Riek:2010py,Petreczky:2010tk,Brambilla:2011sg,Ding:2012sp,Brambilla:2013dpa,Rothkopf:2013kya,Kim:2014iga} and heavy-flavor diffusion coefficients \cite{Francis:2013cva,Riek:2010fk}. The expected sequential melting pattern has been confirmed \cite{Bazavov:2014cta}, but an accurate determination of the in-medium meson properties (masses and widths) will require continued efforts. Implementation of quarkonium \cite{Zhao:2010nk,Strickland:2011aa,Young:2011ug} and open heavy-flavor transport \cite{He:2011qa,Cao:2013ita} into realistic bulk evolution models for nucleus-nucleus collisions is ongoing. Tightening the constraints on both the theoretical input and the evolution models will require extensive dynamical modeling.}

{\bf 7.} An outstanding intellectual challenge in the field is to map out the QCD phase diagram. While the lattice offers an {\it ab initio} approach, its successful implementation is beset by the well known sign problem, which is also experienced in other branches of physics. Nonetheless, approaches employing reweighting and Taylor expansion techniques have become more advanced and are now able to explore the equation of state and freeze-out conditions at baryon chemical potentials $\mu_B/T\le 2$. This covers a large part of the energy range currently explored in the beam energy scan and suggests that a possible critical endpoint may only be found at beam energies less than 20~GeV. Other promising approaches include the complex Langevin approach \cite{Aarts:2009uq,Aarts:2014kja} and the integration over a Lefschetz thimble \cite{Cristoforetti:2012su,Aarts:2014nxa}. There has been considerable work outlining the phenomenological consequences of a critical point in the phase diagram. However, quantitative modeling of how critical fluctuations affect the measured values of the relevant observables will require the concerted theoretical effort sketched in Sec.~\ref{sec:PD}.

\centerline{---------------------------------}

We cannot emphasize strongly enough that the impressive intellectual achievements outlined and the challenges ahead depend strongly on further development of the theory of strongly interacting matter which involves advances in heavy ion phenomenology, perturbative QCD, lattice QCD and effective field theories for QCD as well as the strong synergy with overlapping and related areas in particle physics, condensed matter physics, cold atom physics, string theory and studies of complex dynamical systems. In the case of string theory and condensed matter physics, a strong argument can be made that developments in heavy-ion collision theory have influenced developments in those fields. The depth and extent of interaction of this sub-field of nuclear theory with other branches of physics is perhaps unprecedented in nuclear physics. 

\subsubsection{What the field needs}

Looking ahead, sustaining and expanding the health of this sub-field of nuclear theory will depend on the following key items:

{\bf 1.} Strong continued support of the core nuclear theory program supporting university PI's, national lab groups and the National Institute for Nuclear Theory (INT). Together, they play an essential role in generating and implementing key ideas driving the field, and in training the next generation of students and post-doctoral fellows.

{\bf 2.} Strong continued support of the DOE Early Career Award (ECA) program in Nuclear Theory, as well as the NSF Early Career Development (CAREER) and Presidential Early Career (PECASE) award programs. These provide an important  opportunity for the field to recognize and promote the careers of the most outstanding young nuclear theorists. 

{\bf 3.} Strong support of expanded computational efforts in nuclear physics, as outlined in the Computational Nuclear Physics white paper and reflected in Recommendation \#3 of this document (see Section~\ref{sec2}). Progress in heavy-ion theory is strongly linked to the availability of a diverse and expanding array of computational resources, including both leadership class and capacity class computational resources. 

{\bf 4.} Continuation and expansion of Topical Research Collaboration program. These collaborations are especially valuable where there are several strains of theory developments that need to be coordinated and harnessed to address specific goals. An example of such a successful effort is the JET Topical Collaboration involving the co-ordinated effort of both theorists and experimentalists, as discussed earlier in this document. The field has several outstanding challenges that require the synthesis of a broad range of expertise, and it could strongly benefit from an expansion of such collaborations.

{\bf 5.} Strong support for the conclusions of the Education and Innovation Town Hall meeting. The numbers of scientists from underrepresented minorities and women are particularly low in nuclear theory. We encourage efforts to remedy this situation to a level that at least reflects the diversity of talent seen in other fields of science.


\clearpage
\section[The Electron Ion Collider]{Understanding the glue that binds us all: The Electron Ion Collider}
\label{sec4}

\subsection{The next QCD frontier}

Atomic nuclei are built from protons and neutrons, which themselves are composed of quarks that are bound together by gluons. Quantum Chromo-Dynamics (QCD), the gauge theory of the strong interaction, not only determines the structure of hadrons but also provides the fundamental framework to understand the properties and structure of atomic nuclei at all energy scales in the universe.  QCD is based on the exchange of massless gauge bosons called gluons between the constituents of hadrons, quarks. Without gluons there would be no protons, no neutrons, and no atomic nuclei. Matter as we know it would not exist. Understanding the interior structure and interactions of nucleons and nuclei in terms of the properties and dynamics of the quarks and gluons as dictated by QCD is thus a fundamental and central goal of modern nuclear physics. 

Gluons do not carry an electric charge and are thus not directly visible to electrons, photons, and other common probes of the structure of matter. An understanding of their role in forming the visible matter in the universe thus remains elusive. The Electron Ion Collider (EIC) with its unique capability to collide polarized electrons with polarized protons and light ions at unprecedented luminosity, and with heavy nuclei at high energy, will be the first precision microscope able to explore how gluons bind quarks to form protons and heavier atomic nuclei. 

By precisely imaging gluons and sea quarks inside the proton and nuclei, the EIC will address some of the deepest fundamental and puzzling questions nuclear physicists ask:
\begin{itemize}
\item
Where are the gluons and sea quarks, and how are their spins, distributed in space and momentum inside the nucleon? What is the role of the orbital motion of sea quarks and gluons in building the nucleon spin?
\item
What happens to the gluon density in nuclei at high energy? Does it saturate? Does this mechanism give rise to a universal component of matter in all nuclei, even the proton, when viewed at close to the speed of light?
\item
How does the nuclear environment affect the distributions of quarks and gluons and their interactions in nuclei? How does nuclear matter respond to a fast moving color charge passing through it? How do quarks dress themselves to become hadrons?
\end{itemize}

A full understanding of QCD, in a regime relevant to the structure and properties of hadrons and nuclei, demands a new era at the EIC of precision measurements that can probe them in their full complexity. Theoretical advances over the past decade have resulted in the development of a powerful formalism that provides quantitative links between such measurements and the above questions physicists are trying to answer. A second important advance in recent years is the increasing precision and reach of {\sl ab initio} calculations performed with lattice QCD techniques. Using the experimental data from an EIC, physicists will, for the first time, be able to undertake the detailed comparative study between experimental measurements and the predictions made by continuum- and lattice-QCD theory, as well as elucidate aspects of the structure of hadrons and nuclei whose investigation still requires more phenomenological theoretical methods. 

The experimental study of how hadrons and nuclei emerge from the laws of QCD is a global scientific priority. Two world-leading facilities in the U.S., CEBAF at Jefferson Lab and RHIC at BNL, are international centers for the study of QCD.  With the increase of its beam energy to 12 GeV, Jefferson Lab will soon operate a unique electron microscope, which will systematically map the structure of protons and other nuclei in the valence quark region. In addition to its discovery and continuing exploration of the strongly coupled quark gluon plasma (QGP), RHIC has used its unique capability as a polarized proton collider to make a first direct determination of the contribution of the gluons to the proton's spin. 

The high energy, high luminosity polarized EIC will unite and extend the current scientific programs at CEBAF and RHIC in dramatic and fundamentally important ways. It will enable us to image the transverse momentum and position distributions of quarks and gluons inside fast moving hadrons. When hadrons move at nearly the speed of light, the low-momentum gluons contained in their wave functions become experimentally accessible. The EIC will study the way in which gluons interact with each other by splitting and fusing in addition to providing three-dimensional images of the confined motion of quarks and gluons and their spatial distribution. By colliding electrons with heavy nuclei moving with the light speed, the EIC will provide access to a so far unconfirmed regime of matter where abundant gluons dominate its behavior.  Such universal cold gluon matter is an emergent phenomenon of QCD dynamics and of high scientific interest and curiosity. Furthermore, its properties and its underlying QCD dynamics are critically important for understanding the dynamical origin of the creation of the QGP from colliding two relativistic heavy ions, and the QGPs almost perfect liquid behavior. Enabling these research activities all in one place, the EIC will be a true ``QCD Laboratory'', unique of its kind in the world.

\subsection{Science highlights and deliverables at the EIC}

The high energy, high luminosity polarized EIC will unite and extend the scientific programs at CEBAF, RHIC and the LHC in dramatic and fundamentally important ways.

\begin{figure}[h]
\begin{center}
\includegraphics[height=63.5mm]{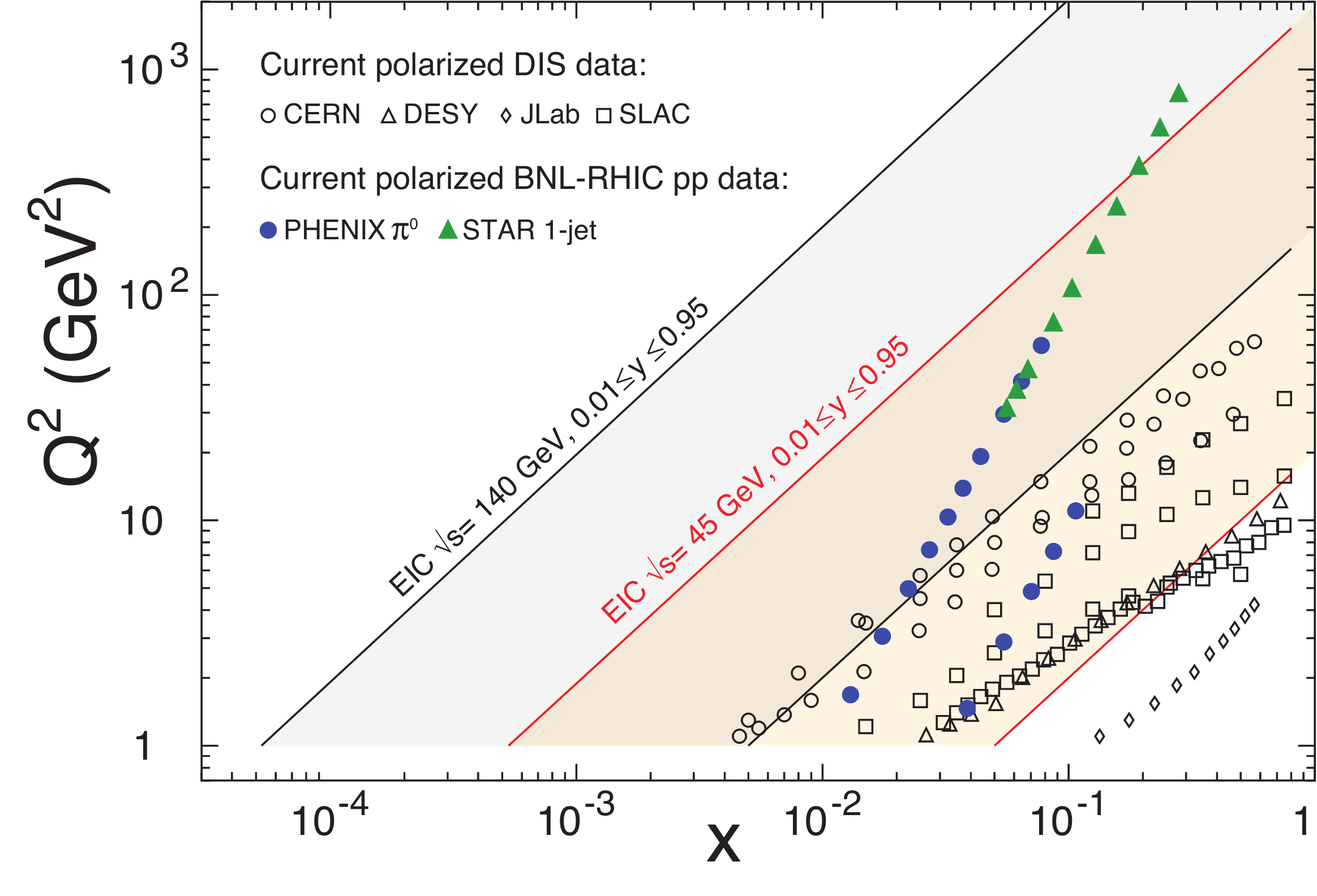}
\hspace*{2mm}
\includegraphics[height=65mm]{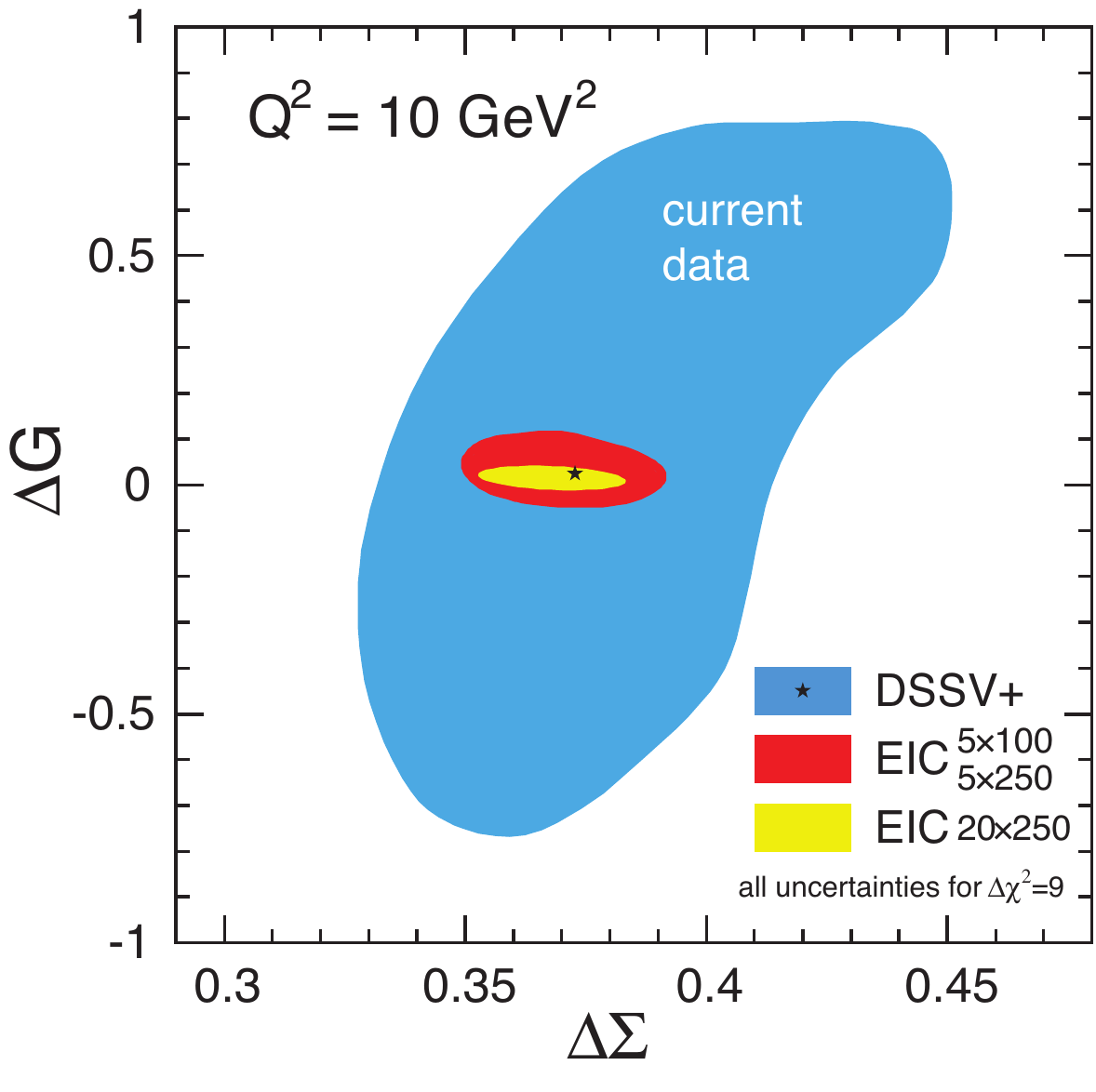}
\end{center}
\caption{{\bf Left panel:} The increase in the proton momentum fraction $x$ vs. the square of the momentum transferred by the electron to proton, $Q^2$, accessible to the EIC in e+p collisions. {\bf Right panel:} The projected reduction in the uncertainties of the gluons' ($\Delta G$) and quarks' ($\Delta\Sigma$) contributions to the proton's spin.
\label{F-EIC1}
}
\end{figure}

{\bf Proton Spin:}  Recent measurements at RHIC along with state-of-the-art perturbative QCD analyses have shown that gluons carry approximately 20-30\% of the proton's helicity, similar to the contribution from quarks and anti-quarks. The blue band in the right panel of Figure~\ref{F-EIC1} shows the current level of uncertainties. The knowledge is limited by the x-range explored so far.  The EIC would greatly increase the kinematic coverage in $x$ and $Q^2$, as shown in the left panel of Figure \ref{F-EIC1}, and hence reduce this uncertainty dramatically, to the level depicted by the red and yellow bands in the right panel.

\begin{figure}[h!]
\begin{center}
\includegraphics[height=58mm]{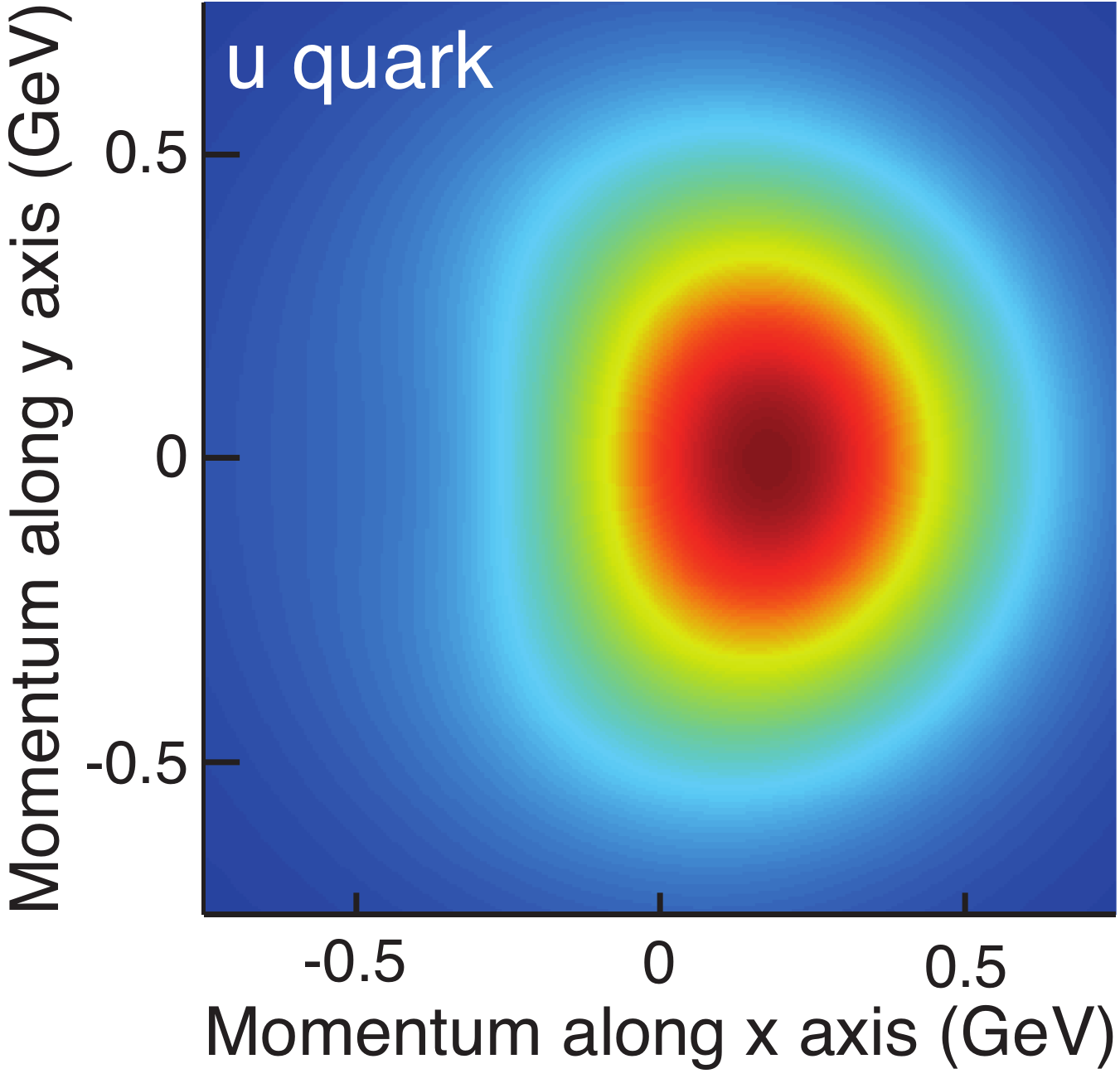}
\includegraphics[height=62mm]{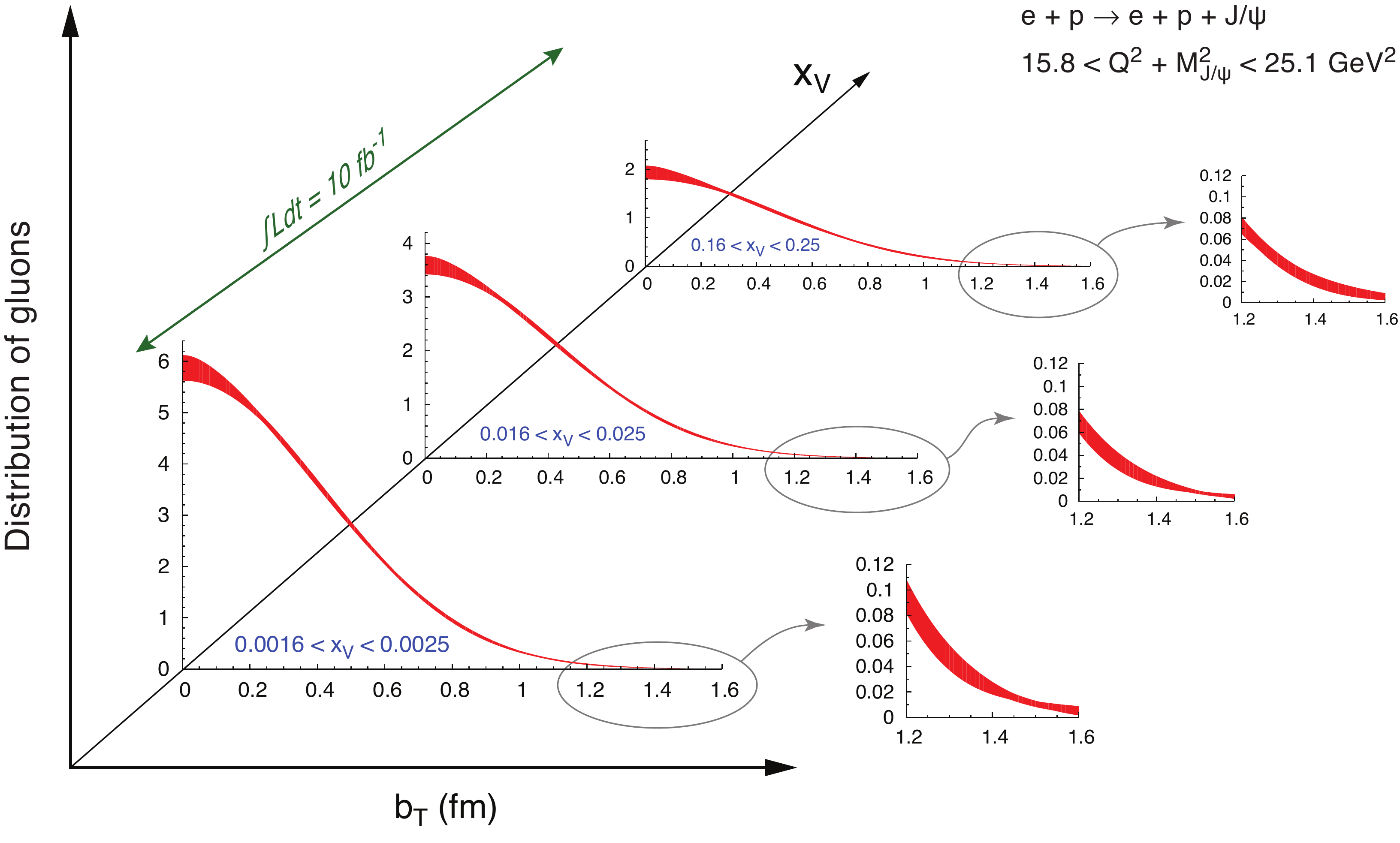}
\end{center}
\caption{{\bf Left panel:} Transverse momentum distribution of a sea $u$-quark with longitudinal momentum fraction $x{\,=\,}0.1$ in a transversely polarized proton moving in the $z$-direction, while polarized in $y$-direction. The color code indicates the probability of finding the up quarks, with dark red indicating the highest probability. {\bf Right panel:} Projected precision of the transverse spatial distribution of gluons obtained from exclusive $J/\psi$ production at the EIC.
\label{F-EIC2}
}
\end{figure}

{\bf Motion of quarks and gluons in a proton:} Semi-inclusive measurements with polarized proton beams would enable us to selectively and precisely probe the correlations between the spin of a fast moving proton and the confined transverse motion of both the quarks and gluons within it. Images in momentum space as shown in the left panel of Figure~\ref{F-EIC2} are simply unattainable without the polarized electron and proton beams of the proposed EIC.

\begin{figure}[t!]
\begin{center}
\includegraphics[height=60mm]{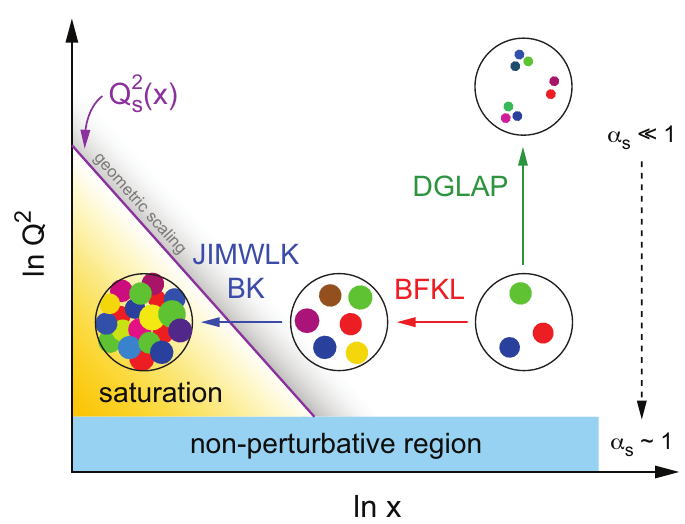}
\includegraphics[height=58mm]{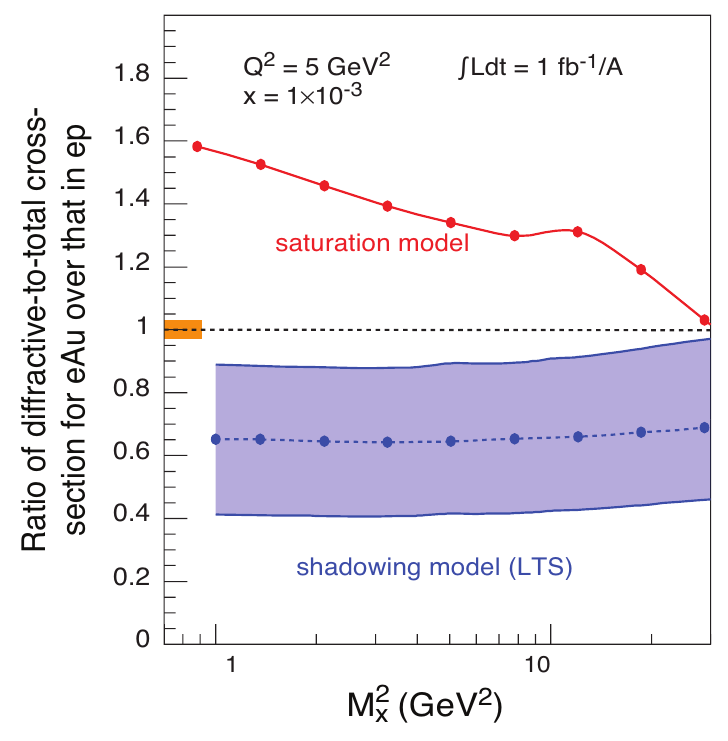}
\end{center}
\caption{{\bf Left panel:} Schematic landscape of probe resolution vs. energy, indicating regions of non-perturbative and perturbative QCD, including low to high parton density and the transition region. {\bf Right panel:} Ratio of diffractive over total cross section for deep-inelastic scattering (DIS) of electrons on gold, normalized to DIS on the proton, for different values of the square of the invariant mass of the hadrons produced in the collisions, with and without saturation.
\label{F-EIC3}
}
\end{figure}

{\bf Tomographic images of the proton:} By choosing particular final states in $e{+}p$ scattering, the EIC, with its unprecedented luminosity and detector coverage, will create detailed images of the proton's gluon matter distribution, as shown in the right panel of Figure~\ref{F-EIC2}. Such measurements would reveal aspects of proton structure that are intimately connected with QCD dynamics at large distances.

{\bf QCD matter at extreme gluon density:} When fast moving hadrons are probed at high energy, the low-momentum gluons contained in their wave functions become experimentally accessible. By colliding electrons with heavy nuclei moving at light-speed, the EIC will provide access to a so far unconfirmed regime of matter where abundant gluons dominate its behavior as shown in the left panel of Figure~\ref{F-EIC3}. Such cold gluon matter is an emergent phenomenon of QCD dynamics and of high scientific interest and curiosity. Furthermore, its underlying QCD dynamics and its predicted universal properties are critically important for understanding the dynamical origin of the creation of the QGP from colliding two relativistic heavy ions.

\begin{figure}[b!]
\begin{center}
\includegraphics[height=70mm]{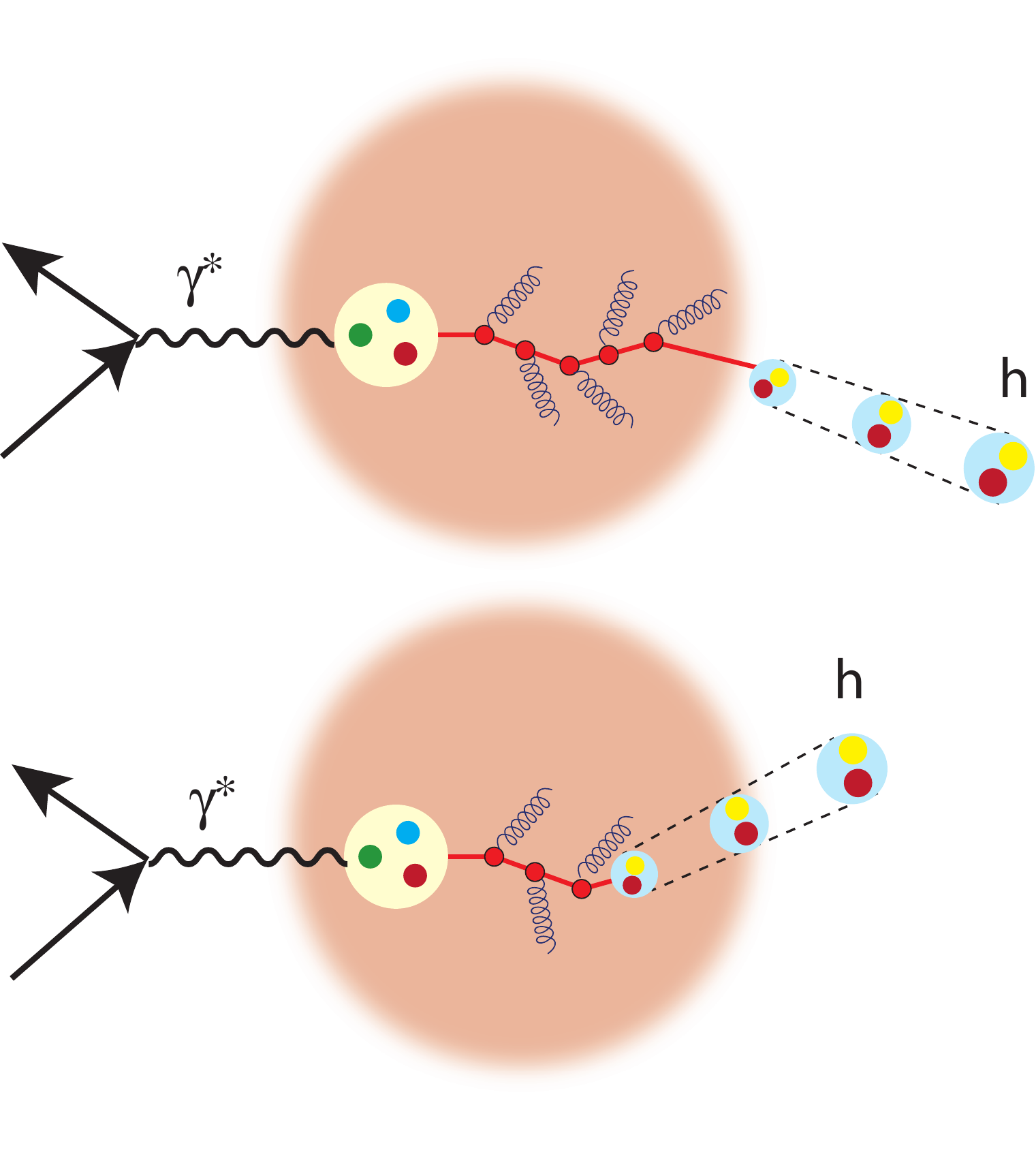}
\includegraphics[height=68mm]{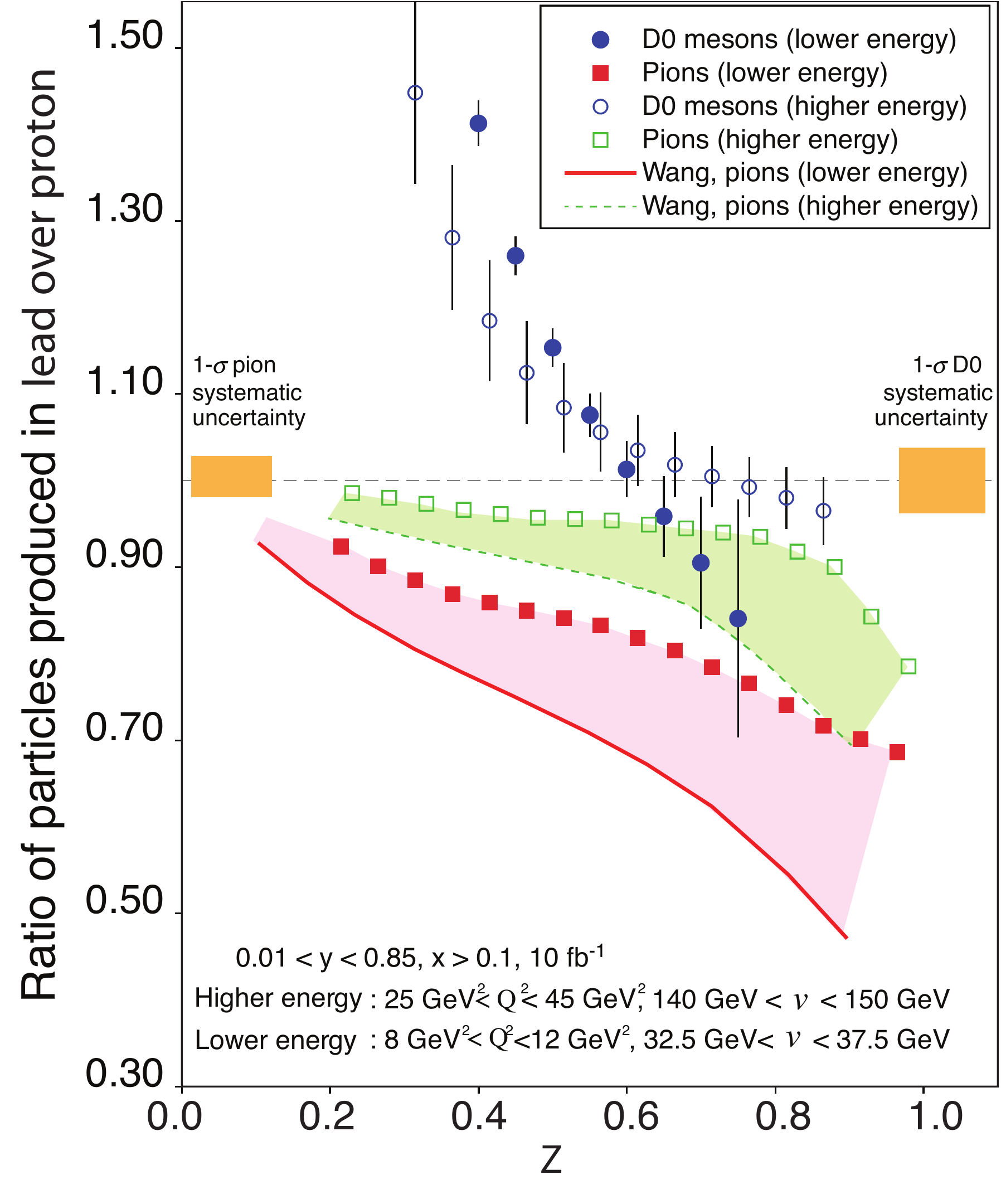}
\end{center}
\caption{{\bf Left panel:} A schematic illustrating the interaction of a parton moving through cold nuclear matter. where the hadron is formed outside (top) or inside the nucleus (bottom). {\bf Right panel:} The ratio of the semi-inclusive cross sections for producing a pion (light quarks, red) and $D^0$ mesons (heavy quarks, blue) in $e$+lead collisions to $e$+deuteron collisions, plotted as a function of the ratio $z$ of the momentum carried by the produced hadron to that of a virtual photon.
\label{F-EIC4}
}
\end{figure}

By measuring diffractive cross-sections together with the total deep-inelastic scattering (DIS) cross-sections in $e{+}p$ and $e{+}A$ collisions, shown in the right panel of Figure~\ref{F-EIC3}, the EIC would provide the first unambiguous evidence for this novel state of saturated gluon matter in QCD. The planned EIC is capable of exploring with precision the new field of collective dynamics of saturated gluons at high energies.

{\bf Hadronization and energy loss:} The mechanism by which colored partons pass through colored media, both cold nuclei and hot QGP matter, and color-singlet hadrons finally emerge from the colored partons is not understood. A nucleus in the EIC would provide an invaluable femtometer filter to explore and expose how colored partons interact and hadronize in nuclear matter, as illustrated in the left panel of Figure~\ref{F-EIC4}. By measuring $\pi$ and $D^0$ meson production in both $e{+}p$ and $e{+}A$ collisions, the EIC would provide the first measurement of the quark mass dependence of the response of nuclear matter to a fast moving quark. The dramatic difference between them, shown in the right panel of Figure~\ref{F-EIC4}, would be readily discernible. The color bands reflect the limitations on our current knowledge of hadronization -- the emergence of a pion from a colored quark. Enabling all such studies  in one place, the EIC will be a true “QCD Laboratory”, a unique facility in the world.

\subsection{EIC machine parameters and designs}

Two independent designs for a future EIC have evolved over the past few years. Both use existing infrastructure and facilities available to the US nuclear scientists. At Brookhaven National Laboratory (BNL), the eRHIC concept adds a new electron beam facility, based on an Energy Recovery Linac (ERL) to be  built inside the RHIC tunnel, to collide electrons with one of the existing RHIC beams.  At Jefferson Laboratory the Medium Energy Electron Ion Collider (MEIC) concept envisions a new electron and ion collider ring complex, together with the 12 GeV upgraded CEBAF, in order to achieve similar collision parameters.
The machine designs aim to reach the following goals and parameters: 
\begin{itemize}
\item
Polarized ($\sim70\%$) beams of electrons, protons and light nuclei;
\item
Ion beams from deuteron to the heaviest nuclei (uranium or lead);
\item
Variable center of mass energies from $\sim20$ to $\sim 100$\,GeV, upgradable to $\sim 140$\,GeV;
\item
High collision luminosity $\sim10^{33}-10^{34}$\,cm$^{-2}$sec$^{-1}$;
\item
Capacity to have more than one interaction region.
\end{itemize}

\subsection{Why now?}

Today, a set of compelling physics questions related to role of gluons in QCD has been formulated, and a corresponding set of measurements at the EIC identified. A powerful formalism that connects those measurements to the QCD structure of hadrons and nuclei has been developed. The EIC was designated in the 2007 Nuclear Physics Long Range Plan as {\it ``embodying the vision for reaching the next QCD frontier.''}  In 2013 the NSAC Subcommittee report on Future Scientific Facilities declared an EIC to be {\it ``absolutely essential in its ability to contribute to the world-leading science in the next decade.''} Accelerator technology has recently developed so that an EIC with the versatile range of kinematics, beam species and polarization, crucial to addressing the most central questions in QCD, can now be constructed at an affordable cost. Realizing the EIC will be essential to maintain U.S. leadership in the important fields of nuclear physics and accelerator science.


\bigskip


\section*{\fp{Acknowledgements}}

\fp{The authors of this Town Meeting summary gratefully acknowledge guidance and input from the other authors of the Hot QCD White Paper \cite{HotQCD_WP} (Yasuyuki Akiba, Aaron Angerami, Helen Caines, Anthony Frawley, Barbara Jacak, Jiangyong Jia, Wei Li, Abhijit Majumder, Mateusz Plosion, Joern Putschke, Ralf Rapp, Urs Wiedemann, Nu Xu, and Bill Zajc). In addition, we thank the following colleagues for helpful and constructive comments on a first draft of this document: Elke Aschenauer, Zhongbo Kang, Christine Nattrass, Peter Petreczky, Paul Romatschke, Lijuan Ruan, Michael Strickland, Mike Tannenbaum, Ivan Vitev, Xin-Nian Wang, and Zhangbu Xu. Last but not least, we want to thank Jim Napolitano and Bernd Surrow (Co-Chairs), together with Jeff Martoff, Andreas Metz and Nikos Spaveris, for their very effective organization and hosting of the QCD Town Meeting at Temple University in September 2014.}  
\clearpage

\section{Bibliography}

\begingroup
\renewcommand{\section}[2]{}%

\bibliographystyle{atlasnote}
\bibliography{HotQCD_TM_Summary}

\providecommand{\href}[2]{#2}\begingroup\raggedright\begin{thebibliography}{100}

\bibitem{Akiba:2015jwa}
Y.~Akiba, A.~Angerami, H.~Caines, A.~Frawley, U.~Heinz, et al., {\em {The Hot
  QCD White Paper: Exploring the Phases of QCD at RHIC and the LHC}\/},
\href{http://arxiv.org/abs/1502.02730}{{\tt arXiv:1502.02730 [nucl-ex]}}.

\bibitem{Accardi:2012qut}
A.~Accardi, J.~Albacete, M.~Anselmino, N.~Armesto, E.~Aschenauer, et al., {\em
  {Electron Ion Collider: The Next QCD Frontier - Understanding the glue that
  binds us all}\/},
\href{http://arxiv.org/abs/1212.1701}{{\tt arXiv:1212.1701 [nucl-ex]}}.

\bibitem{Mishra:2007tw}
A.~P. Mishra, R.~K. Mohapatra, P.~Saumia, and A.~M. Srivastava, {\em
  {Super-horizon fluctuations and acoustic oscillations in relativistic
  heavy-ion collisions}\/},
  \href{http://dx.doi.org/10.1103/PhysRevC.77.064902}{Phys. Rev. {\bf C77}
  (2008)  064902},
\href{http://arxiv.org/abs/0711.1323}{{\tt arXiv:0711.1323 [hep-ph]}}.

\bibitem{Voloshin:2003ud}
S.~A. Voloshin, {\em {Transverse radial expansion in nuclear collisions and two
  particle correlations}\/},
  \href{http://dx.doi.org/10.1016/j.physletb.2005.11.024}{Phys. Lett. {\bf
  B632} (2006)  490--494},
\href{http://arxiv.org/abs/nucl-th/0312065}{{\tt arXiv:nucl-th/0312065
  [nucl-th]}}.

\bibitem{Takahashi:2009na}
J.~Takahashi, B.~Tavares, W.~Qian, R.~Andrade, F.~Grassi, et al., {\em
  {Topology studies of hydrodynamics using two particle correlation
  analysis}\/},  \href{http://dx.doi.org/10.1103/PhysRevLett.103.242301}{Phys.
  Rev. Lett. {\bf 103} (2009)  242301},
\href{http://arxiv.org/abs/0902.4870}{{\tt arXiv:0902.4870 [nucl-th]}}.

\bibitem{Sorensen:2010zq}
P.~Sorensen, {\em {Implications of space-momentum correlations and geometric
  fluctuations in heavy-ion collisions}\/},
  \href{http://dx.doi.org/10.1088/0954-3899/37/9/094011}{J. Phys. {\bf G37}
  (2010)  094011},
\href{http://arxiv.org/abs/1002.4878}{{\tt arXiv:1002.4878 [nucl-ex]}}.

\bibitem{Alver:2010gr}
B.~Alver and G.~Roland, {\em {Collision geometry fluctuations and triangular
  flow in heavy-ion collisions}\/},
  \href{http://dx.doi.org/10.1103/PhysRevC.82.039903,
  10.1103/PhysRevC.81.054905 10.1103/PhysRevC.82.039903,
  10.1103/PhysRevC.81.054905}{Phys. Rev. {\bf C81} (2010)  054905},
\href{http://arxiv.org/abs/1003.0194}{{\tt arXiv:1003.0194 [nucl-th]}}.

\bibitem{ATLAS:2012at}
{ATLAS Collaboration}, G.~Aad et al., {\em {Measurement of the azimuthal
  anisotropy for charged particle production in $\sqrt{s}_\mathrm{NN}=2.76$ TeV
  lead-lead collisions with the ATLAS detector}\/},
  \href{http://dx.doi.org/10.1103/PhysRevC.86.014907}{Phys. Rev. {\bf C86}
  (2012)  014907},
\href{http://arxiv.org/abs/1203.3087}{{\tt arXiv:1203.3087 [hep-ex]}}.

\bibitem{ALICE:2011ab}
{ALICE Collaboration}, K.~Aamodt et al., {\em {Higher harmonic anisotropic flow
  measurements of charged particles in Pb-Pb collisions at $\energy =2.76$
  TeV}\/},  \href{http://dx.doi.org/10.1103/PhysRevLett.107.032301}{Phys. Rev.
  Lett. {\bf 107} (2011)  032301},
\href{http://arxiv.org/abs/1105.3865}{{\tt arXiv:1105.3865 [nucl-ex]}}.

\bibitem{Adare:2011tg}
{PHENIX Collaboration}, A.~Adare et al., {\em {Measurements of Higher-Order
  Flow Harmonics in Au+Au Collisions at $\sqrt{s_{NN}} = 200$ GeV}\/},
  \href{http://dx.doi.org/10.1103/PhysRevLett.107.252301}{Phys. Rev. Lett. {\bf
  107} (2011)  252301},
\href{http://arxiv.org/abs/1105.3928}{{\tt arXiv:1105.3928 [nucl-ex]}}.

\bibitem{Adamczyk:2013waa}
{STAR Collaboration}, L.~Adamczyk et al., {\em {Third Harmonic Flow of Charged
  Particles in Au+Au Collisions at sqrtsNN = 200 GeV}\/},
  \href{http://dx.doi.org/10.1103/PhysRevC.88.014904}{Phys. Rev. {\bf C88}
  (2013)  014904},
\href{http://arxiv.org/abs/1301.2187}{{\tt arXiv:1301.2187 [nucl-ex]}}.

\bibitem{Adler:2002pu}
{STAR Collaboration}, C.~Adler et al., {\em {Elliptic flow from two and four
  particle correlations in Au+Au collisions at s(NN)**(1/2) = 130-GeV}\/},
  \href{http://dx.doi.org/10.1103/PhysRevC.66.034904}{Phys. Rev. {\bf C66}
  (2002)  034904},
\href{http://arxiv.org/abs/nucl-ex/0206001}{{\tt arXiv:nucl-ex/0206001
  [nucl-ex]}}.

\bibitem{Miller:2003kd}
M.~Miller and R.~Snellings, {\em {Eccentricity fluctuations and its possible
  effect on elliptic flow measurements}\/},
\href{http://arxiv.org/abs/nucl-ex/0312008}{{\tt arXiv:nucl-ex/0312008
  [nucl-ex]}}.

\bibitem{Alver:2008zza}
B.~Alver, B.~Back, M.~Baker, M.~Ballintijn, D.~Barton, et al., {\em {Importance
  of correlations and fluctuations on the initial source eccentricity in
  high-energy nucleus-nucleus collisions}\/},
  \href{http://dx.doi.org/10.1103/PhysRevC.77.014906}{Phys. Rev. {\bf C77}
  (2008)  014906},
\href{http://arxiv.org/abs/0711.3724}{{\tt arXiv:0711.3724 [nucl-ex]}}.

\bibitem{Heinz:2013th}
U.~Heinz and R.~Snellings, {\em {Collective flow and viscosity in relativistic
  heavy-ion collisions}\/},
  \href{http://dx.doi.org/10.1146/annurev-nucl-102212-170540}{Annu. Rev. Nucl.
  Part. Sci. {\bf 63} (2013)  123--151},
\href{http://arxiv.org/abs/1301.2826}{{\tt arXiv:1301.2826 [nucl-th]}}.

\bibitem{Teaney:2003kp}
D.~Teaney, {\em {The effects of viscosity on spectra, elliptic flow, and HBT
  radii}\/},  \href{http://dx.doi.org/10.1103/PhysRevC.68.034913}{Phys. Rev.
  {\bf C68} (2003)  034913},
\href{http://arxiv.org/abs/nucl-th/0301099}{{\tt arXiv:nucl-th/0301099
  [nucl-th]}}.

\bibitem{Romatschke:2007mq}
P.~Romatschke and U.~Romatschke, {\em {Viscosity information from relativistic
  nuclear collisions: how perfect is the fluid observed at RHIC?}\/},
  \href{http://dx.doi.org/10.1103/PhysRevLett.99.172301}{Phys. Rev. Lett. {\bf
  99} (2007)  172301},
\href{http://arxiv.org/abs/0706.1522}{{\tt arXiv:0706.1522 [nucl-th]}}.

\bibitem{Song:2010mg}
H.~Song, S.~A. Bass, U.~Heinz, T.~Hirano, and C.~Shen, {\em {200 A GeV Au+Au
  collisions serve a nearly perfect quark-gluon liquid}\/},
  \href{http://dx.doi.org/10.1103/PhysRevLett.106.192301,
  10.1103/PhysRevLett.109.139904}{Phys. Rev. Lett. {\bf 106} (2011)  192301},
\href{http://arxiv.org/abs/1011.2783}{{\tt arXiv:1011.2783 [nucl-th]}}.

\bibitem{Kovtun:2004de}
P.~Kovtun, D.~T. Son, and A.~O. Starinets, {\em {Viscosity in strongly
  interacting quantum field theories from black hole physics}\/},
  \href{http://dx.doi.org/10.1103/PhysRevLett.94.111601}{Phys. Rev. Lett. {\bf
  94} (2005)  111601},
\href{http://arxiv.org/abs/hep-th/0405231}{{\tt arXiv:hep-th/0405231
  [hep-th]}}.

\bibitem{Heinz:2013wva}
U.~W. Heinz, {\em {Towards the Little Bang Standard Model}\/},
  \href{http://dx.doi.org/10.1088/1742-6596/455/1/012044}{J. Phys. Conf. Ser.
  {\bf 455} (2013)  012044},
\href{http://arxiv.org/abs/1304.3634}{{\tt arXiv:1304.3634 [nucl-th]}}.

\bibitem{Schenke:2010rr}
B.~Schenke, S.~Jeon, and C.~Gale, {\em {Elliptic and triangular flow in
  event-by-event (3+1)D viscous hydrodynamics}\/},
  \href{http://dx.doi.org/10.1103/PhysRevLett.106.042301}{Phys. Rev. Lett. {\bf
  106} (2011)  042301},
\href{http://arxiv.org/abs/1009.3244}{{\tt arXiv:1009.3244 [hep-ph]}}.

\bibitem{Song:2010aq}
H.~Song, S.~A. Bass, and U.~Heinz, {\em {Viscous QCD matter in a hybrid
  hydrodynamic+Boltzmann approach}\/},
  \href{http://dx.doi.org/10.1103/PhysRevC.83.024912}{Phys. Rev. {\bf C83}
  (2011)  024912},
\href{http://arxiv.org/abs/1012.0555}{{\tt arXiv:1012.0555 [nucl-th]}}.

\bibitem{Schenke:2011bn}
B.~Schenke, S.~Jeon, and C.~Gale, {\em {Higher flow harmonics from (3+1)D
  event-by-event viscous hydrodynamics}\/},
  \href{http://dx.doi.org/10.1103/PhysRevC.85.024901}{Phys. Rev. {\bf C85}
  (2012)  024901},
\href{http://arxiv.org/abs/1109.6289}{{\tt arXiv:1109.6289 [hep-ph]}}.

\bibitem{Schenke:2012wb}
B.~Schenke, P.~Tribedy, and R.~Venugopalan, {\em {Fluctuating Glasma initial
  conditions and flow in heavy ion collisions}\/},
  \href{http://dx.doi.org/10.1103/PhysRevLett.108.252301}{Phys. Rev. Lett. {\bf
  108} (2012)  252301},
\href{http://arxiv.org/abs/1202.6646}{{\tt arXiv:1202.6646 [nucl-th]}}.

\bibitem{Gale:2012rq}
C.~Gale, S.~Jeon, B.~Schenke, P.~Tribedy, and R.~Venugopalan, {\em
  {Event-by-event anisotropic flow in heavy-ion collisions from combined
  Yang-Mills and viscous fluid dynamics}\/},
  \href{http://dx.doi.org/10.1103/PhysRevLett.110.012302}{Phys. Rev. Lett. {\bf
  110} (2013)  012302},
\href{http://arxiv.org/abs/1209.6330}{{\tt arXiv:1209.6330 [nucl-th]}}.

\bibitem{Song:2013tpa}
H.~Song, F.~Meng, X.~Xin, and Y.-X. Liu, {\em {Elliptic flow of $\Lambda, \Xi$
  and $\Omega$ in 2.76 A TeV Pb+Pb collisions}\/},
  \href{http://dx.doi.org/10.1088/1742-6596/509/1/012089}{J. Phys. Conf. Ser.
  {\bf 509} (2014)  012089},
\href{http://arxiv.org/abs/1310.3462}{{\tt arXiv:1310.3462 [nucl-th]}}.

\bibitem{Song:2013qma}
H.~Song, S.~Bass, and U.~W. Heinz, {\em {Spectra and elliptic flow for
  identified hadrons in 2.76A TeV Pb + Pb collisions}\/},
  \href{http://dx.doi.org/10.1103/PhysRevC.89.034919}{Phys. Rev. {\bf C89}
  (2014)  034919},
\href{http://arxiv.org/abs/1311.0157}{{\tt arXiv:1311.0157 [nucl-th]}}.

\bibitem{vanderSchee:2013pia}
W.~van~der Schee, P.~Romatschke, and S.~Pratt, {\em {Fully Dynamical Simulation
  of Central Nuclear Collisions}\/},
  \href{http://dx.doi.org/10.1103/PhysRevLett.111.222302}{Phys. Rev. Lett. {\bf
  111} (2013)  222302},
\href{http://arxiv.org/abs/1307.2539}{{\tt arXiv:1307.2539}}.

\bibitem{Habich:2014jna}
M.~Habich, J.~Nagle, and P.~Romatschke, {\em {Particle spectra and HBT radii
  for simulated central nuclear collisions of C+C, Al+Al, Cu+Cu, Au+Au, and
  Pb+Pb from Sqrt(s)=62.4-2760 GeV}\/},
\href{http://arxiv.org/abs/1409.0040}{{\tt arXiv:1409.0040 [nucl-th]}}.

\bibitem{Bazavov:2009zn}
A.~Bazavov, T.~Bhattacharya, M.~Cheng, N.~Christ, C.~DeTar, et al., {\em
  {Equation of state and QCD transition at finite temperature}\/},
  \href{http://dx.doi.org/10.1103/PhysRevD.80.014504}{Phys. Rev. {\bf D80}
  (2009)  014504},
\href{http://arxiv.org/abs/0903.4379}{{\tt arXiv:0903.4379 [hep-lat]}}.

\bibitem{Borsanyi:2010cj}
S.~Borsanyi, G.~Endrodi, Z.~Fodor, A.~Jakovac, S.~D. Katz, et al., {\em {The
  QCD equation of state with dynamical quarks}\/},
  \href{http://dx.doi.org/10.1007/JHEP11(2010)077}{JHEP {\bf 1011} (2010)
  077},
\href{http://arxiv.org/abs/1007.2580}{{\tt arXiv:1007.2580 [hep-lat]}}.

\bibitem{Borsanyi:2013bia}
S.~Borsanyi, Z.~Fodor, C.~Hoelbling, S.~D. Katz, S.~Krieg, et al., {\em {Full
  result for the QCD equation of state with 2+1 flavors}\/},
  \href{http://dx.doi.org/10.1016/j.physletb.2014.01.007}{Phys. Lett. {\bf
  B730} (2014)  99--104},
\href{http://arxiv.org/abs/1309.5258}{{\tt arXiv:1309.5258 [hep-lat]}}.

\bibitem{Bazavov:2014pvz}
{HotQCD Collaboration}, A.~Bazavov et al., {\em {Equation of state in ( 2+1
  )-flavor QCD}\/},  \href{http://dx.doi.org/10.1103/PhysRevD.90.094503}{Phys.
  Rev. {\bf D90} (2014)  094503},
\href{http://arxiv.org/abs/1407.6387}{{\tt arXiv:1407.6387 [hep-lat]}}.

\bibitem{Gale:2013da}
C.~Gale, S.~Jeon, and B.~Schenke, {\em {Hydrodynamic Modeling of Heavy-Ion
  Collisions}\/},  \href{http://dx.doi.org/10.1142/S0217751X13400113}{Int. J.
  Mod. Phys. {\bf A28} (2013)  1340011},
\href{http://arxiv.org/abs/1301.5893}{{\tt arXiv:1301.5893 [nucl-th]}}.

\bibitem{Bhalerao:2011yg}
R.~S. Bhalerao, M.~Luzum, and J.-Y. Ollitrault, {\em {Determining initial-state
  fluctuations from flow measurements in heavy-ion collisions}\/},
  \href{http://dx.doi.org/10.1103/PhysRevC.84.034910}{Phys.~Rev. {\bf C84}
  (2011)  034910},
\href{http://arxiv.org/abs/1104.4740}{{\tt arXiv:1104.4740 [nucl-th]}}.

\bibitem{Jia:2014jca}
J.~Jia, {\em {Event-shape fluctuations and flow correlations in
  ultra-relativistic heavy-ion collisions}\/},
  \href{http://dx.doi.org/10.1088/0954-3899/41/12/124003}{J.~Phys.~G {\bf 41}
  (2014)  124003},
\href{http://arxiv.org/abs/1407.6057}{{\tt arXiv:1407.6057 [nucl-ex]}}.

\bibitem{Bhalerao:2014xra}
R.~S. Bhalerao, J.-Y. Ollitrault, and S.~Pal, {\em {Characterizing flow
  fluctuations with moments}\/},
\href{http://arxiv.org/abs/1411.5160}{{\tt arXiv:1411.5160 [nucl-th]}}.

\bibitem{Aad:2013xma}
{ATLAS}, G.~Aad et al., {\em {Measurement of the distributions of
  event-by-event flow harmonics in lead-lead collisions at = 2.76 TeV with the
  ATLAS detector at the LHC}\/},
  \href{http://dx.doi.org/10.1007/JHEP11(2013)183}{JHEP {\bf 1311} (2013)
  183},
\href{http://arxiv.org/abs/1305.2942}{{\tt arXiv:1305.2942 [hep-ex]}}.

\bibitem{Aad:2014fla}
{ATLAS Collaboration}, G.~Aad et al., {\em {Measurement of event-plane
  correlations in $\sqrt{s_{NN}}=2.76$ TeV lead-lead collisions with the ATLAS
  detector}\/},  \href{http://dx.doi.org/10.1103/PhysRevC.90.024905}{Phys.~Rev.
  {\bf C90} (2014)  024905},
\href{http://arxiv.org/abs/1403.0489}{{\tt arXiv:1403.0489 [hep-ex]}}.

\bibitem{CMS:2013bza}
{CMS Collaboration}, S.~Chatrchyan et al., {\em {Studies of azimuthal dihadron
  correlations in ultra-central PbPb collisions at $\sqrt{s_{NN}}$ = 2.76
  TeV}\/},
\href{http://arxiv.org/abs/1312.1845}{{\tt arXiv:1312.1845 [nucl-ex]}}.

\bibitem{Heinz:2013bua}
U.~Heinz, Z.~Qiu, and C.~Shen, {\em {Fluctuating flow angles and anisotropic
  flow measurements}\/},
  \href{http://dx.doi.org/10.1103/PhysRevC.87.034913}{Phys. Rev. {\bf C87}
  (2013)  034913},
\href{http://arxiv.org/abs/1302.3535}{{\tt arXiv:1302.3535 [nucl-th]}}.

\bibitem{Qiu:2012uy}
Z.~Qiu and U.~Heinz, {\em {Hydrodynamic event-plane correlations in Pb+Pb
  collisions at $\sqrt{s}=2.76$ATeV}\/},
  \href{http://dx.doi.org/10.1016/j.physletb.2012.09.030}{Phys.~Lett. {\bf
  B717} (2012)  261},
\href{http://arxiv.org/abs/1208.1200}{{\tt arXiv:1208.1200 [nucl-th]}}.

\bibitem{Bhalerao:2013ina}
R.~S. Bhalerao, J.-Y. Ollitrault, and S.~Pal, {\em {Event-plane
  correlators}\/},
  \href{http://dx.doi.org/10.1103/PhysRevC.88.024909}{Phys.~Rev. {\bf C88}
  (2013)  024909},
\href{http://arxiv.org/abs/1307.0980}{{\tt arXiv:1307.0980 [nucl-th]}}.

\bibitem{Kovtun:2011np}
P.~Kovtun, G.~D. Moore, and P.~Romatschke, {\em {The stickiness of sound: An
  absolute lower limit on viscosity and the breakdown of second order
  relativistic hydrodynamics}\/},
  \href{http://dx.doi.org/10.1103/PhysRevD.84.025006}{Phys. Rev. {\bf D84}
  (2011)  025006},
\href{http://arxiv.org/abs/1104.1586}{{\tt arXiv:1104.1586 [hep-ph]}}.

\bibitem{Kapusta:2011gt}
J.~Kapusta, B.~Muller, and M.~Stephanov, {\em {Relativistic theory of
  hydrodynamic fluctuations with applications to heavy ion collisions}\/},
  \href{http://dx.doi.org/10.1103/PhysRevC.85.054906}{Phys. Rev. {\bf C85}
  (2012)  054906},
\href{http://arxiv.org/abs/1112.6405}{{\tt arXiv:1112.6405 [nucl-th]}}.

\bibitem{Murase:2013tma}
K.~Murase and T.~Hirano, {\em {Relativistic fluctuating hydrodynamics with
  memory functions and colored noises}\/},
\href{http://arxiv.org/abs/1304.3243}{{\tt arXiv:1304.3243 [nucl-th]}}.

\bibitem{Young:2013fka}
C.~Young, {\em {Numerical integration of thermal noise in relativistic
  hydrodynamics}\/},  \href{http://dx.doi.org/10.1103/PhysRevC.89.024913}{Phys.
  Rev. {\bf C89} (2014)  024913},
\href{http://arxiv.org/abs/1306.0472}{{\tt arXiv:1306.0472 [nucl-th]}}.

\bibitem{Novak:2013bqa}
J.~Novak, K.~Novak, S.~Pratt, J.~Vredevoogd, C.~Coleman-Smith, et al., {\em
  {Determining Fundamental Properties of Matter Created in Ultrarelativistic
  Heavy-Ion Collisions}\/},
  \href{http://dx.doi.org/10.1103/PhysRevC.89.034917}{Phys. Rev. {\bf C89}
  (2014)  034917},
\href{http://arxiv.org/abs/1303.5769}{{\tt arXiv:1303.5769 [nucl-th]}}.

\bibitem{Pratt:2015zsa}
S.~Pratt, E.~Sangaline, P.~Sorensen, and H.~Wang, {\em {Constraining the Eq. of
  State of Super-Hadronic Matter from Heavy-Ion Collisions}\/},
\href{http://arxiv.org/abs/1501.04042}{{\tt arXiv:1501.04042 [nucl-th]}}.

\bibitem{Shen:2013vja}
C.~Shen, U.~W. Heinz, J.-F. Paquet, and C.~Gale, {\em {Thermal photons as a
  quark-gluon plasma thermometer reexamined}\/},
  \href{http://dx.doi.org/10.1103/PhysRevC.89.044910}{Phys. Rev. {\bf C89}
  (2014)  044910},
\href{http://arxiv.org/abs/1308.2440}{{\tt arXiv:1308.2440 [nucl-th]}}.

\bibitem{Shen:2014cga}
C.~Shen, U.~Heinz, J.-F. Paquet, and C.~Gale, {\em {Thermal photon anisotropic
  flow serves as a quark-gluon plasma viscometer}\/},
\href{http://arxiv.org/abs/1403.7558}{{\tt arXiv:1403.7558 [nucl-th]}}.

\bibitem{Adare:2011zr}
{PHENIX Collaboration}, A.~Adare et al., {\em {Observation of direct-photon
  collective flow in $\sqrt{s_{NN}}=200$ GeV Au+Au collisions}\/},
  \href{http://dx.doi.org/10.1103/PhysRevLett.109.122302}{Phys. Rev. Lett. {\bf
  109} (2012)  122302},
\href{http://arxiv.org/abs/1105.4126}{{\tt arXiv:1105.4126 [nucl-ex]}}.

\bibitem{Lohner:2012ct}
{ALICE Collaboration}, D.~Lohner, {\em {Measurement of Direct-Photon Elliptic
  Flow in Pb-Pb Collisions at $\sqrt{s_{NN}} = 2.76$ TeV}\/},
  \href{http://dx.doi.org/10.1088/1742-6596/446/1/012028}{J. Phys. Conf. Ser.
  {\bf 446} (2013)  012028},
\href{http://arxiv.org/abs/1212.3995}{{\tt arXiv:1212.3995 [hep-ex]}}.

\bibitem{vanHees:2014ida}
H.~van Hees, M.~He, and R.~Rapp, {\em {Pseudo-Critical Enhancement of Thermal
  Photons in Relativistic Heavy-Ion Collisions}\/},
  \href{http://dx.doi.org/10.1016/j.nuclphysa.2014.09.009}{Nucl. Phys. {\bf
  A933} (2014)  256},
\href{http://arxiv.org/abs/1404.2846}{{\tt arXiv:1404.2846 [nucl-th]}}.

\bibitem{Khachatryan:2010gv}
{CMS Collaboration}, V.~Khachatryan et al., {\em {Observation of Long-Range
  Near-Side Angular Correlations in Proton-Proton Collisions at the LHC}\/},
  \href{http://dx.doi.org/10.1007/JHEP09(2010)091}{JHEP {\bf 1009} (2010)
  091},
\href{http://arxiv.org/abs/1009.4122}{{\tt arXiv:1009.4122 [hep-ex]}}.

\bibitem{CMS:2012qk}
{CMS Collaboration}, S.~Chatrchyan et al., {\em {Observation of long-range
  near-side angular correlations in proton-lead collisions at the LHC}\/},
  \href{http://dx.doi.org/10.1016/j.physletb.2012.11.025}{Phys. Lett. {\bf
  B718} (2013)  795--814},
\href{http://arxiv.org/abs/1210.5482}{{\tt arXiv:1210.5482 [nucl-ex]}}.

\bibitem{Abelev:2012ola}
{ALICE Collaboration}, B.~Abelev et al., {\em {Long-range angular correlations
  on the near and away side in $p$-Pb collisions at $\sqrt{s_{NN}}=5.02$
  TeV}\/},  \href{http://dx.doi.org/10.1016/j.physletb.2013.01.012}{Phys. Lett.
  {\bf B719} (2013)  29--41},
\href{http://arxiv.org/abs/1212.2001}{{\tt arXiv:1212.2001 [nucl-ex]}}.

\bibitem{Aad:2012gla}
{ATLAS Collaboration}, G.~Aad et al., {\em {Observation of Associated Near-Side
  and Away-Side Long-Range Correlations in $\sqrt{s_{NN}}$=5.02  TeV
  Proton-Lead Collisions with the ATLAS Detector}\/},
  \href{http://dx.doi.org/10.1103/PhysRevLett.110.182302}{Phys. Rev. Lett. {\bf
  110} (2013)  182302},
\href{http://arxiv.org/abs/1212.5198}{{\tt arXiv:1212.5198 [hep-ex]}}.

\bibitem{Adare:2013piz}
{PHENIX Collaboration}, A.~Adare et al., {\em {Quadrupole Anisotropy in
  Dihadron Azimuthal Correlations in Central $d$$+$Au Collisions at
  $\sqrt{s_{_{NN}}}$=200 GeV}\/},
  \href{http://dx.doi.org/10.1103/PhysRevLett.111.212301}{Phys. Rev. Lett. {\bf
  111} (2013)  212301},
\href{http://arxiv.org/abs/1303.1794}{{\tt arXiv:1303.1794 [nucl-ex]}}.

\bibitem{Chatrchyan:2013eya}
{CMS Collaboration}, S.~Chatrchyan et al., {\em {Study of the production of
  charged pions, kaons, and protons in pPb collisions at $\sqrt{s_{NN}} =\ $
  5.02 $\,\text {TeV}$}\/},
  \href{http://dx.doi.org/10.1140/epjc/210052-014-2847-x,
  10.1140/epjc/s10052-014-2847-x}{Eur. Phys. J. {\bf C74} (2014)  2847},
\href{http://arxiv.org/abs/1307.3442}{{\tt arXiv:1307.3442 [hep-ex]}}.

\bibitem{Abelev:2013haa}
{ALICE Collaboration}, B.~B. Abelev et al., {\em {Multiplicity Dependence of
  Pion, Kaon, Proton and Lambda Production in p-Pb Collisions at
  $\sqrt{s_{NN}}$ = 5.02 TeV}\/},
  \href{http://dx.doi.org/10.1016/j.physletb.2013.11.020}{Phys. Lett. {\bf
  B728} (2014)  25--38},
\href{http://arxiv.org/abs/1307.6796}{{\tt arXiv:1307.6796 [nucl-ex]}}.

\bibitem{Aad:2013fja}
{ATLAS Collaboration}, G.~Aad et al., {\em {Measurement with the ATLAS detector
  of multi-particle azimuthal correlations in p+Pb collisions at
  $\sqrt{s_{NN}}$=5.02 TeV}\/},
  \href{http://dx.doi.org/10.1016/j.physletb.2013.06.057}{Phys. Lett. {\bf
  B725} (2013)  60--78},
\href{http://arxiv.org/abs/1303.2084}{{\tt arXiv:1303.2084 [hep-ex]}}.

\bibitem{Chatrchyan:2013nka}
{CMS Collaboration}, S.~Chatrchyan et al., {\em {Multiplicity and transverse
  momentum dependence of two- and four-particle correlations in pPb and PbPb
  collisions}\/},
  \href{http://dx.doi.org/10.1016/j.physletb.2013.06.028}{Phys. Lett. {\bf
  B724} (2013)  213--240},
\href{http://arxiv.org/abs/1305.0609}{{\tt arXiv:1305.0609 [nucl-ex]}}.

\bibitem{ABELEV:2013wsa}
{ALICE Collaboration}, B.~B. Abelev et al., {\em {Long-range angular
  correlations of $\rm \pi$, K and p in p-Pb collisions at $\sqrt{s_{\rm NN}}$
  = 5.02 TeV}\/},
  \href{http://dx.doi.org/10.1016/j.physletb.2013.08.024}{Phys. Lett. {\bf
  B726} (2013)  164--177},
\href{http://arxiv.org/abs/1307.3237}{{\tt arXiv:1307.3237 [nucl-ex]}}.

\bibitem{Abelev:2014mda}
{ALICE Collaboration}, B.~B. Abelev et al., {\em {Multiparticle azimuthal
  correlations in p -Pb and Pb-Pb collisions at the CERN Large Hadron
  Collider}\/},  \href{http://dx.doi.org/10.1103/PhysRevC.90.054901}{Phys. Rev.
  {\bf C90} (2014)  054901},
\href{http://arxiv.org/abs/1406.2474}{{\tt arXiv:1406.2474 [nucl-ex]}}.

\bibitem{Aad:2014lta}
{ATLAS Collaboration}, G.~Aad et al., {\em {Measurement of long-range
  pseudorapidity correlations and azimuthal harmonics in $\sqrt{s_{NN}}=5.02$
  TeV proton-lead collisions with the ATLAS detector}\/},
  \href{http://dx.doi.org/10.1103/PhysRevC.90.044906}{Phys. Rev. {\bf C90}
  (2014)  044906},
\href{http://arxiv.org/abs/1409.1792}{{\tt arXiv:1409.1792 [hep-ex]}}.

\bibitem{Khachatryan:2014jra}
{CMS Collaboration}, V.~Khachatryan et al., {\em {Long-range two-particle
  correlations of strange hadrons with charged particles in pPb and PbPb
  collisions at LHC energies}\/},
\href{http://arxiv.org/abs/1409.3392}{{\tt arXiv:1409.3392 [nucl-ex]}}.

\bibitem{CMS-PAS-HIN-14-006}
{CMS Collaboration}, {\em {Multiplicity dependence of multiparticle
  correlations in pPb and PbPb collisions}\/},  CMS-PAS-HIN-14-006.

\bibitem{CMS_cumulants_pPb}
{CMS Collaboration}, R.~Granier~de Cassagnac et al., {\em CMS heavy-ion
  overview\/},
  \href{http://dx.doi.org/doi:10.1016/j.nuclphysa.2014.10.043}{Nucl. Phys. {\bf
  A931} (2014)  13--21}.

\bibitem{Niemi:2014wta}
H.~Niemi and G.~Denicol, {\em {How large is the Knudsen number reached in fluid
  dynamical simulations of ultrarelativistic heavy ion collisions?}\/},
\href{http://arxiv.org/abs/1404.7327}{{\tt arXiv:1404.7327 [nucl-th]}}.

\bibitem{Aoki:2006we}
Y.~Aoki, G.~Endrodi, Z.~Fodor, S.~Katz, and K.~Szabo, {\em {The Order of the
  quantum chromodynamics transition predicted by the standard model of particle
  physics}\/},  \href{http://dx.doi.org/10.1038/nature05120}{Nature {\bf 443}
  (2006)  675--678},
\href{http://arxiv.org/abs/hep-lat/0611014}{{\tt arXiv:hep-lat/0611014
  [hep-lat]}}.

\bibitem{Aoki:2009sc}
Y.~Aoki, S.~Borsanyi, S.~Durr, Z.~Fodor, S.~D. Katz, et al., {\em {The QCD
  transition temperature: results with physical masses in the continuum limit
  II.}\/},  \href{http://dx.doi.org/10.1088/1126-6708/2009/06/088}{JHEP {\bf
  0906} (2009)  088},
\href{http://arxiv.org/abs/0903.4155}{{\tt arXiv:0903.4155 [hep-lat]}}.

\bibitem{Bazavov:2011nk}
A.~Bazavov, T.~Bhattacharya, M.~Cheng, C.~DeTar, H.~Ding, et al., {\em {The
  chiral and deconfinement aspects of the QCD transition}\/},
  \href{http://dx.doi.org/10.1103/PhysRevD.85.054503}{Phys. Rev. {\bf D85}
  (2012)  054503},
\href{http://arxiv.org/abs/1111.1710}{{\tt arXiv:1111.1710 [hep-lat]}}.

\bibitem{Stephanov:1998dy}
M.~A. Stephanov, K.~Rajagopal, and E.~V. Shuryak, {\em {Signatures of the
  tricritical point in QCD}\/},
  \href{http://dx.doi.org/10.1103/PhysRevLett.81.4816}{Phys. Rev. Lett. {\bf
  81} (1998)  4816--4819},
\href{http://arxiv.org/abs/hep-ph/9806219}{{\tt arXiv:hep-ph/9806219
  [hep-ph]}}.

\bibitem{Fodor:2004nz}
Z.~Fodor and S.~Katz, {\em {Critical point of QCD at finite T and mu, lattice
  results for physical quark masses}\/},
  \href{http://dx.doi.org/10.1088/1126-6708/2004/04/050}{JHEP {\bf 0404} (2004)
   050},
\href{http://arxiv.org/abs/hep-lat/0402006}{{\tt arXiv:hep-lat/0402006
  [hep-lat]}}.

\bibitem{Allton:2005gk}
C.~Allton, M.~Doring, S.~Ejiri, S.~Hands, O.~Kaczmarek, et al., {\em
  {Thermodynamics of two flavor QCD to sixth order in quark chemical
  potential}\/},  \href{http://dx.doi.org/10.1103/PhysRevD.71.054508}{Phys.
  Rev. {\bf D71} (2005)  054508},
\href{http://arxiv.org/abs/hep-lat/0501030}{{\tt arXiv:hep-lat/0501030
  [hep-lat]}}.

\bibitem{Gavai:2008zr}
R.~Gavai and S.~Gupta, {\em {QCD at finite chemical potential with six time
  slices}\/},  \href{http://dx.doi.org/10.1103/PhysRevD.78.114503}{Phys. Rev.
  {\bf D78} (2008)  114503},
\href{http://arxiv.org/abs/0806.2233}{{\tt arXiv:0806.2233 [hep-lat]}}.

\bibitem{deForcrand:2008zi}
P.~de~Forcrand and O.~Philipsen, {\em {The curvature of the critical surface
  (m(u,d),m(s))**crit(mu): A Progress report}\/},  PoS {\bf LATTICE2008} (2008)
   208,
\href{http://arxiv.org/abs/0811.3858}{{\tt arXiv:0811.3858 [hep-lat]}}.

\bibitem{Datta:2012pj}
S.~Datta, R.~V. Gavai, and S.~Gupta, {\em {The QCD Critical Point : marching
  towards continuum}\/},
  \href{http://dx.doi.org/10.1016/j.nuclphysa.2013.02.156}{Nucl. Phys. {\bf
  A904-905} (2013)  883c--886c},
\href{http://arxiv.org/abs/1210.6784}{{\tt arXiv:1210.6784 [hep-lat]}}.

\bibitem{BESII}
{\em Studying the Phase Diagram of QCD Matter at RHIC}, 2014.
\newblock
  \url{https://drupal.star.bnl.gov/STAR/files/BES_WPII_ver6.9_Cover.pdf}.

\bibitem{Kumar:2012fb}
{STAR Collaboration}, L.~Kumar, {\em {STAR Results from the RHIC Beam Energy
  Scan-I}\/},  \href{http://dx.doi.org/10.1016/j.nuclphysa.2013.01.070}{Nucl.
  Phys. A {\bf 904-905} (2013)  256c--263c},
\href{http://arxiv.org/abs/1211.1350}{{\tt arXiv:1211.1350 [nucl-ex]}}.

\bibitem{Adamczyk:2014mxp}
{STAR Collaboration}, L.~Adamczyk et al., {\em {Beam energy dependent two-pion
  interferometry and the freeze-out eccentricity of pions in heavy ion
  collisions at STAR}\/},
\href{http://arxiv.org/abs/1403.4972}{{\tt arXiv:1403.4972 [nucl-ex]}}.

\bibitem{Adare:2014qvs}
{PHENIX Collaboration}, A.~Adare et al., {\em {Beam-energy and system-size
  dependence of the space-time extent of the pion emission source produced in
  heavy ion collisions}\/},
\href{http://arxiv.org/abs/1410.2559}{{\tt arXiv:1410.2559 [nucl-ex]}}.

\bibitem{Aamodt:2011mr}
{ALICE Collaboration}, K.~Aamodt et al., {\em {Two-pion Bose-Einstein
  correlations in central Pb-Pb collisions at $sqrt(s_NN)$ = 2.76 TeV}\/},
  \href{http://dx.doi.org/10.1016/j.physletb.2010.12.053}{Phys. Lett. {\bf
  B696} (2011)  328--337},
\href{http://arxiv.org/abs/1012.4035}{{\tt arXiv:1012.4035 [nucl-ex]}}.

\bibitem{Rischke:1996em}
D.~H. Rischke and M.~Gyulassy, {\em {The Time delay signature of quark - gluon
  plasma formation in relativistic nuclear collisions}\/},
  \href{http://dx.doi.org/10.1016/0375-9474(96)00259-X}{Nucl. Phys. {\bf A608}
  (1996)  479--512},
\href{http://arxiv.org/abs/nucl-th/9606039}{{\tt arXiv:nucl-th/9606039
  [nucl-th]}}.

\bibitem{Adamczyk:2013dal}
{STAR Collaboration}, L.~Adamczyk et al., {\em {Energy Dependence of Moments of
  Net-proton Multiplicity Distributions at RHIC}\/},
  \href{http://dx.doi.org/10.1103/PhysRevLett.112.032302}{Phys. Rev. Lett. {\bf
  112} (2014)  032302},
\href{http://arxiv.org/abs/1309.5681}{{\tt arXiv:1309.5681 [nucl-ex]}}.

\bibitem{CPODKurtosis}
{STAR Collaboration}, X.~Luo, {\em Search for the QCD Critical Point: Energy
  Dependence of Higher Moments of Net-proton and Net-charge Distributions at
  RHIC\/}, .
  \url{http://www.physik.uni-bielefeld.de/~cpod2014/CPOD2014_LuoXiaofeng_ver5.pdf}.
  Talk at 2014 CPOD Conference.

\bibitem{Stephanov:2011pb}
M.~Stephanov, {\em {On the sign of kurtosis near the QCD critical point}\/},
  \href{http://dx.doi.org/10.1103/PhysRevLett.107.052301}{Phys. Rev. Lett. {\bf
  107} (2011)  052301},
\href{http://arxiv.org/abs/1104.1627}{{\tt arXiv:1104.1627 [hep-ph]}}.

\bibitem{Adamczyk:2014ipa}
{STAR Collaboration}, L.~Adamczyk et al., {\em {Beam-Energy Dependence of the
  Directed Flow of Protons, Antiprotons, and Pions in Au+Au Collisions}\/},
  \href{http://dx.doi.org/10.1103/PhysRevLett.112.162301(2014),
  10.1103/PhysRevLett.112.162301}{Phys. Rev. Lett. {\bf 112} (2014)  162301},
\href{http://arxiv.org/abs/1401.3043}{{\tt arXiv:1401.3043 [nucl-ex]}}.

\bibitem{Brachmann:1999mp}
J.~Brachmann, A.~Dumitru, H.~Stoecker, and W.~Greiner, {\em {The Directed flow
  maximum near c(s) = 0}\/},
  \href{http://dx.doi.org/10.1007/s100500070077}{Eur. Phys. J. {\bf A8} (2000)
  549--552},
\href{http://arxiv.org/abs/nucl-th/9912014}{{\tt arXiv:nucl-th/9912014
  [nucl-th]}}.

\bibitem{Stoecker:2004qu}
H.~Stoecker, {\em {Collective flow signals the quark gluon plasma}\/},
  \href{http://dx.doi.org/10.1016/j.nuclphysa.2004.12.074}{Nucl. Phys. {\bf
  A750} (2005)  121--147},
\href{http://arxiv.org/abs/nucl-th/0406018}{{\tt arXiv:nucl-th/0406018
  [nucl-th]}}.

\bibitem{Huovinen:2014woa}
P.~Huovinen, P.~Petreczky, and C.~Schmidt, {\em Equation of state at finite
  net-baryon density using Taylor coefficients up to sixth order\/},
  \href{http://dx.doi.org/http://dx.doi.org/10.1016/j.nuclphysa.2014.08.069}{Nucl.
  Phys. {\bf A931} (2014)  769 -- 773},
  \href{http://arxiv.org/abs/1407.8532}{{\tt arXiv:1407.8532 [nucl-th]}}.

\bibitem{Abelev:2007rw}
{STAR Collaboration}, B.~Abelev et al., {\em {Partonic flow and phi-meson
  production in Au + Au collisions at s(NN)**(1/2) = 200-GeV}\/},
  \href{http://dx.doi.org/10.1103/PhysRevLett.99.112301}{Phys. Rev. Lett. {\bf
  99} (2007)  112301},
\href{http://arxiv.org/abs/nucl-ex/0703033}{{\tt arXiv:nucl-ex/0703033
  [NUCL-EX]}}.

\bibitem{Abelev:2009ac}
{STAR Collaboration}, B.~Abelev et al., {\em {Azimuthal Charged-Particle
  Correlations and Possible Local Strong Parity Violation}\/},
  \href{http://dx.doi.org/10.1103/PhysRevLett.103.251601}{Phys. Rev. Lett. {\bf
  103} (2009)  251601},
\href{http://arxiv.org/abs/0909.1739}{{\tt arXiv:0909.1739 [nucl-ex]}}.

\bibitem{Abelev:2012pa}
{ALICE Collaboration}, B.~Abelev et al., {\em {Charge separation relative to
  the reaction plane in Pb-Pb collisions at $\sqrt{s_{NN}}= 2.76$ TeV}\/},
  \href{http://dx.doi.org/10.1103/PhysRevLett.110.012301}{Phys. Rev. Lett. {\bf
  110} (2013)  012301},
\href{http://arxiv.org/abs/1207.0900}{{\tt arXiv:1207.0900 [nucl-ex]}}.

\bibitem{Adamczyk:2013kcb}
{STAR Collaboration}, L.~Adamczyk et al., {\em {Measurement of Charge
  Multiplicity Asymmetry Correlations in High Energy Nucleus-Nucleus Collisions
  at 200 GeV}\/},  \href{http://dx.doi.org/10.1103/PhysRevC.89.044908}{Phys.
  Rev. {\bf C89} (2014)  044908},
\href{http://arxiv.org/abs/1303.0901}{{\tt arXiv:1303.0901 [nucl-ex]}}.

\bibitem{Fukushima:2008xe}
K.~Fukushima, D.~E. Kharzeev, and H.~J. Warringa, {\em {The Chiral Magnetic
  Effect}\/},  \href{http://dx.doi.org/10.1103/PhysRevD.78.074033}{Phys. Rev.
  {\bf D78} (2008)  074033},
\href{http://arxiv.org/abs/0808.3382}{{\tt arXiv:0808.3382 [hep-ph]}}.

\bibitem{Kharzeev:2010gr}
D.~E. Kharzeev and D.~T. Son, {\em {Testing the chiral magnetic and chiral
  vortical effects in heavy ion collisions}\/},
  \href{http://dx.doi.org/10.1103/PhysRevLett.106.062301}{Phys. Rev. Lett. {\bf
  106} (2011)  062301},
\href{http://arxiv.org/abs/1010.0038}{{\tt arXiv:1010.0038 [hep-ph]}}.

\bibitem{Bzdak:2009fc}
A.~Bzdak, V.~Koch, and J.~Liao, {\em {Remarks on possible local parity
  violation in heavy ion collisions}\/},
  \href{http://dx.doi.org/10.1103/PhysRevC.81.031901}{Phys. Rev. {\bf C81}
  (2010)  031901},
\href{http://arxiv.org/abs/0912.5050}{{\tt arXiv:0912.5050 [nucl-th]}}.

\bibitem{Pratt:2010zn}
S.~Pratt, S.~Schlichting, and S.~Gavin, {\em {Effects of Momentum Conservation
  and Flow on Angular Correlations at RHIC}\/},
  \href{http://dx.doi.org/10.1103/PhysRevC.84.024909}{Phys. Rev. {\bf C84}
  (2011)  024909},
\href{http://arxiv.org/abs/1011.6053}{{\tt arXiv:1011.6053 [nucl-th]}}.

\bibitem{Adamczyk:2014mzf}
{STAR Collaboration}, L.~Adamczyk et al., {\em {Beam-energy dependence of
  charge separation along the magnetic field in Au+Au collisions at RHIC}\/},
  \href{http://dx.doi.org/10.1103/PhysRevLett.113.052302}{Phys. Rev. Lett. {\bf
  113} (2014)  052302},
\href{http://arxiv.org/abs/1404.1433}{{\tt arXiv:1404.1433 [nucl-ex]}}.

\bibitem{DElia:2010nq}
M.~D'Elia, S.~Mukherjee, and F.~Sanfilippo, {\em {QCD Phase Transition in a
  Strong Magnetic Background}\/},
  \href{http://dx.doi.org/10.1103/PhysRevD.82.051501}{Phys. Rev. {\bf D82}
  (2010)  051501},
\href{http://arxiv.org/abs/1005.5365}{{\tt arXiv:1005.5365 [hep-lat]}}.

\bibitem{Bali:2011qj}
G.~Bali, F.~Bruckmann, G.~Endrodi, Z.~Fodor, S.~Katz, et al., {\em {The QCD
  phase diagram for external magnetic fields}\/},
  \href{http://dx.doi.org/10.1007/JHEP02(2012)044}{JHEP {\bf 1202} (2012)
  044},
\href{http://arxiv.org/abs/1111.4956}{{\tt arXiv:1111.4956 [hep-lat]}}.

\bibitem{Bali:2014kia}
G.~Bali, F.~Bruckmann, G.~Endr{\"o}di, S.~Katz, and A.~Sch{\"a}fer, {\em {The
  QCD equation of state in background magnetic fields}\/},
  \href{http://dx.doi.org/10.1007/JHEP08(2014)177}{JHEP {\bf 1408} (2014)
  177},
\href{http://arxiv.org/abs/1406.0269}{{\tt arXiv:1406.0269 [hep-lat]}}.

\bibitem{Kharzeev:2010gd}
D.~E. Kharzeev and H.-U. Yee, {\em {Chiral Magnetic Wave}\/},
  \href{http://dx.doi.org/10.1103/PhysRevD.83.085007}{Phys. Rev. {\bf D83}
  (2011)  085007},
\href{http://arxiv.org/abs/1012.6026}{{\tt arXiv:1012.6026 [hep-th]}}.

\bibitem{Burnier:2011bf}
Y.~Burnier, D.~E. Kharzeev, J.~Liao, and H.-U. Yee, {\em {Chiral magnetic wave
  at finite baryon density and the electric quadrupole moment of quark-gluon
  plasma in heavy ion collisions}\/},
  \href{http://dx.doi.org/10.1103/PhysRevLett.107.052303}{Phys. Rev. Lett. {\bf
  107} (2011)  052303},
\href{http://arxiv.org/abs/1103.1307}{{\tt arXiv:1103.1307 [hep-ph]}}.

\bibitem{Wang:2012qs}
{STAR Collaboration}, G.~Wang, {\em {Search for Chiral Magnetic Effects in
  High-Energy Nuclear Collisions}\/},
  \href{http://dx.doi.org/10.1016/j.nuclphysa.2013.01.069}{Nucl.Phys. {\bf
  A904-905} (2013)  248c--255c},
\href{http://arxiv.org/abs/1210.5498}{{\tt arXiv:1210.5498 [nucl-ex]}}.

\bibitem{Belmont:2014lta}
{ALICE Collaboration}, R.~Belmont, {\em {Charge-dependent anisotropic flow
  studies and the search for the Chiral Magnetic Wave in ALICE}\/},
\href{http://arxiv.org/abs/1408.1043}{{\tt arXiv:1408.1043 [nucl-ex]}}.

\bibitem{Stephanov:2008qz}
M.~Stephanov, {\em {Non-Gaussian fluctuations near the QCD critical point}\/},
  \href{http://dx.doi.org/10.1103/PhysRevLett.102.032301}{Phys. Rev. Lett. {\bf
  102} (2009)  032301},
\href{http://arxiv.org/abs/0809.3450}{{\tt arXiv:0809.3450 [hep-ph]}}.

\bibitem{Athanasiou:2010kw}
C.~Athanasiou, K.~Rajagopal, and M.~Stephanov, {\em {Using Higher Moments of
  Fluctuations and their Ratios in the Search for the QCD Critical Point}\/},
  \href{http://dx.doi.org/10.1103/PhysRevD.82.074008}{Phys. Rev. {\bf D82}
  (2010)  074008},
\href{http://arxiv.org/abs/1006.4636}{{\tt arXiv:1006.4636 [hep-ph]}}.

\bibitem{Ejiri:2005wq}
S.~Ejiri, F.~Karsch, and K.~Redlich, {\em {Hadronic fluctuations at the QCD
  phase transition}\/},
  \href{http://dx.doi.org/10.1016/j.physletb.2005.11.083}{Phys. Lett. {\bf
  B633} (2006)  275--282},
\href{http://arxiv.org/abs/hep-ph/0509051}{{\tt arXiv:hep-ph/0509051
  [hep-ph]}}.

\bibitem{Karsch:2010ck}
F.~Karsch and K.~Redlich, {\em {Probing freeze-out conditions in heavy ion
  collisions with moments of charge fluctuations}\/},
  \href{http://dx.doi.org/10.1016/j.physletb.2010.10.046}{Phys. Lett. {\bf
  B695} (2011)  136--142},
\href{http://arxiv.org/abs/1007.2581}{{\tt arXiv:1007.2581 [hep-ph]}}.

\bibitem{Adamczyk:2014fia}
{STAR Collaboration}, L.~Adamczyk et al., {\em {Beam energy dependence of
  moments of the net-charge multiplicity distributions in Au+Au collisions at
  RHIC}\/},  \href{http://dx.doi.org/10.1103/PhysRevLett.113.092301}{Phys. Rev.
  Lett. {\bf 113} (2014)  092301},
\href{http://arxiv.org/abs/1402.1558}{{\tt arXiv:1402.1558 [nucl-ex]}}.

\bibitem{Berdnikov:1999ph}
B.~Berdnikov and K.~Rajagopal, {\em {Slowing out-of-equilibrium near the QCD
  critical point}\/},
  \href{http://dx.doi.org/10.1103/PhysRevD.61.105017}{Phys. Rev. {\bf D61}
  (2000)  105017},
\href{http://arxiv.org/abs/hep-ph/9912274}{{\tt arXiv:hep-ph/9912274
  [hep-ph]}}.

\bibitem{Karsch:2012wm}
F.~Karsch, {\em {Determination of Freeze-out Conditions from Lattice QCD
  Calculations}\/},  \href{http://dx.doi.org/10.2478/s11534-012-0074-3}{Central
  Eur. J. Phys. {\bf 10} (2012)  1234--1237},
\href{http://arxiv.org/abs/1202.4173}{{\tt arXiv:1202.4173 [hep-lat]}}.

\bibitem{Bazavov:2012vg}
A.~Bazavov, H.~Ding, P.~Hegde, O.~Kaczmarek, F.~Karsch, et al., {\em
  {Freeze-out Conditions in Heavy Ion Collisions from QCD Thermodynamics}\/},
  \href{http://dx.doi.org/10.1103/PhysRevLett.109.192302}{Phys. Rev. Lett. {\bf
  109} (2012)  192302},
\href{http://arxiv.org/abs/1208.1220}{{\tt arXiv:1208.1220 [hep-lat]}}.

\bibitem{Mukherjee:2013lsa}
S.~Mukherjee and M.~Wagner, {\em {Deconfinement of strangeness and freeze-out
  from charge fluctuations}\/},  PoS {\bf CPOD2013} (2013)  039,
\href{http://arxiv.org/abs/1307.6255}{{\tt arXiv:1307.6255 [nucl-th]}}.

\bibitem{Borsanyi:2013hza}
S.~Borsanyi, Z.~Fodor, S.~Katz, S.~Krieg, C.~Ratti, et al., {\em {Freeze-out
  parameters: lattice meets experiment}\/},
  \href{http://dx.doi.org/10.1103/PhysRevLett.111.062005}{Phys. Rev. Lett. {\bf
  111} (2013)  062005},
\href{http://arxiv.org/abs/1305.5161}{{\tt arXiv:1305.5161 [hep-lat]}}.

\bibitem{Borsanyi:2014ewa}
S.~Borsanyi, Z.~Fodor, S.~Katz, S.~Krieg, C.~Ratti, et al., {\em {Freeze-out
  parameters from electric charge and baryon number fluctuations: is there
  consistency?}\/},
  \href{http://dx.doi.org/10.1103/PhysRevLett.113.052301}{Phys. Rev. Lett. {\bf
  113} (2014)  052301},
\href{http://arxiv.org/abs/1403.4576}{{\tt arXiv:1403.4576 [hep-lat]}}.

\bibitem{Rapp:2009yu}
R.~Rapp, J.~Wambach, and H.~van Hees, {\em {The Chiral Restoration Transition
  of QCD and Low Mass Dileptons}\/},
\href{http://arxiv.org/abs/0901.3289}{{\tt arXiv:0901.3289 [hep-ph]}}.

\bibitem{Hohler:2013eba}
P.~M. Hohler and R.~Rapp, {\em {Is $\rho$-Meson Melting Compatible with Chiral
  Restoration?}\/},
  \href{http://dx.doi.org/10.1016/j.physletb.2014.02.021}{Phys. Lett. {\bf
  B731} (2014)  103--109},
\href{http://arxiv.org/abs/1311.2921}{{\tt arXiv:1311.2921 [hep-ph]}}.

\bibitem{Huck:2014mfa}
P.~Huck, {\em Beam energy dependence of dielectron production in Au+Au
  collisions from \{STAR\} at \{RHIC\}\/},
  \href{http://dx.doi.org/http://dx.doi.org/10.1016/j.nuclphysa.2014.09.090}{Nuclear
  Physics A {\bf 931} (2014)  659 -- 664},
  \href{http://arxiv.org/abs/1409.5675}{{\tt arXiv:1409.5675 [nucl-ex]}}.

\bibitem{Adamczyk:2013caa}
{STAR Collaboration}, L.~Adamczyk et al., {\em {Dielectron Mass Spectra from
  Au+Au Collisions at $\sqrt{s_{\rm NN}}$ = 200 GeV}\/},
  \href{http://dx.doi.org/10.1103/PhysRevLett.113.022301,
  10.1103/PhysRevLett.113.049903}{Phys. Rev. Lett. {\bf 113} (2014)  022301},
\href{http://arxiv.org/abs/1312.7397}{{\tt arXiv:1312.7397 [hep-ex]}}.

\bibitem{Adamczyk:2015bha}
{STAR Collaboration}, L.~Adamczyk et al., {\em {Energy dependence of
  acceptance-corrected dielectron excess mass spectrum at mid-rapidity in Au+Au
  collisions at $\sqrt{s_{NN}} = 19.6$ and 200 GeV}\/},
\href{http://arxiv.org/abs/1501.05341}{{\tt arXiv:1501.05341 [hep-ex]}}.

\bibitem{Rapp:2013nxa}
R.~Rapp, {\em {Dilepton Spectroscopy of QCD Matter at Collider Energies}\/},
  \href{http://dx.doi.org/10.1155/2013/148253}{Adv. High Energy Phys. {\bf
  2013} (2013)  148253},
\href{http://arxiv.org/abs/1304.2309}{{\tt arXiv:1304.2309 [hep-ph]}}.

\bibitem{Rapp:2014hha}
R.~Rapp and H.~van Hees, {\em {Thermal Dileptons as Fireball Thermometer and
  Chronometer}\/},
\href{http://arxiv.org/abs/1411.4612}{{\tt arXiv:1411.4612 [hep-ph]}}.

\bibitem{Borsanyi:2011sw}
S.~Borsanyi, Z.~Fodor, S.~D. Katz, S.~Krieg, C.~Ratti, et al., {\em
  {Fluctuations of conserved charges at finite temperature from lattice
  QCD}\/},  \href{http://dx.doi.org/10.1007/JHEP01(2012)138}{JHEP {\bf 1201}
  (2012)  138},
\href{http://arxiv.org/abs/1112.4416}{{\tt arXiv:1112.4416 [hep-lat]}}.

\bibitem{Bazavov:2012jq}
{HotQCD Collaboration}, A.~Bazavov et al., {\em {Fluctuations and Correlations
  of net baryon number, electric charge, and strangeness: A comparison of
  lattice QCD results with the hadron resonance gas model}\/},
  \href{http://dx.doi.org/10.1103/PhysRevD.86.034509}{Phys. Rev. {\bf D86}
  (2012)  034509},
\href{http://arxiv.org/abs/1203.0784}{{\tt arXiv:1203.0784 [hep-lat]}}.

\bibitem{Borsanyi:2012cr}
S.~Borsanyi, G.~Endrodi, Z.~Fodor, S.~Katz, S.~Krieg, et al., {\em {QCD
  equation of state at nonzero chemical potential: continuum results with
  physical quark masses at order $mu^2$}\/},
  \href{http://dx.doi.org/10.1007/JHEP08(2012)053}{JHEP {\bf 1208} (2012)
  053},
\href{http://arxiv.org/abs/1204.6710}{{\tt arXiv:1204.6710 [hep-lat]}}.

\bibitem{Hegde:2014wga}
{for the BNL-Bielefeld-CCNU collaboration}, P.~Hegde, {\em {The QCD equation of
  state to $\mathcal{O}(\mu_B^4)$}\/},
\href{http://arxiv.org/abs/1412.6727}{{\tt arXiv:1412.6727 [hep-lat]}}.

\bibitem{Sterman:1977wj}
G.~F. Sterman and S.~Weinberg, {\em {Jets from Quantum Chromodynamics}\/},
\href{http://dx.doi.org/10.1103/PhysRevLett.39.1436}{Phys. Rev. Lett. {\bf 39}
  (1977)  1436}.

\bibitem{Feynman:1978dt}
R.~Feynman, R.~Field, and G.~Fox, {\em {A Quantum Chromodynamic Approach for
  the Large Transverse Momentum Production of Particles and Jets}\/},
\href{http://dx.doi.org/10.1103/PhysRevD.18.3320}{Phys. Rev. {\bf D18} (1978)
  3320}.

\bibitem{Field:1977fa}
R.~Field and R.~Feynman, {\em {A Parametrization of the Properties of Quark
  Jets}\/},
\href{http://dx.doi.org/10.1016/0550-3213(78)90015-9}{Nucl. Phys. {\bf B136}
  (1978)  1}.

\bibitem{Adcox:2001jp}
{PHENIX Collaboration}, K.~Adcox et al., {\em {Suppression of hadrons with
  large transverse momentum in central Au+Au collisions at $\sqrt{s_{NN}}$ =
  130-GeV}\/},  \href{http://dx.doi.org/10.1103/PhysRevLett.88.022301}{Phys.
  Rev. Lett. {\bf 88} (2002)  022301},
\href{http://arxiv.org/abs/nucl-ex/0109003}{{\tt arXiv:nucl-ex/0109003
  [nucl-ex]}}.

\bibitem{Adler:2002tq}
{STAR Collaboration}, C.~Adler et al., {\em {Disappearance of back-to-back high
  $p_{T}$ hadron correlations in central Au+Au collisions at $\sqrt{s_{NN}}$ =
  200-GeV}\/},  \href{http://dx.doi.org/10.1103/PhysRevLett.90.082302}{Phys.
  Rev. Lett. {\bf 90} (2003)  082302},
\href{http://arxiv.org/abs/nucl-ex/0210033}{{\tt arXiv:nucl-ex/0210033
  [nucl-ex]}}.

\bibitem{Bjorken:1982tu}
J.~Bjorken,
{\em {Energy Loss of Energetic Partons in Quark - Gluon Plasma: Possible
  Extinction of High p(t) Jets in Hadron - Hadron Collisions}\/}, .

\bibitem{Gyulassy:1993hr}
M.~Gyulassy and X.-N. Wang, {\em {Multiple collisions and induced gluon
  Bremsstrahlung in QCD}\/},
  \href{http://dx.doi.org/10.1016/0550-3213(94)90079-5}{Nucl. Phys. {\bf B420}
  (1994)  583--614},
\href{http://arxiv.org/abs/nucl-th/9306003}{{\tt arXiv:nucl-th/9306003
  [nucl-th]}}.

\bibitem{Wang:1991xy}
X.-N. Wang and M.~Gyulassy, {\em {Gluon shadowing and jet quenching in A + A
  collisions at s**(1/2) = 200-GeV}\/},
\href{http://dx.doi.org/10.1103/PhysRevLett.68.1480}{Phys. Rev. Lett. {\bf 68}
  (1992)  1480--1483}.

\bibitem{Baier:1996kr}
R.~Baier, Y.~L. Dokshitzer, A.~H. Mueller, S.~Peigne, and D.~Schiff, {\em
  {Radiative energy loss of high-energy quarks and gluons in a finite volume
  quark - gluon plasma}\/},
  \href{http://dx.doi.org/10.1016/S0550-3213(96)00553-6}{Nucl. Phys. {\bf B483}
  (1997)  291--320},
\href{http://arxiv.org/abs/hep-ph/9607355}{{\tt arXiv:hep-ph/9607355
  [hep-ph]}}.

\bibitem{Baier:1996sk}
R.~Baier, Y.~L. Dokshitzer, A.~H. Mueller, S.~Peigne, and D.~Schiff, {\em
  {Radiative energy loss and p(T) broadening of high-energy partons in
  nuclei}\/},  \href{http://dx.doi.org/10.1016/S0550-3213(96)00581-0}{Nucl.
  Phys. {\bf B484} (1997)  265--282},
\href{http://arxiv.org/abs/hep-ph/9608322}{{\tt arXiv:hep-ph/9608322
  [hep-ph]}}.

\bibitem{Zakharov:1996fv}
B.~Zakharov, {\em {Fully quantum treatment of the Landau-Pomeranchuk-Migdal
  effect in QED and QCD}\/},  \href{http://dx.doi.org/10.1134/1.567126}{JETP
  Lett. {\bf 63} (1996)  952--957},
\href{http://arxiv.org/abs/hep-ph/9607440}{{\tt arXiv:hep-ph/9607440
  [hep-ph]}}.

\bibitem{Zakharov:1997uu}
B.~Zakharov, {\em {Radiative energy loss of high-energy quarks in finite size
  nuclear matter and quark - gluon plasma}\/},
  \href{http://dx.doi.org/10.1134/1.567389}{JETP Lett. {\bf 65} (1997)
  615--620},
\href{http://arxiv.org/abs/hep-ph/9704255}{{\tt arXiv:hep-ph/9704255
  [hep-ph]}}.

\bibitem{Gyulassy:2000er}
M.~Gyulassy, P.~Levai, and I.~Vitev, {\em {Reaction operator approach to
  nonAbelian energy loss}\/},
  \href{http://dx.doi.org/10.1016/S0550-3213(00)00652-0}{Nucl. Phys. {\bf B594}
  (2001)  371--419},
\href{http://arxiv.org/abs/nucl-th/0006010}{{\tt arXiv:nucl-th/0006010
  [nucl-th]}}.

\bibitem{Wang:2001ifa}
X.-N. Wang and X.-F. Guo, {\em {Multiple parton scattering in nuclei: Parton
  energy loss}\/},
  \href{http://dx.doi.org/10.1016/S0375-9474(01)01130-7}{Nucl. Phys. {\bf A696}
  (2001)  788--832},
\href{http://arxiv.org/abs/hep-ph/0102230}{{\tt arXiv:hep-ph/0102230
  [hep-ph]}}.

\bibitem{Arnold:2002ja}
P.~B. Arnold, G.~D. Moore, and L.~G. Yaffe, {\em {Photon and gluon emission in
  relativistic plasmas}\/},
  \href{http://dx.doi.org/10.1088/1126-6708/2002/06/030}{JHEP {\bf 0206} (2002)
   030},
\href{http://arxiv.org/abs/hep-ph/0204343}{{\tt arXiv:hep-ph/0204343
  [hep-ph]}}.

\bibitem{Majumder:2009ge}
A.~Majumder, {\em {Hard collinear gluon radiation and multiple scattering in a
  medium}\/},  \href{http://dx.doi.org/10.1103/PhysRevD.85.014023}{Phys. Rev.
  {\bf D85} (2012)  014023},
\href{http://arxiv.org/abs/0912.2987}{{\tt arXiv:0912.2987 [nucl-th]}}.

\bibitem{Chesler:2008uy}
P.~M. Chesler, K.~Jensen, A.~Karch, and L.~G. Yaffe, {\em {Light quark energy
  loss in strongly-coupled N = 4 supersymmetric Yang-Mills plasma}\/},
  \href{http://dx.doi.org/10.1103/PhysRevD.79.125015}{Phys. Rev. {\bf D79}
  (2009)  125015},
\href{http://arxiv.org/abs/0810.1985}{{\tt arXiv:0810.1985 [hep-th]}}.

\bibitem{Chesler:2008wd}
P.~M. Chesler, K.~Jensen, and A.~Karch, {\em {Jets in strongly-coupled N = 4
  super Yang-Mills theory}\/},
  \href{http://dx.doi.org/10.1103/PhysRevD.79.025021}{Phys. Rev. {\bf D79}
  (2009)  025021},
\href{http://arxiv.org/abs/0804.3110}{{\tt arXiv:0804.3110 [hep-th]}}.

\bibitem{Friess:2006aw}
J.~J. Friess, S.~S. Gubser, and G.~Michalogiorgakis, {\em {Dissipation from a
  heavy quark moving through N=4 super-Yang-Mills plasma}\/},
  \href{http://dx.doi.org/10.1088/1126-6708/2006/09/072}{JHEP {\bf 0609} (2006)
   072},
\href{http://arxiv.org/abs/hep-th/0605292}{{\tt arXiv:hep-th/0605292
  [hep-th]}}.

\bibitem{CasalderreySolana:2006rq}
J.~Casalderrey-Solana and D.~Teaney, {\em {Heavy quark diffusion in strongly
  coupled N=4 Yang-Mills}\/},
  \href{http://dx.doi.org/10.1103/PhysRevD.74.085012}{Phys. Rev. {\bf D74}
  (2006)  085012},
\href{http://arxiv.org/abs/hep-ph/0605199}{{\tt arXiv:hep-ph/0605199
  [hep-ph]}}.

\bibitem{Baier:2002tc}
R.~Baier, {\em {Jet quenching}\/},
  \href{http://dx.doi.org/10.1016/S0375-9474(02)01429-X}{Nucl. Phys. {\bf A715}
  (2003)  209--218},
\href{http://arxiv.org/abs/hep-ph/0209038}{{\tt arXiv:hep-ph/0209038
  [hep-ph]}}.

\bibitem{Majumder:2008zg}
A.~Majumder, {\em {Elastic energy loss and longitudinal straggling of a hard
  jet}\/},  \href{http://dx.doi.org/10.1103/PhysRevC.80.031902}{Phys. Rev. {\bf
  C80} (2009)  031902},
\href{http://arxiv.org/abs/0810.4967}{{\tt arXiv:0810.4967 [nucl-th]}}.

\bibitem{Burke:2013yra}
{JET Collaboration}, K.~M. Burke et al., {\em {Extracting the jet transport
  coefficient from jet quenching in high-energy heavy-ion collisions}\/},
  \href{http://dx.doi.org/10.1103/PhysRevC.90.014909}{Phys. Rev. {\bf C90}
  (2014)  014909},
\href{http://arxiv.org/abs/1312.5003}{{\tt arXiv:1312.5003 [nucl-th]}}.

\bibitem{Bass:2008rv}
S.~A. Bass, C.~Gale, A.~Majumder, C.~Nonaka, G.-Y. Qin, et al., {\em
  {Systematic Comparison of Jet Energy-Loss Schemes in a realistic hydrodynamic
  medium}\/},  \href{http://dx.doi.org/10.1103/PhysRevC.79.024901}{Phys. Rev.
  {\bf C79} (2009)  024901},
\href{http://arxiv.org/abs/0808.0908}{{\tt arXiv:0808.0908 [nucl-th]}}.

\bibitem{CMS:2012aa}
{CMS Collaboration}, S.~Chatrchyan et al., {\em {Study of high-pT charged
  particle suppression in PbPb compared to $pp$ collisions at
  $\sqrt{s_{NN}}=2.76$ TeV}\/},
  \href{http://dx.doi.org/10.1140/epjc/s10052-012-1945-x}{Eur. Phys. J. {\bf
  C72} (2012)  1945},
\href{http://arxiv.org/abs/1202.2554}{{\tt arXiv:1202.2554 [nucl-ex]}}.

\bibitem{sPHENIX}
{\em sPHENIX Science Proposal}, 2014.
\newblock
  \url{http://www.phenix.bnl.gov/phenix/WWW/publish/documents/sPHENIX_proposal_19112014.pdf}.

\bibitem{Aad:2010bu}
{ATLAS Collaboration}, G.~Aad et al., {\em {Observation of a
  Centrality-Dependent Dijet Asymmetry in Lead-Lead Collisions at
  $\sqrt{s_{NN}}=2.77$ TeV with the ATLAS Detector at the LHC}\/},
  \href{http://dx.doi.org/10.1103/PhysRevLett.105.252303}{Phys. Rev. Lett. {\bf
  105} (2010)  252303},
\href{http://arxiv.org/abs/1011.6182}{{\tt arXiv:1011.6182 [hep-ex]}}.

\bibitem{Chatrchyan:2011sx}
{CMS Collaboration}, S.~Chatrchyan et al., {\em {Observation and studies of jet
  quenching in PbPb collisions at nucleon-nucleon center-of-mass energy = 2.76
  TeV}\/},  \href{http://dx.doi.org/10.1103/PhysRevC.84.024906}{Phys. Rev. C
  {\bf 84} (2011)  024906},
\href{http://arxiv.org/abs/1102.1957}{{\tt arXiv:1102.1957 [nucl-ex]}}.

\bibitem{Qin:2010mn}
G.-Y. Qin and B.~Muller, {\em {Explanation of Di-jet asymmetry in Pb+Pb
  collisions at the Large Hadron Collider}\/},
  \href{http://dx.doi.org/10.1103/PhysRevLett.108.189904,
  10.1103/PhysRevLett.106.162302}{Phys. Rev. Lett. {\bf 106} (2011)  162302},
\href{http://arxiv.org/abs/1012.5280}{{\tt arXiv:1012.5280 [hep-ph]}}.

\bibitem{Lokhtin:2011qq}
I.~Lokhtin, A.~Belyaev, and A.~Snigirev, {\em {Jet quenching pattern at LHC in
  PYQUEN model}\/},
  \href{http://dx.doi.org/10.1140/epjc/s10052-011-1650-1}{Eur. Phys. J. {\bf
  C71} (2011)  1650},
\href{http://arxiv.org/abs/1103.1853}{{\tt arXiv:1103.1853 [hep-ph]}}.

\bibitem{Young:2011qx}
C.~Young, B.~Schenke, S.~Jeon, and C.~Gale, {\em {Dijet asymmetry at the
  energies available at the CERN Large Hadron Collider}\/},
  \href{http://dx.doi.org/10.1103/PhysRevC.84.024907}{Phys. Rev. {\bf C84}
  (2011)  024907},
\href{http://arxiv.org/abs/1103.5769}{{\tt arXiv:1103.5769 [nucl-th]}}.

\bibitem{Renk:2012cb}
T.~Renk, {\em {Energy dependence of the dijet imbalance in Pb-Pb collisions at
  2.76 ATeV}\/},  \href{http://dx.doi.org/10.1103/PhysRevC.86.061901}{Phys.
  Rev. {\bf C86} (2012)  061901},
\href{http://arxiv.org/abs/1204.5572}{{\tt arXiv:1204.5572 [hep-ph]}}.

\bibitem{Dai:2012am}
W.~Dai, I.~Vitev, and B.-W. Zhang, {\em {Momentum imbalance of isolated
  photon-tagged jet production at RHIC and LHC}\/},
  \href{http://dx.doi.org/10.1103/PhysRevLett.110.142001}{Phys. Rev. Lett. {\bf
  110} (2013)  142001},
\href{http://arxiv.org/abs/1207.5177}{{\tt arXiv:1207.5177 [hep-ph]}}.

\bibitem{Qin:2012gp}
G.-Y. Qin, {\em {Parton shower evolution in medium and nuclear modification of
  photon-tagged jets in Pb+Pb collisions at the LHC}\/},
  \href{http://dx.doi.org/10.1140/epjc/s10052-014-2959-3}{Eur. Phys. J. {\bf
  C74} (2014)  2959},
\href{http://arxiv.org/abs/1210.6610}{{\tt arXiv:1210.6610 [hep-ph]}}.

\bibitem{Wang:2013cia}
X.-N. Wang and Y.~Zhu, {\em {Medium Modification of $\gamma$-jets in
  High-energy Heavy-ion Collisions}\/},
  \href{http://dx.doi.org/10.1103/PhysRevLett.111.062301}{Phys. Rev. Lett. {\bf
  111} (2013)  062301},
\href{http://arxiv.org/abs/1302.5874}{{\tt arXiv:1302.5874 [hep-ph]}}.

\bibitem{Chatrchyan:2012gt}
{CMS Collaboration}, S.~Chatrchyan et al., {\em {Studies of jet quenching using
  isolated-photon+jet correlations in PbPb and pp collisions at $\sqrt{s_{NN}}$
  = 2.76 TeV}\/},
  \href{http://dx.doi.org/10.1016/j.physletb.2012.11.003}{Phys. Lett. B {\bf
  718} (2013)  773},
\href{http://arxiv.org/abs/1205.0206}{{\tt arXiv:1205.0206 [nucl-ex]}}.

\bibitem{Steinberg:2014xha}
{ATLAS Collaboration}, P.~Steinberg, {\em {Centrality, rapidity and pT
  dependence of isolated prompt photon production in lead--lead collisions at
  sNN=2.76 TeV with the ATLAS detector at the LHC}\/},
\href{http://dx.doi.org/10.1016/j.nuclphysa.2014.10.038}{Nucl. Phys. {\bf A931}
  (2014)  422--427}.

\bibitem{Adare:2012qi}
{PHENIX Collaboration}, A.~Adare et al., {\em {Medium modification of jet
  fragmentation in Au + Au collisions at $\sqrt{s_{NN}}= 200$ GeV measured in
  direct photon-hadron correlations}\/},
  \href{http://dx.doi.org/10.1103/PhysRevLett.111.032301}{Phys. Rev. Lett. {\bf
  111} (2013)  032301},
\href{http://arxiv.org/abs/1212.3323}{{\tt arXiv:1212.3323 [nucl-ex]}}.

\bibitem{Adamczyk:2013jei}
{STAR Collaboration}, L.~Adamczyk et al., {\em {Jet-hadron correlations in
  $\sqrt{s_{NN}}$ = 200 GeV p+p and central Au+Au collisions}\/},
  \href{http://dx.doi.org/10.1103/PhysRevLett.112.122301}{Phys. Rev. Lett. {\bf
  112} (2014)  122301},
\href{http://arxiv.org/abs/1302.6184}{{\tt arXiv:1302.6184 [nucl-ex]}}.

\bibitem{Vitev:2008rz}
I.~Vitev, S.~Wicks, and B.-W. Zhang, {\em {A Theory of jet shapes and cross
  sections: From hadrons to nuclei}\/},
  \href{http://dx.doi.org/10.1088/1126-6708/2008/11/093}{JHEP {\bf 0811} (2008)
   093},
\href{http://arxiv.org/abs/0810.2807}{{\tt arXiv:0810.2807 [hep-ph]}}.

\bibitem{Vitev:2009rd}
I.~Vitev and B.-W. Zhang, {\em {Jet tomography of high-energy nucleus-nucleus
  collisions at next-to-leading order}\/},
  \href{http://dx.doi.org/10.1103/PhysRevLett.104.132001}{Phys. Rev. Lett. {\bf
  104} (2010)  132001},
\href{http://arxiv.org/abs/0910.1090}{{\tt arXiv:0910.1090 [hep-ph]}}.

\bibitem{Renk:2014lza}
T.~Renk, {\em {A study of the constraining power of high-$p_T$ observables in
  heavy-ion collisions}\/},
\href{http://arxiv.org/abs/1408.6684}{{\tt arXiv:1408.6684 [hep-ph]}}.

\bibitem{He:2011pd}
Y.~He, I.~Vitev, and B.-W. Zhang, {\em {${\cal O}(\alpha_s^3)$ Analysis of
  Inclusive Jet and di-Jet Production in Heavy Ion Reactions at the Large
  Hadron Collider}\/},
  \href{http://dx.doi.org/10.1016/j.physletb.2012.05.054}{Phys. Lett. {\bf
  B713} (2012)  224--232},
\href{http://arxiv.org/abs/1105.2566}{{\tt arXiv:1105.2566 [hep-ph]}}.

\bibitem{Renk:2012hz}
T.~Renk, {\em {Theoretical assessment of jet-hadron correlations}\/},
  \href{http://dx.doi.org/10.1103/PhysRevC.87.024905}{Phys. Rev. {\bf C87}
  (2013)  024905},
\href{http://arxiv.org/abs/1210.1330}{{\tt arXiv:1210.1330 [hep-ph]}}.

\bibitem{Aad:2012vca}
{ATLAS Collaboration}, G.~Aad et al., {\em {Measurement of the jet radius and
  transverse momentum dependence of inclusive jet suppression in lead-lead
  collisions at $\sqrt{s_{NN}}$= 2.76 TeV with the ATLAS detector}\/},
  \href{http://dx.doi.org/10.1016/j.physletb.2013.01.024}{Phys. Lett. {\bf
  B719} (2013)  220--241},
\href{http://arxiv.org/abs/1208.1967}{{\tt arXiv:1208.1967 [hep-ex]}}.

\bibitem{Chatrchyan:2012gw}
{CMS Collaboration}, S.~Chatrchyan et al., {\em {Measurement of jet
  fragmentation into charged particles in $pp$ and PbPb collisions at
  $\sqrt{s_{NN}}=2.76$ TeV}\/},
  \href{http://dx.doi.org/10.1007/JHEP10(2012)087}{JHEP {\bf 1210} (2012)
  087},
\href{http://arxiv.org/abs/1205.5872}{{\tt arXiv:1205.5872 [nucl-ex]}}.

\bibitem{ALICE:2012ab}
{ALICE Collaboration}, B.~Abelev et al., {\em {Suppression of high transverse
  momentum D mesons in central Pb-Pb collisions at $\sqrt{s_{NN}}=2.76$
  TeV}\/},  \href{http://dx.doi.org/10.1007/JHEP09(2012)112}{JHEP {\bf 1209}
  (2012)  112},
\href{http://arxiv.org/abs/1203.2160}{{\tt arXiv:1203.2160 [nucl-ex]}}.

\bibitem{Chatrchyan:2012np}
{CMS Collaboration}, S.~Chatrchyan et al., {\em {Suppression of non-prompt
  $J/\psi$, prompt $J/\psi$, and Y(1S) in PbPb collisions at
  $\sqrt{s_{NN}}=2.76$ TeV}\/},
  \href{http://dx.doi.org/10.1007/JHEP05(2012)063}{JHEP {\bf 1205} (2012)
  063},
\href{http://arxiv.org/abs/1201.5069}{{\tt arXiv:1201.5069 [nucl-ex]}}.

\bibitem{Adare:2006nq}
{PHENIX Collaboration}, A.~Adare et al., {\em {Energy Loss and Flow of Heavy
  Quarks in Au+Au Collisions at s(NN)**(1/2) = 200-GeV}\/},
  \href{http://dx.doi.org/10.1103/PhysRevLett.98.172301}{Phys. Rev. Lett. {\bf
  98} (2007)  172301},
\href{http://arxiv.org/abs/nucl-ex/0611018}{{\tt arXiv:nucl-ex/0611018
  [nucl-ex]}}.

\bibitem{Adamczyk:2014uip}
{STAR Collaboration}, L.~Adamczyk et al., {\em {Observation of $D^0$ Meson
  Nuclear Modifications in Au+Au Collisions at
  $\sqrt{s_{NN}}=200$  GeV}\/},
  \href{http://dx.doi.org/10.1103/PhysRevLett.113.142301}{Phys. Rev. Lett. {\bf
  113} (2014)  142301},
\href{http://arxiv.org/abs/1404.6185}{{\tt arXiv:1404.6185 [nucl-ex]}}.

\bibitem{Adare:2009ic}
{PHENIX Collaboration}, A.~Adare et al., {\em {Measurement of Bottom versus
  Charm as a Function of Transverse Momentum with Electron-Hadron Correlations
  in $p^+ p$ Collisions at $\sqrt{s}=200$ GeV}\/},
  \href{http://dx.doi.org/10.1103/PhysRevLett.103.082002}{Phys. Rev. Lett. {\bf
  103} (2009)  082002},
\href{http://arxiv.org/abs/0903.4851}{{\tt arXiv:0903.4851 [hep-ex]}}.

\bibitem{Aggarwal:2010xp}
{STAR Collaboration}, M.~Aggarwal et al., {\em {Measurement of the Bottom
  contribution to non-photonic electron production in $p+p$ collisions at
  $\sqrt{s} $=200 GeV}\/},
  \href{http://dx.doi.org/10.1103/PhysRevLett.105.202301}{Phys. Rev. Lett. {\bf
  105} (2010)  202301},
\href{http://arxiv.org/abs/1007.1200}{{\tt arXiv:1007.1200 [nucl-ex]}}.

\bibitem{Brambilla:2010cs}
N.~Brambilla, S.~Eidelman, B.~K. Heltsley, R.~Vogt, G.~T. Bodwin, et al., {\em
  {Heavy quarkonium: progress, puzzles, and opportunities}\/},
  \href{http://arxiv.org/abs/1010.5827}{{\tt arXiv:1010.5827 [hep-ph]}}.

\bibitem{Brambilla:2010vq}
N.~Brambilla, M.~A. Escobedo, J.~Ghiglieri, J.~Soto, and A.~Vairo, {\em {Heavy
  Quarkonium in a weakly-coupled quark-gluon plasma below the melting
  temperature}\/},  \href{http://dx.doi.org/10.1007/JHEP09(2010)038}{JHEP {\bf
  1009} (2010)  038},
\href{http://arxiv.org/abs/1007.4156}{{\tt arXiv:1007.4156 [hep-ph]}}.

\bibitem{Karsch:1987pv}
F.~Karsch, M.~Mehr, and H.~Satz, {\em {Color Screening and Deconfinement for
  Bound States of Heavy Quarks}\/},
\href{http://dx.doi.org/10.1007/BF01549722}{Z. Phys. {\bf C37} (1988)  617}.

\bibitem{Strickland:2011mw}
M.~Strickland, {\em {Thermal $\Upsilon_{1s}$ and $\chi_{b1}$ suppression in
  $\sqrt{s_{NN}}=2.76$ TeV Pb-Pb collisions at the LHC}\/},
  \href{http://dx.doi.org/10.1103/PhysRevLett.107.132301}{Phys. Rev. Lett. {\bf
  107} (2011)  132301},
\href{http://arxiv.org/abs/1106.2571}{{\tt arXiv:1106.2571 [hep-ph]}}.

\bibitem{Strickland:2011aa}
M.~Strickland and D.~Bazow, {\em {Thermal bottomonium suppression at RHIC and
  LHC}\/},  \href{http://dx.doi.org/10.1016/j.nuclphysa.2012.02.003}{Nucl.
  Phys. {\bf A879} (2012)  25--58},
\href{http://arxiv.org/abs/1112.2761}{{\tt arXiv:1112.2761 [nucl-th]}}.

\bibitem{Grandchamp:2005yw}
L.~Grandchamp, S.~Lumpkins, D.~Sun, H.~van Hees, and R.~Rapp, {\em {Bottomonium
  production at RHIC and CERN LHC}\/},
  \href{http://dx.doi.org/10.1103/PhysRevC.73.064906}{Phys. Rev. {\bf C73}
  (2006)  064906},
\href{http://arxiv.org/abs/hep-ph/0507314}{{\tt arXiv:hep-ph/0507314
  [hep-ph]}}.

\bibitem{Emerick:2011xu}
A.~Emerick, X.~Zhao, and R.~Rapp, {\em {Bottomonia in the quark-gluon plasma
  and their production at RHIC and LHC}\/},
  \href{http://dx.doi.org/10.1140/epja/i2012-12072-y}{Eur. Phys. J. {\bf A48}
  (2012)  72},
\href{http://arxiv.org/abs/1111.6537}{{\tt arXiv:1111.6537 [hep-ph]}}.

\bibitem{Chatrchyan:2011pe}
{CMS Collaboration}, S.~Chatrchyan et al., {\em {Indications of suppression of
  excited $\Upsilon$ states in PbPb collisions at $\sqrt{S_{NN}}$ = 2.76
  TeV}\/},  \href{http://dx.doi.org/10.1103/PhysRevLett.107.052302}{Phys. Rev.
  Lett. {\bf 107} (2011)  052302},
\href{http://arxiv.org/abs/1105.4894}{{\tt arXiv:1105.4894 [nucl-ex]}}.

\bibitem{Adare:2014hje}
{PHENIX Collaboration}, A.~Adare et al., {\em {Measurement of
  $\Upsilon$(1S+2S+3S) production in $p$$+$$p$ and Au$+$Au collisions at
  $\sqrt{s_{_{NN}}}=200$ GeV}\/},
\href{http://arxiv.org/abs/1404.2246}{{\tt arXiv:1404.2246 [nucl-ex]}}.

\bibitem{Adamczyk:2013poh}
{STAR Collaboration}, L.~Adamczyk et al., {\em {Suppression of Upsilon
  Production in d+Au and Au+Au Collisions at $\sqrt{s_{NN}}$ = 200 GeV}\/},
  \href{http://dx.doi.org/10.1016/j.physletb.2014.06.028}{Phys. Lett. {\bf
  B735} (2014)  127},
\href{http://arxiv.org/abs/1312.3675}{{\tt arXiv:1312.3675 [nucl-ex]}}.

\bibitem{Aschenauer:2015eha}
E.-C. Aschenauer, A.~Bazilevsky, M.~Diehl, J.~Drachenberg, K.~O. Eyser, et al.,
  {\em {The RHIC SPIN Program: Achievements and Future Opportunities}\/},
\href{http://arxiv.org/abs/1501.01220}{{\tt arXiv:1501.01220 [nucl-ex]}}.

\bibitem{Gelis:2010nm}
F.~Gelis, E.~Iancu, J.~Jalilian-Marian, and R.~Venugopalan, {\em {The Color
  Glass Condensate}\/},
  \href{http://dx.doi.org/10.1146/annurev.nucl.010909.083629}{Annu. Rev. Nucl.
  Part. Sci. {\bf 60} (2010)  463--489},
\href{http://arxiv.org/abs/1002.0333}{{\tt arXiv:1002.0333 [hep-ph]}}.

\bibitem{Dusling:2013oia}
K.~Dusling and R.~Venugopalan, {\em {Comparison of the color glass condensate
  to dihadron correlations in proton-proton and proton-nucleus collisions}\/},
  \href{http://dx.doi.org/10.1103/PhysRevD.87.094034}{Phys. Rev. {\bf D87}
  (2013)  094034},
\href{http://arxiv.org/abs/1302.7018}{{\tt arXiv:1302.7018 [hep-ph]}}.

\bibitem{starloi}
{STAR Collaboration}, {\em A Polarized p+p and p+A Progrom for the Next Years},
  2014.
\newblock
  \url{https://drupal.star.bnl.gov/STAR/files/pp.pA_.LoI_.pp_.pA_.v7.pdf}.

\bibitem{PHENIXpppA}
{PHENIX Collaboration}, {\em Future Opportunities in p+p and p+A Collisions at
  RHIC with the Forward sPHENIX Detector\/}, .
  \url{http://www.phenix.bnl.gov/phenix/WWW/publish/dave/sPHENIX/pp_pA_whitepaper.pdf}.

\bibitem{Dominguez:2011wm}
F.~Dominguez, C.~Marquet, B.-W. Xiao, and F.~Yuan, {\em {Universality of
  Unintegrated Gluon Distributions at small x}\/},
  \href{http://dx.doi.org/10.1103/PhysRevD.83.105005}{Phys. Rev. {\bf D83}
  (2011)  105005},
\href{http://arxiv.org/abs/1101.0715}{{\tt arXiv:1101.0715 [hep-ph]}}.

\bibitem{Kang:2011ni}
Z.-B. Kang and F.~Yuan, {\em {Single Spin Asymmetry Scaling in the Forward
  Rapidity Region at RHIC}\/},
  \href{http://dx.doi.org/10.1103/PhysRevD.84.034019}{Phys. Rev. {\bf D84}
  (2011)  034019},
\href{http://arxiv.org/abs/1106.1375}{{\tt arXiv:1106.1375 [hep-ph]}}.

\bibitem{Kovchegov:2013cva}
Y.~V. Kovchegov and M.~D. Sievert, {\em {Sivers Function in the Quasi-Classical
  Approximation}\/},  \href{http://dx.doi.org/10.1103/PhysRevD.89.054035}{Phys.
  Rev. {\bf D89} (2014)  054035},
\href{http://arxiv.org/abs/1310.5028}{{\tt arXiv:1310.5028 [hep-ph]}}.

\bibitem{Schenke:2012fw}
B.~Schenke, P.~Tribedy, and R.~Venugopalan, {\em {Event-by-event gluon
  multiplicity, energy density, and eccentricities in ultrarelativistic
  heavy-ion collisions}\/},
  \href{http://dx.doi.org/10.1103/PhysRevC.86.034908}{Phys.~Rev. {\bf C86}
  (2012)  034908},
\href{http://arxiv.org/abs/1206.6805}{{\tt arXiv:1206.6805 [hep-ph]}}.

\bibitem{Paatelainen:2013eea}
R.~Paatelainen, K.~Eskola, H.~Niemi, and K.~Tuominen, {\em {Fluid dynamics with
  saturated minijet initial conditions in ultrarelativistic heavy-ion
  collisions}\/},
  \href{http://dx.doi.org/10.1016/j.physletb.2014.02.018}{Phys.~Lett. {\bf
  B731} (2014)  126--130},
\href{http://arxiv.org/abs/1310.3105}{{\tt arXiv:1310.3105 [hep-ph]}}.

\bibitem{Aoki:2006br}
Y.~Aoki, Z.~Fodor, S.~Katz, and K.~Szabo, {\em {The QCD transition temperature:
  Results with physical masses in the continuum limit}\/},
  \href{http://dx.doi.org/10.1016/j.physletb.2006.10.021}{Phys. Lett. {\bf
  B643} (2006)  46--54},
\href{http://arxiv.org/abs/hep-lat/0609068}{{\tt arXiv:hep-lat/0609068
  [hep-lat]}}.

\bibitem{Kang:2013raa}
Z.-B. Kang, E.~Wang, X.-N. Wang, and H.~Xing, {\em {Next-to-Leading QCD
  Factorization for Semi-Inclusive Deep Inelastic Scattering at Twist-4}\/},
  \href{http://dx.doi.org/10.1103/PhysRevLett.112.102001}{Phys. Rev. Lett. {\bf
  112} (2014)  102001},
\href{http://arxiv.org/abs/1310.6759}{{\tt arXiv:1310.6759 [hep-ph]}}.

\bibitem{Liou:2013qya}
T.~Liou, A.~Mueller, and B.~Wu, {\em {Radiative $p_\bot$-broadening of
  high-energy quarks and gluons in QCD matter}\/},
  \href{http://dx.doi.org/10.1016/j.nuclphysa.2013.08.005}{Nucl. Phys. {\bf
  A916} (2013)  102--125},
\href{http://arxiv.org/abs/1304.7677}{{\tt arXiv:1304.7677 [hep-ph]}}.

\bibitem{Blaizot:2014bha}
J.-P. Blaizot and Y.~Mehtar-Tani, {\em {Renormalization of the jet-quenching
  parameter}\/},
  \href{http://dx.doi.org/10.1016/j.nuclphysa.2014.05.018}{Nucl. Phys. {\bf
  A929} (2014)  202--229},
\href{http://arxiv.org/abs/1403.2323}{{\tt arXiv:1403.2323 [hep-ph]}}.

\bibitem{Blaizot:2012fh}
J.-P. Blaizot, F.~Dominguez, E.~Iancu, and Y.~Mehtar-Tani, {\em {Medium-induced
  gluon branching}\/},  \href{http://dx.doi.org/10.1007/JHEP01(2013)143}{JHEP
  {\bf 01} (2013)  143},
\href{http://arxiv.org/abs/1209.4585}{{\tt arXiv:1209.4585 [hep-ph]}}.

\bibitem{Blaizot:2013hx}
J.-P. Blaizot, E.~Iancu, and Y.~Mehtar-Tani, {\em {Medium-induced QCD cascade:
  democratic branching and wave turbulence}\/},
  \href{http://dx.doi.org/10.1103/PhysRevLett.111.052001}{Phys. Rev. Lett. {\bf
  111} (2013)  052001},
\href{http://arxiv.org/abs/1301.6102}{{\tt arXiv:1301.6102 [hep-ph]}}.

\bibitem{Blaizot:2014rla}
J.-P. Blaizot, L.~Fister, and Y.~Mehtar-Tani, {\em {Angular distribution of
  medium-induced QCD cascades}\/},
\href{http://arxiv.org/abs/1409.6202}{{\tt arXiv:1409.6202 [hep-ph]}}.

\bibitem{Fister:2014zxa}
L.~Fister and E.~Iancu, {\em {Medium-induced jet evolution: wave turbulence and
  energy loss}\/},
\href{http://arxiv.org/abs/1409.2010}{{\tt arXiv:1409.2010 [hep-ph]}}.

\bibitem{Almeida:2014uva}
L.~G. Almeida, S.~D. Ellis, C.~Lee, G.~Sterman, I.~Sung, et al., {\em
  {Comparing and counting logs in direct and effective methods of QCD
  resummation}\/},  \href{http://dx.doi.org/10.1007/JHEP04(2014)174}{JHEP {\bf
  1404} (2014)  174},
\href{http://arxiv.org/abs/1401.4460}{{\tt arXiv:1401.4460 [hep-ph]}}.

\bibitem{Idilbi:2008vm}
A.~Idilbi and A.~Majumder, {\em {Extending soft-collinear-effective-theory to
  describe hard jets in dense QCD media}\/},
  \href{http://dx.doi.org/10.1103/PhysRevD.80.054022}{Phys. Rev. D {\bf 80}
  (2009)  054022},
\href{http://arxiv.org/abs/0808.1087}{{\tt arXiv:0808.1087 [hep-ph]}}.

\bibitem{Ovanesyan:2011xy}
G.~Ovanesyan and I.~Vitev, {\em {An effective theory for jet propagation in
  dense QCD matter: jet broadening and medium-induced bremsstrahlung}\/},
  \href{http://dx.doi.org/10.1007/JHEP06(2011)080}{JHEP {\bf 1106} (2011)
  080},
\href{http://arxiv.org/abs/1103.1074}{{\tt arXiv:1103.1074 [hep-ph]}}.

\bibitem{DEramo:2010ak}
F.~D'Eramo, H.~Liu, and K.~Rajagopal, {\em {Transverse Momentum Broadening and
  the Jet Quenching Parameter, Redux}\/},
  \href{http://dx.doi.org/10.1103/PhysRevD.84.065015}{Phys. Rev. {\bf D84}
  (2011)  065015},
\href{http://arxiv.org/abs/1006.1367}{{\tt arXiv:1006.1367 [hep-ph]}}.

\bibitem{Majumder:2012sh}
A.~Majumder, {\em {Calculating the jet quenching parameter q̂ in lattice gauge
  theory}\/},  \href{http://dx.doi.org/10.1103/PhysRevC.87.034905}{Phys. Rev.
  {\bf C87} (2013)  034905},
\href{http://arxiv.org/abs/1202.5295}{{\tt arXiv:1202.5295 [nucl-th]}}.

\bibitem{Panero:2013pla}
M.~Panero, K.~Rummukainen, and A.~Sch{\"a}fer, {\em {A lattice study of the jet
  quenching parameter}\/},
  \href{http://dx.doi.org/10.1103/PhysRevLett.112.162001}{Phys. Rev. Lett. {\bf
  112} (2014)  162001},
\href{http://arxiv.org/abs/1307.5850}{{\tt arXiv:1307.5850 [hep-ph]}}.

\bibitem{Huang:2013vaa}
J.~Huang, Z.-B. Kang, and I.~Vitev, {\em {Inclusive b-jet production in heavy
  ion collisions at the LHC}\/},
  \href{http://dx.doi.org/10.1016/j.physletb.2013.08.009}{Phys. Lett. {\bf B
  726} (2013)  251--256},
\href{http://arxiv.org/abs/1306.0909}{{\tt arXiv:1306.0909 [hep-ph]}}.

\bibitem{Moore:2004tg}
G.~D. Moore and D.~Teaney, {\em {How much do heavy quarks thermalize in a heavy
  ion collision?}\/},
  \href{http://dx.doi.org/10.1103/PhysRevC.71.064904}{Phys. Rev. {\bf C71}
  (2005)  064904},
\href{http://arxiv.org/abs/hep-ph/0412346}{{\tt arXiv:hep-ph/0412346
  [hep-ph]}}.

\bibitem{vanHees:2004gq}
H.~van Hees and R.~Rapp, {\em {Thermalization of heavy quarks in the
  quark-gluon plasma}\/},
  \href{http://dx.doi.org/10.1103/PhysRevC.71.034907}{Phys. Rev. {\bf C71}
  (2005)  034907},
\href{http://arxiv.org/abs/nucl-th/0412015}{{\tt arXiv:nucl-th/0412015
  [nucl-th]}}.

\bibitem{Adil:2006ra}
A.~Adil and I.~Vitev, {\em {Collisional dissociation of heavy mesons in dense
  QCD matter}\/},
  \href{http://dx.doi.org/10.1016/j.physletb.2007.03.050}{Phys. Lett. {\bf B
  649} (2007)  139--146},
\href{http://arxiv.org/abs/hep-ph/0611109}{{\tt arXiv:hep-ph/0611109
  [hep-ph]}}.

\bibitem{Sharma:2009hn}
R.~Sharma, I.~Vitev, and B.-W. Zhang, {\em {Light-cone wave function approach
  to open heavy flavor dynamics in QCD matter}\/},
  \href{http://dx.doi.org/10.1103/PhysRevC.80.054902}{Phys. Rev. {\bf C80}
  (2009)  054902},
\href{http://arxiv.org/abs/0904.0032}{{\tt arXiv:0904.0032 [hep-ph]}}.

\bibitem{Kang:2011rt}
Z.-B. Kang and I.~Vitev, {\em {Photon-tagged heavy meson production in high
  energy nuclear collisions}\/},
  \href{http://dx.doi.org/10.1103/PhysRevD.84.014034}{Phys. Rev. {\bf D84}
  (2011)  014034},
\href{http://arxiv.org/abs/1106.1493}{{\tt arXiv:1106.1493 [hep-ph]}}.

\bibitem{Djordjevic:2013pba}
M.~Djordjevic, {\em {Heavy flavor puzzle at LHC: a serendipitous interplay of
  jet suppression and fragmentation}\/},
  \href{http://dx.doi.org/10.1103/PhysRevLett.112.042302}{Phys. Rev. Lett. {\bf
  112} (2014)  042302},
\href{http://arxiv.org/abs/1307.4702}{{\tt arXiv:1307.4702 [nucl-th]}}.

\bibitem{Kang:2013hta}
Z.-B. Kang, Y.-Q. Ma, and R.~Venugopalan, {\em {Quarkonium production in high
  energy proton-nucleus collisions: CGC meets NRQCD}\/},
  \href{http://dx.doi.org/10.1007/JHEP01(2014)056}{JHEP {\bf 1401} (2014)
  056},
\href{http://arxiv.org/abs/1309.7337}{{\tt arXiv:1309.7337 [hep-ph]}}.

\bibitem{Ma:2014mri}
Y.-Q. Ma and R.~Venugopalan, {\em {Comprehensive Description of J/ψ Production
  in Proton-Proton Collisions at Collider Energies}\/},
  \href{http://dx.doi.org/10.1103/PhysRevLett.113.192301}{Phys. Rev. Lett. {\bf
  113} (2014)  192301},
\href{http://arxiv.org/abs/1408.4075}{{\tt arXiv:1408.4075 [hep-ph]}}.

\bibitem{Liu:2009nb}
Y.-p. Liu, Z.~Qu, N.~Xu, and P.-f. Zhuang, {\em {J/psi Transverse Momentum
  Distribution in High Energy Nuclear Collisions at RHIC}\/},
  \href{http://dx.doi.org/10.1016/j.physletb.2009.06.006}{Phys. Lett. {\bf
  B678} (2009)  72--76},
\href{http://arxiv.org/abs/0901.2757}{{\tt arXiv:0901.2757 [nucl-th]}}.

\bibitem{Sharma:2012dy}
R.~Sharma and I.~Vitev, {\em {High transverse momentum quarkonium production
  and dissociation in heavy ion collisions}\/},
  \href{http://dx.doi.org/10.1103/PhysRevC.87.044905}{Phys. Rev. {\bf C87}
  (2013)  044905},
\href{http://arxiv.org/abs/1203.0329}{{\tt arXiv:1203.0329 [hep-ph]}}.

\bibitem{Kang:2011zza}
Z.-B. Kang, J.-W. Qiu, and G.~Sterman, {\em {Factorization and quarkonium
  production}\/},
\href{http://dx.doi.org/10.1016/j.nuclphysbps.2011.03.054}{Nucl. Phys. Proc.
  Suppl. {\bf 214} (2011)  39--43}.

\bibitem{Kang:2011mg}
Z.-B. Kang, J.-W. Qiu, and G.~Sterman, {\em {Heavy quarkonium production and
  polarization}\/},
  \href{http://dx.doi.org/10.1103/PhysRevLett.108.102002}{Phys. Rev. Lett. {\bf
  108} (2012)  102002},
\href{http://arxiv.org/abs/1109.1520}{{\tt arXiv:1109.1520 [hep-ph]}}.

\bibitem{Kang:2014tta}
Z.-B. Kang, Y.-Q. Ma, J.-W. Qiu, and G.~Sterman, {\em {Heavy Quarkonium
  Production at Collider Energies: Factorization and Evolution}\/},
  \href{http://dx.doi.org/10.1103/PhysRevD.90.034006}{Phys. Rev. {\bf D90}
  (2014)  034006},
\href{http://arxiv.org/abs/1401.0923}{{\tt arXiv:1401.0923 [hep-ph]}}.

\bibitem{Kang:2014pya}
Z.-B. Kang, Y.-Q. Ma, J.-W. Qiu, and G.~Sterman, {\em {Heavy Quarkonium
  Production at Collider Energies: Partonic Cross Section and Polarization}\/},
\href{http://arxiv.org/abs/1411.2456}{{\tt arXiv:1411.2456 [hep-ph]}}.

\bibitem{Fleming:2012wy}
S.~Fleming, A.~K. Leibovich, T.~Mehen, and I.~Z. Rothstein, {\em {The
  Systematics of Quarkonium Production at the LHC and Double Parton
  Fragmentation}\/},  \href{http://dx.doi.org/10.1103/PhysRevD.86.094012}{Phys.
  Rev. {\bf D86} (2012)  094012},
\href{http://arxiv.org/abs/1207.2578}{{\tt arXiv:1207.2578 [hep-ph]}}.

\bibitem{Fleming:2013qu}
S.~Fleming, A.~K. Leibovich, T.~Mehen, and I.~Z. Rothstein, {\em {Anomalous
  dimensions of the double parton fragmentation functions}\/},
  \href{http://dx.doi.org/10.1103/PhysRevD.87.074022}{Phys. Rev. {\bf D87}
  (2013)  074022},
\href{http://arxiv.org/abs/1301.3822}{{\tt arXiv:1301.3822 [hep-ph]}}.

\bibitem{Laine:2006ns}
M.~Laine, O.~Philipsen, P.~Romatschke, and M.~Tassler, {\em {Real-time static
  potential in hot QCD}\/},
  \href{http://dx.doi.org/10.1088/1126-6708/2007/03/054}{JHEP {\bf 0703} (2007)
   054},
\href{http://arxiv.org/abs/hep-ph/0611300}{{\tt arXiv:hep-ph/0611300
  [hep-ph]}}.

\bibitem{Brambilla:2008cx}
N.~Brambilla, J.~Ghiglieri, A.~Vairo, and P.~Petreczky, {\em {Static
  quark-antiquark pairs at finite temperature}\/},
  \href{http://dx.doi.org/10.1103/PhysRevD.78.014017}{Phys. Rev. {\bf D78}
  (2008)  014017},
\href{http://arxiv.org/abs/0804.0993}{{\tt arXiv:0804.0993 [hep-ph]}}.

\bibitem{Riek:2010py}
F.~Riek and R.~Rapp, {\em {Selfconsistent evaluation of charm and charmonium in
  the quark-gluon plasma}\/},
  \href{http://dx.doi.org/10.1088/1367-2630/13/4/045007}{New J. Phys. {\bf 13}
  (2011)  045007},
\href{http://arxiv.org/abs/1012.0019}{{\tt arXiv:1012.0019 [nucl-th]}}.

\bibitem{Petreczky:2010tk}
P.~Petreczky, C.~Miao, and A.~Mocsy, {\em {Quarkonium spectral functions with
  complex potential}\/},
  \href{http://dx.doi.org/10.1016/j.nuclphysa.2011.02.028}{Nucl. Phys. {\bf
  A855} (2011)  125--132},
\href{http://arxiv.org/abs/1012.4433}{{\tt arXiv:1012.4433 [hep-ph]}}.

\bibitem{Brambilla:2011sg}
N.~Brambilla, M.~A. Escobedo, J.~Ghiglieri, and A.~Vairo, {\em {Thermal width
  and gluo-dissociation of quarkonium in pNRQCD}\/},
  \href{http://dx.doi.org/10.1007/JHEP12(2011)116}{JHEP {\bf 1112} (2011)
  116},
\href{http://arxiv.org/abs/1109.5826}{{\tt arXiv:1109.5826 [hep-ph]}}.

\bibitem{Ding:2012sp}
H.~Ding, A.~Francis, O.~Kaczmarek, F.~Karsch, H.~Satz, et al., {\em {Charmonium
  properties in hot quenched lattice QCD}\/},
  \href{http://dx.doi.org/10.1103/PhysRevD.86.014509}{Phys. Rev. {\bf D86}
  (2012)  014509},
\href{http://arxiv.org/abs/1204.4945}{{\tt arXiv:1204.4945 [hep-lat]}}.

\bibitem{Brambilla:2013dpa}
N.~Brambilla, M.~A. Escobedo, J.~Ghiglieri, and A.~Vairo, {\em {Thermal width
  and quarkonium dissociation by inelastic parton scattering}\/},
  \href{http://dx.doi.org/10.1007/JHEP05(2013)130}{JHEP {\bf 1305} (2013)
  130},
\href{http://arxiv.org/abs/1303.6097}{{\tt arXiv:1303.6097 [hep-ph]}}.

\bibitem{Rothkopf:2013kya}
A.~Rothkopf, {\em {A first look at Bottomonium melting via a stochastic
  potential}\/},  \href{http://dx.doi.org/10.1007/JHEP04(2014)085}{JHEP {\bf
  1404} (2014)  085},
\href{http://arxiv.org/abs/1312.3246}{{\tt arXiv:1312.3246 [hep-ph]}}.

\bibitem{Kim:2014iga}
S.~Kim, P.~Petreczky, and A.~Rothkopf, {\em {Lattice NRQCD study of S- and
  P-wave bottomonium states in a thermal medium with $N_f=2+1$ light
  flavors}\/},
\href{http://arxiv.org/abs/1409.3630}{{\tt arXiv:1409.3630 [hep-lat]}}.

\bibitem{Francis:2013cva}
A.~Francis, O.~Kaczmarek, M.~Laine, M.~M{\"u}ller, T.~Neuhaus, et al., {\em
  {Towards the continuum limit in transport coefficient computations}\/},  PoS
  {\bf LATTICE2013} (2014)  453,
\href{http://arxiv.org/abs/1311.3759}{{\tt arXiv:1311.3759 [hep-lat]}}.

\bibitem{Riek:2010fk}
F.~Riek and R.~Rapp, {\em {Quarkonia and Heavy-Quark Relaxation Times in the
  Quark-Gluon Plasma}\/},
  \href{http://dx.doi.org/10.1103/PhysRevC.82.035201}{Phys. Rev. {\bf C82}
  (2010)  035201},
\href{http://arxiv.org/abs/1005.0769}{{\tt arXiv:1005.0769 [hep-ph]}}.

\bibitem{Bazavov:2014cta}
A.~Bazavov, F.~Karsch, Y.~Maezawa, S.~Mukherjee, and P.~Petreczky, {\em
  {In-medium modifications of open and hidden strange-charm mesons from spatial
  correlation functions}\/},
\href{http://arxiv.org/abs/1411.3018}{{\tt arXiv:1411.3018 [hep-lat]}}.

\bibitem{Zhao:2010nk}
X.~Zhao and R.~Rapp, {\em {Charmonium in Medium: From Correlators to
  Experiment}\/},  \href{http://dx.doi.org/10.1103/PhysRevC.82.064905}{Phys.
  Rev. {\bf C82} (2010)  064905},
\href{http://arxiv.org/abs/1008.5328}{{\tt arXiv:1008.5328 [hep-ph]}}.

\bibitem{Young:2011ug}
C.~Young, B.~Schenke, S.~Jeon, and C.~Gale, {\em {MARTINI event generator for
  heavy quarks: Initialization, parton evolution, and hadronization}\/},
  \href{http://dx.doi.org/10.1103/PhysRevC.86.034905}{Phys. Rev. {\bf C86}
  (2012)  034905},
\href{http://arxiv.org/abs/1111.0647}{{\tt arXiv:1111.0647 [nucl-th]}}.

\bibitem{He:2011qa}
M.~He, R.~J. Fries, and R.~Rapp, {\em {Heavy-Quark Diffusion and Hadronization
  in Quark-Gluon Plasma}\/},
  \href{http://dx.doi.org/10.1103/PhysRevC.86.014903}{Phys. Rev. {\bf C86}
  (2012)  014903},
\href{http://arxiv.org/abs/1106.6006}{{\tt arXiv:1106.6006 [nucl-th]}}.

\bibitem{Cao:2013ita}
S.~Cao, G.-Y. Qin, and S.~A. Bass, {\em {Heavy-quark dynamics and hadronization
  in ultrarelativistic heavy-ion collisions: Collisional versus radiative
  energy loss}\/},  \href{http://dx.doi.org/10.1103/PhysRevC.88.044907}{Phys.
  Rev. {\bf C88} (2013)  044907},
\href{http://arxiv.org/abs/1308.0617}{{\tt arXiv:1308.0617 [nucl-th]}}.

\bibitem{Aarts:2009uq}
G.~Aarts, E.~Seiler, and I.-O. Stamatescu, {\em {The Complex Langevin method:
  When can it be trusted?}\/},
  \href{http://dx.doi.org/10.1103/PhysRevD.81.054508}{Phys. Rev. {\bf D81}
  (2010)  054508},
\href{http://arxiv.org/abs/0912.3360}{{\tt arXiv:0912.3360 [hep-lat]}}.

\bibitem{Aarts:2014kja}
G.~Aarts, F.~Attanasio, B.~J{\"a}ger, E.~Seiler, D.~Sexty, et al., {\em
  {Exploring the phase diagram of QCD with complex Langevin simulations}\/},
  PoS {\bf LATTICE2014} (2014)  200,
\href{http://arxiv.org/abs/1411.2632}{{\tt arXiv:1411.2632 [hep-lat]}}.

\bibitem{Cristoforetti:2012su}
{AuroraScience Collaboration}, M.~Cristoforetti, F.~Di~Renzo, and L.~Scorzato,
  {\em {New approach to the sign problem in quantum field theories: High
  density QCD on a Lefschetz thimble}\/},
  \href{http://dx.doi.org/10.1103/PhysRevD.86.074506}{Phys. Rev. {\bf D86}
  (2012)  074506},
\href{http://arxiv.org/abs/1205.3996}{{\tt arXiv:1205.3996 [hep-lat]}}.

\bibitem{Aarts:2014nxa}
G.~Aarts, L.~Bongiovanni, E.~Seiler, and D.~Sexty, {\em {Some remarks on
  Lefschetz thimbles and complex Langevin dynamics}\/},
  \href{http://dx.doi.org/10.1007/JHEP10(2014)159}{JHEP {\bf 1410} (2014)
  159},
\href{http://arxiv.org/abs/1407.2090}{{\tt arXiv:1407.2090 [hep-lat]}}.

\bibitem{HotQCD_WP}
Y.~Akiba, A.~Angerami, H.~Caines, A.~Frawley, U.~Heinz, et al., {\em {The Hot
  QCD White Paper: Exploring the Phases of QCD at RHIC and the LHC}\/},
\href{http://arxiv.org/abs/1502.02730}{{\tt arXiv:1502.02730 [nucl-ex]}}.

\end{thebibliography}\endgroup

\endgroup
\clearpage

\section[Appendix: Town Meeting Agenda]{Appendix: Agenda of the ``Phases of QCD Matter'' and joint sessions at the QCD Town Meeting}

\begin{figure}[!h]
\includegraphics[width=\textwidth]{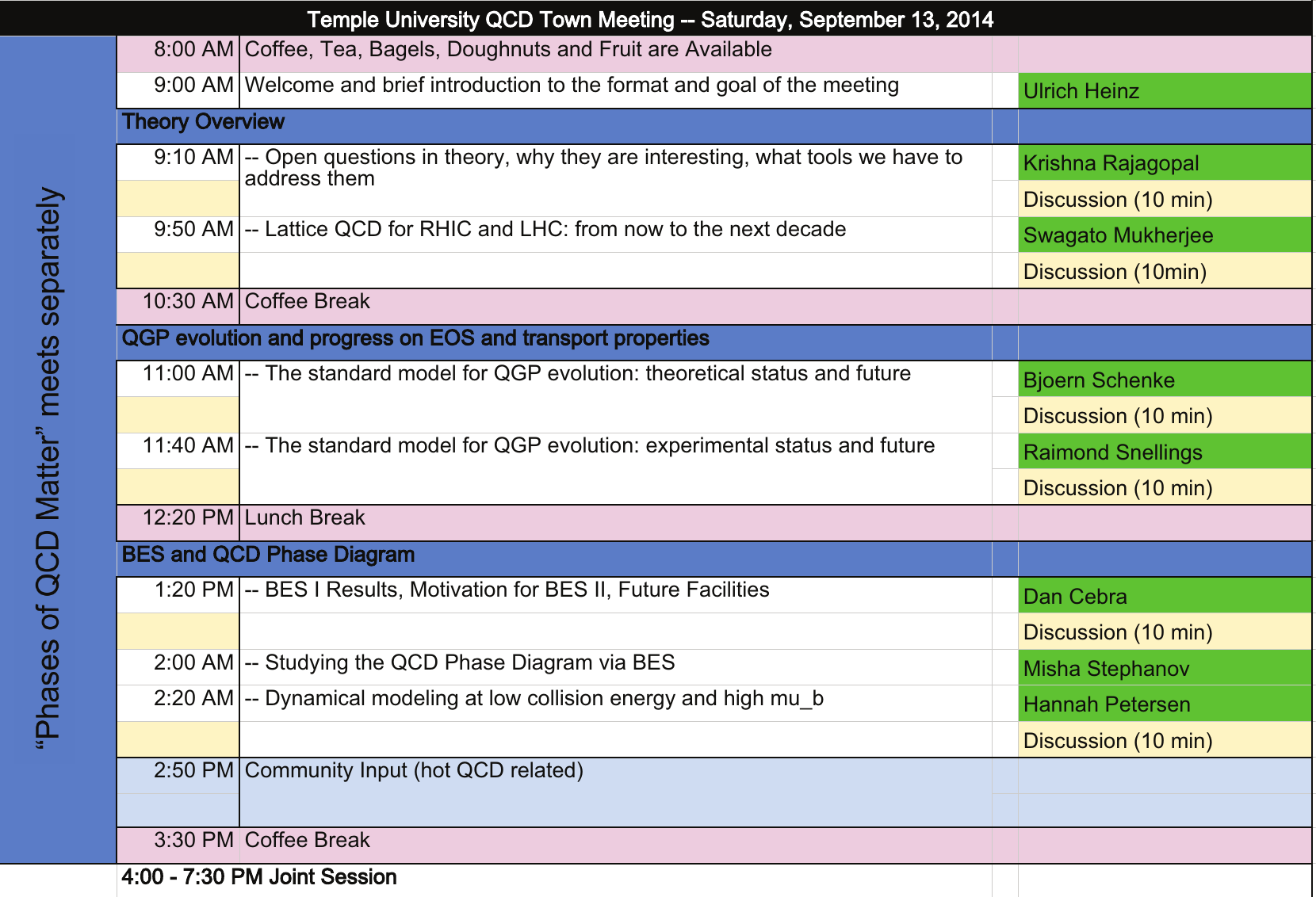}
\caption{Agenda of the ``Phases of QCD Matter'' parallel session, part I (Saturday)}
\end{figure}

\newpage

\begin{figure}[!t]
\includegraphics[width=\textwidth]{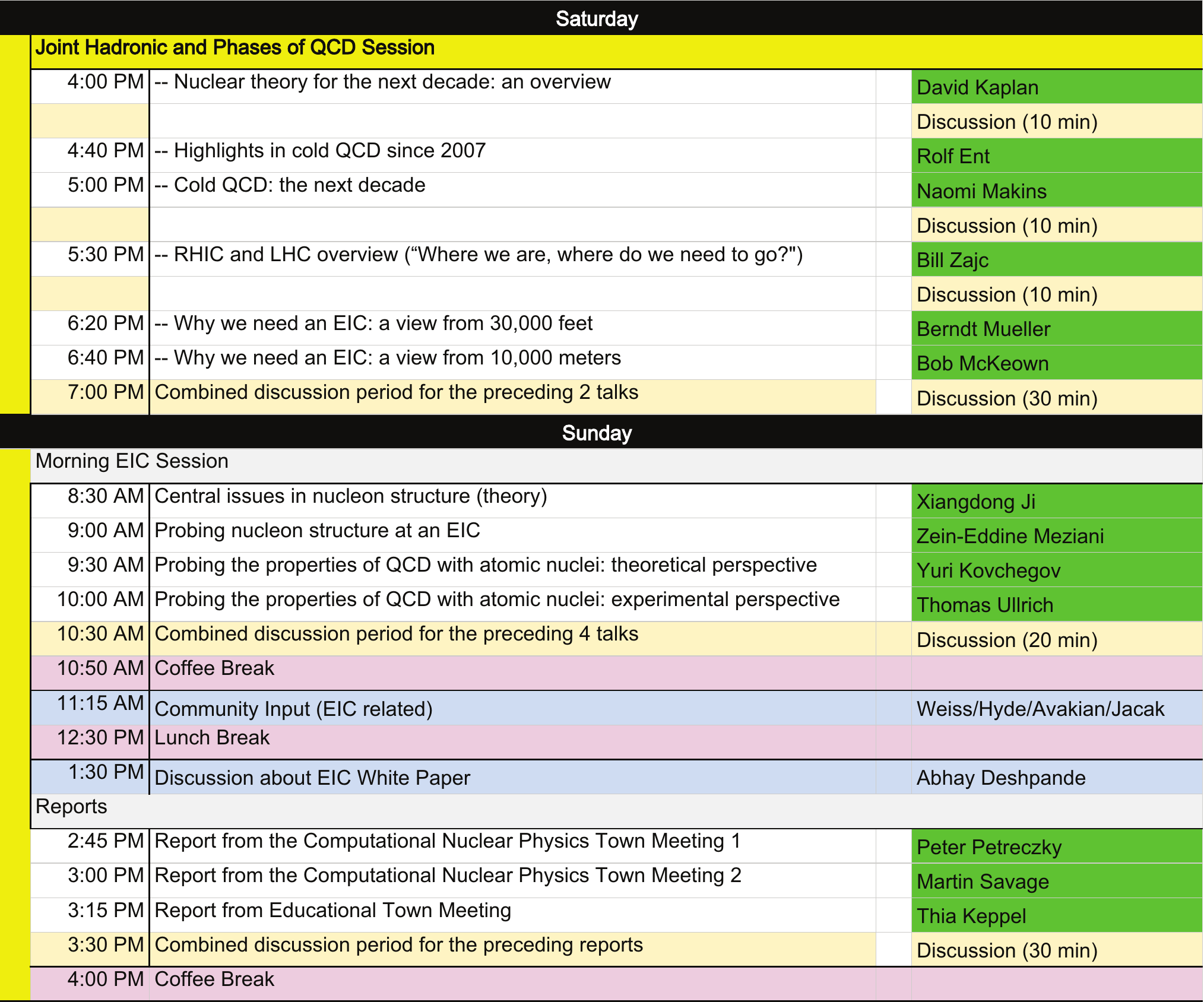}
\caption{Agenda of the joint session (Saturday/Sunday)}
\end{figure}

\newpage

\begin{figure}[!t]
\includegraphics[width=\textwidth]{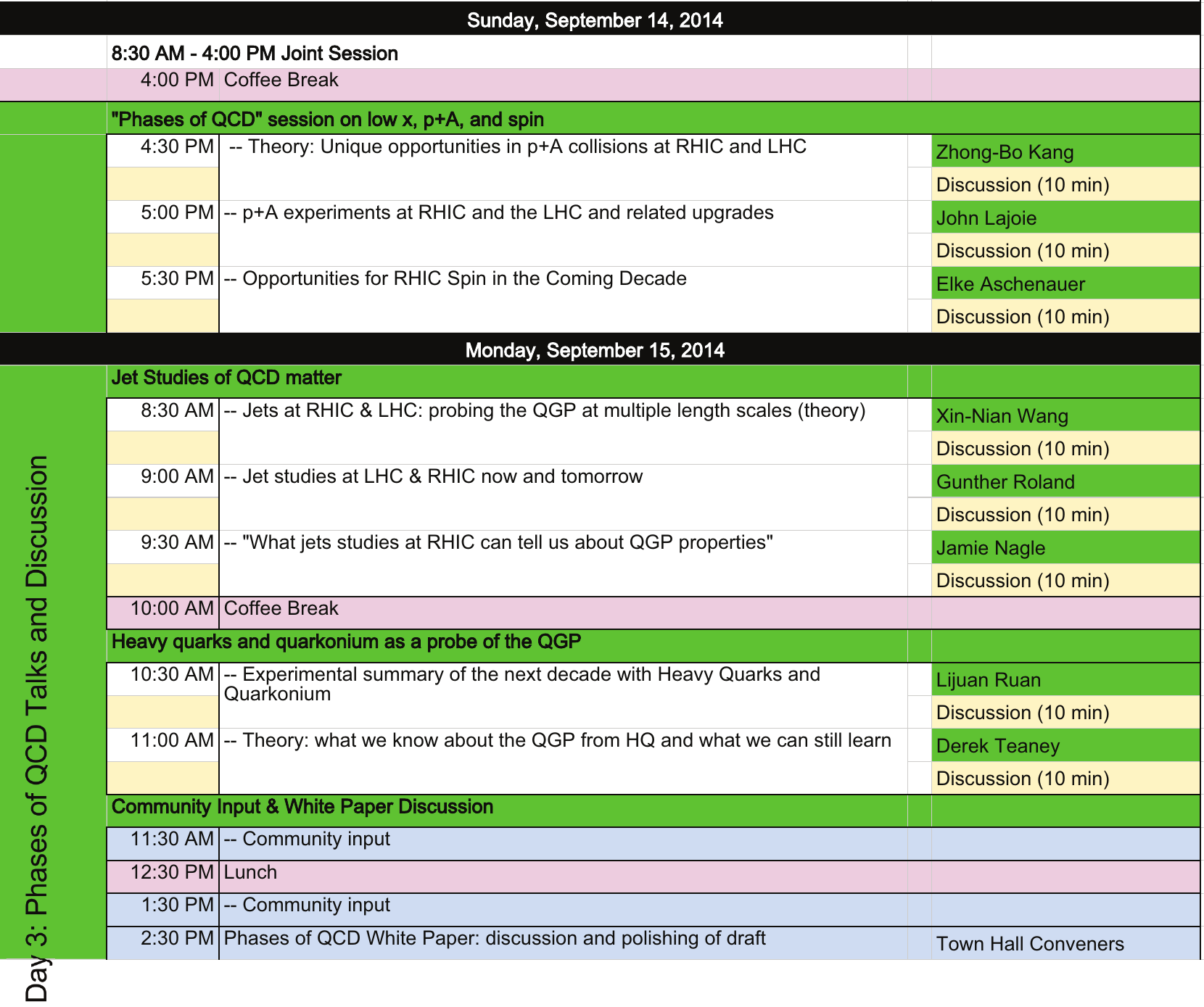}
\includegraphics[width=\textwidth]{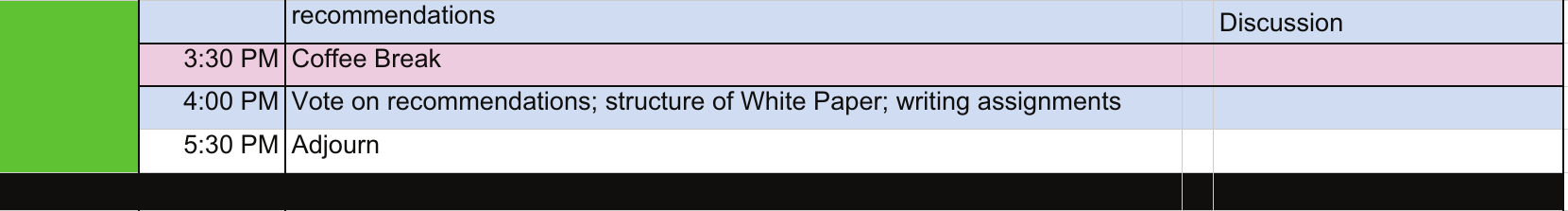}
\caption{Agenda of the ``Phases of QCD Matter'' parallel session, part II (Sunday/Monday)}
\end{figure}
\clearpage



\section*{Supplementary material}

# 6 Appendix: Agenda of the "Phases of QCD Matter" and joint sessions at the QCD Town Meeting

| Temple University QCD Town Meeting -- Saturday, September 13, 2014 | | | |
|---|---|---|---|
| | 8:00 AM | Coffee, Tea, Bagels, Doughnuts and Fruit are Available | |
| | 9:00 AM | Welcome and brief introduction to the format and goal of the meeting | Ulrich Heinz |
| | **Theory Overview** | | |
| | 9:10 AM | -- Open questions in theory, why they are interesting, what tools we have to address them | Krishna Rajagopal |
| | | | Discussion (10 min) |
| | 9:50 AM | -- Lattice QCD for RHIC and LHC: from now to the next decade | Swagato Mukherjee |
| | | | Discussion (10min) |
| | 10:30 AM | Coffee Break | |
| | **QGP evolution and progress on EOS and transport properties** | | |
| "Phases of QCD Matter" meets separately | 11:00 AM | -- The standard model for QGP evolution: theoretical status and future | Bjoern Schenke |
| | | | Discussion (10 min) |
| | 11:40 AM | -- The standard model for QGP evolution: experimental status and future | Raimond Snellings |
| | | | Discussion (10 min) |
| | 12:20 PM | Lunch Break | |
| | **BES and QCD Phase Diagram** | | |
| | 1:20 PM | -- BES I Results, Motivation for BES II, Future Facilities | Dan Cebra |
| | | | Discussion (10 min) |
| | 2:00 AM | -- Studying the QCD Phase Diagram via BES | Misha Stephanov |
| | 2:20 AM | -- Dynamical modeling at low collision energy and high mu_b | Hannah Petersen |
| | | | Discussion (10 min) |
| | 2:50 PM | Community Input (hot QCD related) | |
| | 3:30 PM | Coffee Break | |
| 4:00 - 7:30 PM Joint Session | | | |

Figure 14: Agenda of the "Phases of QCD Matter" parallel session, part I (Saturday)

|  | Saturday | |
|---|---|---|
| **Joint Hadronic and Phases of QCD Session** | | |
| 4:00 PM | -- Nuclear theory for the next decade: an overview | David Kaplan |
| | | Discussion (10 min) |
| 4:40 PM | -- Highlights in cold QCD since 2007 | Rolf Ent |
| 5:00 PM | -- Cold QCD: the next decade | Naomi Makins |
| | | Discussion (10 min) |
| 5:30 PM | -- RHIC and LHC overview ("Where we are, where do we need to go?") | Bill Zajc |
| | | Discussion (10 min) |
| 6:20 PM | -- Why we need an EIC: a view from 30,000 feet | Berndt Mueller |
| 6:40 PM | -- Why we need an EIC: a view from 10,000 meters | Bob McKeown |
| 7:00 PM | Combined discussion period for the preceding 2 talks | Discussion (30 min) |
|  | Sunday | |
| Morning EIC Session | | |
| 8:30 AM | Central issues in nucleon structure (theory) | Xiangdong Ji |
| 9:00 AM | Probing nucleon structure at an EIC | Zein-Eddine Meziani |
| 9:30 AM | Probing the properties of QCD with atomic nuclei: theoretical perspective | Yuri Kovchegov |
| 10:00 AM | Probing the properties of QCD with atomic nuclei: experimental perspective | Thomas Ullrich |
| 10:30 AM | Combined discussion period for the preceding 4 talks | Discussion (20 min) |
| 10:50 AM | Coffee Break | |
| 11:15 AM | Community Input (EIC related) | Weiss/Hyde/Avakian/Jacak |
| 12:30 PM | Lunch Break | |
| 1:30 PM | Discussion about EIC White Paper | Abhay Deshpande |
| Reports | | |
| 2:45 PM | Report from the Computational Nuclear Physics Town Meeting 1 | Peter Petreczky |
| 3:00 PM | Report from the Computational Nuclear Physics Town Meeting 2 | Martin Savage |
| 3:15 PM | Report from Educational Town Meeting | Thia Keppel |
| 3:30 PM | Combined discussion period for the preceding reports | Discussion (30 min) |
| 4:00 PM | Coffee Break | |

Figure 15: Agenda of the joint session (Saturday/Sunday)

| | Sunday, September 14, 2014 | |
|---|---|---|
| | 8:30 AM - 4:00 PM Joint Session | |
| | 4:00 PM Coffee Break | |
| | **"Phases of QCD" session on low x, p+A, and spin** | |
| | 4:30 PM -- Theory: Unique opportunities in p+A collisions at RHIC and LHC | Zhong-Bo Kang |
| | | Discussion (10 min) |
| | 5:00 PM -- p+A experiments at RHIC and the LHC and related upgrades | John Lajoie |
| | | Discussion (10 min) |
| | 5:30 PM -- Opportunities for RHIC Spin in the Coming Decade | Elke Aschenauer |
| | | Discussion (10 min) |
| | **Monday, September 15, 2014** | |
| | **Jet Studies of QCD matter** | |
| | 8:30 AM -- Jets at RHIC & LHC: probing the QGP at multiple length scales (theory) | Xin-Nian Wang |
| | | Discussion (10 min) |
| | 9:00 AM -- Jet studies at LHC & RHIC now and tomorrow | Gunther Roland |
| | | Discussion (10 min) |
| | 9:30 AM -- "What jets studies at RHIC can tell us about QGP properties" | Jamie Nagle |
| | | Discussion (10 min) |
| | 10:00 AM Coffee Break | |
| | **Heavy quarks and quarkonium as a probe of the QGP** | |
| | 10:30 AM -- Experimental summary of the next decade with Heavy Quarks and Quarkonium | Lijuan Ruan |
| | | Discussion (10 min) |
| | 11:00 AM -- Theory: what we know about the QGP from HQ and what we can still learn | Derek Teaney |
| | | Discussion (10 min) |
| | **Community Input & White Paper Discussion** | |
| | 11:30 AM -- Community input | |
| | 12:30 PM Lunch | |
| | 1:30 PM -- Community input | |
| | 2:30 PM Phases of QCD White Paper: discussion and polishing of draft | Town Hall Conveners |
| | recommendations | Discussion |
| | 3:30 PM Coffee Break | |
| | 4:00 PM Vote on recommendations; structure of White Paper; writing assignments | |
| | 5:30 PM Adjourn | |

Day 3: Phases of QCD Talks and Discussion

Figure 16: Agenda of the "Phases of QCD Matter" parallel session, part II (Sunday/Monday)

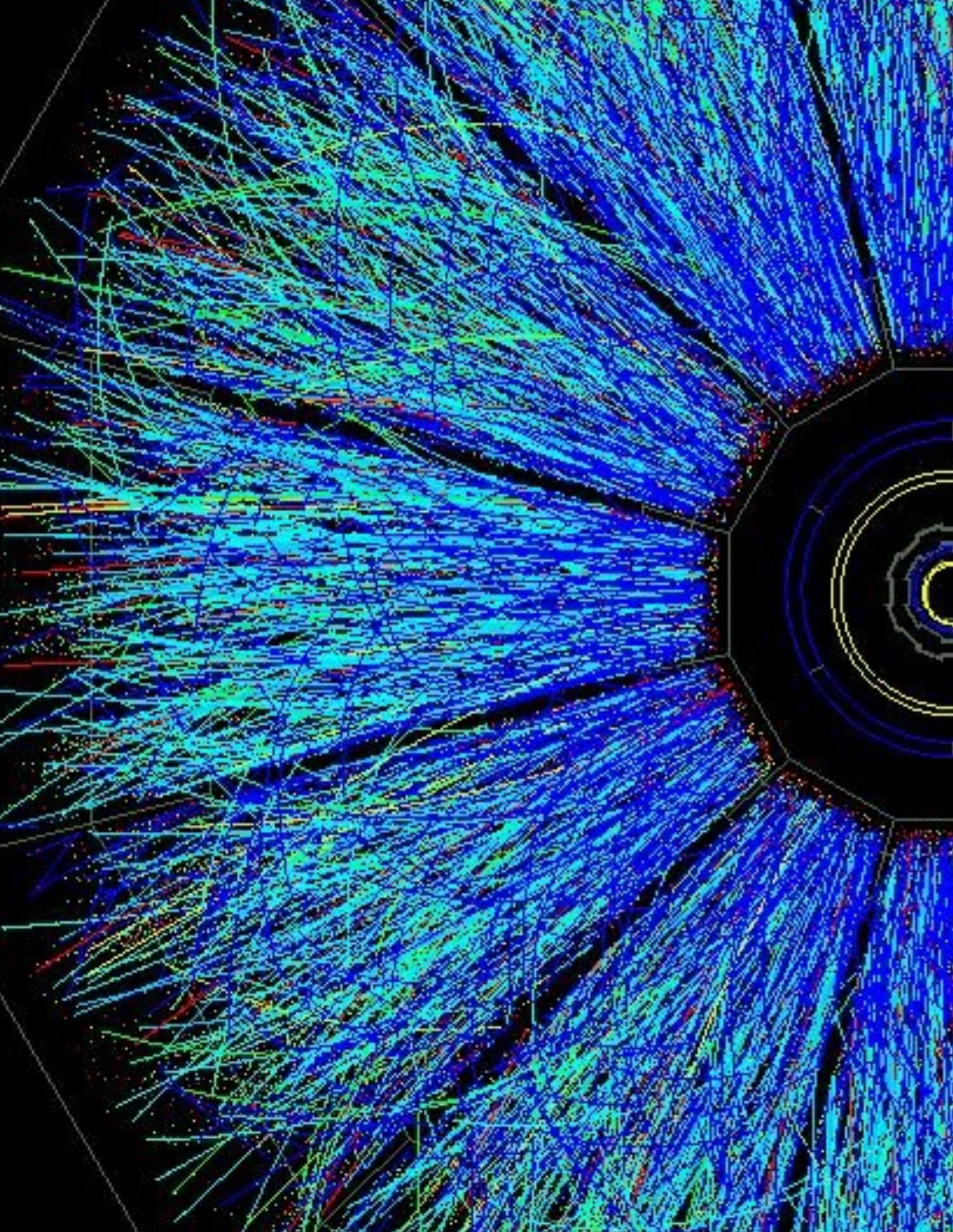



\end{document}